\documentclass[twoside,12pt]{article}

\usepackage{epsfig}
\usepackage{epstopdf}      
\usepackage{amsmath,amssymb}
\usepackage{ulem}
\usepackage{xspace}
\usepackage{slashed}
\usepackage{pbox}
\def\p{\vec p}

\newcommand{\be}{\begin{equation}}
\newcommand{\ee}{\end{equation}}
\newcommand{\bea}{\begin{eqnarray}}
\newcommand{\eea}{\end{eqnarray}}

\newcommand{\add}[1]{\textbf{#1}}

\renewcommand{\emph}[1]{\textit{#1}}

\newcommand{\half}{\frac{1}{2}}

\newcommand{\dd}{\text{d}}

\renewcommand{\Re}{\text{Re}}
\renewcommand{\Im}{\text{Im}}

\newcommand{\kv}{\vec{k}}
\newcommand{\pv}{\vec{p}}
\newcommand{\qv}{\vec{q}}

\newcommand{\calA}{\mathcal{A}} \newcommand{\calC}{\mathcal{C}}
 
\newcommand{\calF}{\mathcal{F}} \newcommand{\calG}{\mathcal{G}}
 \newcommand{\calK}{\mathcal{K}}
\newcommand{\calL}{\mathcal{L}} \newcommand{\calM}{\mathcal{M}}
 \newcommand{\calO}{\mathcal{O}}
 
 \newcommand{\calT}{\mathcal{T}}

\newcommand{\MeV}{\ensuremath{\mathrm{MeV}}}

\newcommand{\eftnopi}{EFT(${\pi\hskip-0.55em /}$)\xspace}
\newcommand{\NXLO}[1]{N\ensuremath{{}^{#1}}LO\xspace}
\newcommand{\NtwoLO}{\NXLO{2}}
\newcommand{\NthreeLO}{\NXLO{3}}

\newcommand{\chiPT}{$\chi$PT\xspace}

\newcommand{\wave}[3]{\ensuremath{{}^{#1}{#2}_{#3}}\xspace}
\newcommand{\oneS}{\wave{1}{S}{0}}
\newcommand{\twoS}{\wave{2}{S}{\half}}
\newcommand{\threeS}{\wave{3}{S}{1}}
\newcommand{\fourS}{\wave{4}{S}{\frac{3}{2}}}
\newcommand{\oneP}{\wave{1}{P}{1}}
\newcommand{\threePzero}{\wave{3}{P}{0}}
\newcommand{\threePone}{\wave{3}{P}{1}}
\newcommand{\twoPone}{\wave{2}{P}{\half}}
\newcommand{\twoPthree}{\wave{2}{P}{\frac{3}{2}}}
\newcommand{\fourPone}{\wave{4}{P}{\half}}
\newcommand{\fourPthree}{\wave{4}{P}{\frac{3}{2}}}

\newcommand{\ND}{N^\dagger}
\newcommand{\CSing}{{\cal C}_0^{(^1 \! S_0)}}
\newcommand{\CTrip}{{\cal C}_0^{(^3 \! S_1)}}

\newcommand{\aSing}{a^{(^1 \! S_0)}}

\newcommand{\rTrip}{r^{(^3 \! S_1)}}
\newcommand{\VS}{\vec{\sigma}}
\newcommand{\VT}{\vec{\tau}}
\newcommand{\LRd}{\overset{\leftrightarrow}{D}}
\newcommand{\CA}{{\cal C}^{(^3 \! S_1-^1 \! P_1)}}
\newcommand{\CB}{{\cal C}^{(^1 \! S_0-^3 \! P_0)}_{(\Delta I=0)}}
\newcommand{\CC}{{\cal C}^{(^1 \! S_0-^3 \! P_0)}_{(\Delta I=1)}}
\newcommand{\CD}{{\cal C}^{(^1 \! S_0-^3 \! P_0)}_{(\Delta I=2)}}
\newcommand{\CE}{{\cal C}^{(^3 \! S_1-^3 \! P_1)}}

\newcommand{\gA}{g^{(^3 \! S_1-^1 \! P_1)}}
\newcommand{\gB}{g^{(^1 \! S_0-^3 \! P_0)}_{(\Delta I=0)}}
\newcommand{\gC}{g^{(^1 \! S_0-^3 \! P_0)}_{(\Delta I=1)}}
\newcommand{\gD}{g^{(^1 \! S_0-^3 \! P_0)}_{(\Delta I=2)}}
\newcommand{\gE}{g^{(^3 \! S_1-^3 \! P_1)}}

\newcommand{\HThree}{\ensuremath{{}^3\text{H}}\xspace}
\newcommand{\HeThree}{\ensuremath{{}^3\text{He}}\xspace}
\newcommand{\HeFour}{\ensuremath{{}^4\text{He}}\xspace}

\begin{document}

\title{ \vspace{1cm} The Theory of Parity Violation in Few-Nucleon Systems}
\author{M. R.\ Schindler$^1$ and R. P.\ Springer$^2$\\
\\
$^1$Department of Physics and Astronomy, University of South Carolina,\\ Columbia, SC, USA\\
$^2$Department of Physics, Duke University, Durham, NC, USA}
\date{January 31, 2013}
\maketitle

\begin{abstract} 

We review recent progress in the theoretical description of hadronic parity violation in few-nucleon systems. 
After introducing the different methods that have been used to study parity-violating observables we discuss the available calculations for reactions with up to five nucleons. 
Particular emphasis is put on effective field theory calculations where they exist, but earlier and complementary approaches are also presented.  We hope this  review will serve as a guide for those who wish to know what calculations are available and what further calculations need to be completed before we can claim to have a comprehensive picture of parity violation in few nucleon systems.

\end{abstract}

\newpage

\section{Introduction}\label{sec:intro}

Important advancements have occurred since the last extensive theoretical review of low energy parity violation in few nucleon systems \cite{RamseyMusolf:2006dz}. For earlier reviews see Refs.~\cite{Adelberger:1985ik,Haeberli:1995uz,Desplanques:1998ak}.
 The seminal analysis of parity-violating (PV) couplings in a meson-exchange model \cite{Desplanques:1979hn} (referred to as ``DDH'' in the following)
in 1980 ushered in several decades in which theorists calculated systems of interest using this model, and experimentalists proposed, designed, and interpreted their experiments in terms of the ``DDH couplings" that characterize the DDH model. 
Meanwhile, in the late 80s and early 90s, a systematic effective field theory (EFT) treatment of  interactions among nucleons was developed \cite{Weinberg:1990rz,Weinberg:1991um,Weinberg:1992yk}.
Using EFT methods it was possible to (i) identify all operators consistent with QCD and (ii) order them in a power counting scheme so that the precision of a given calculation is predictable.  
The EFTs are written in the language of nucleons; since matching to a quark-level QCD calculation is not possible at present, the EFTs contain a set of unknown parameters that must be determined by experiment or lattice data.  
These theories have enjoyed considerable success in the two light-quark sector (SU(2) flavor QCD).  Various versions such as (heavy) baryon chiral perturbation theory ((H)B$\chi$PT), pionless EFT (\eftnopi), and chiral EFTs for two and more nucleons were developed and applied.  For review articles on these developments see, e.g., Refs.~\cite{Georgi:1994qn,Bernard:1995dp,vanKolck:1999mw,Beane:2000fx,Bedaque:2002mn,Bernard:2006gx,Bernard:2007zu,Epelbaum:2008ga,Platter:2009gz,Scherer:2009bt,Machleidt:2011zz}. In the realm of parity violation, a compendium of PV operators in the single nucleon sector was provided in  Ref.~\cite{Kaplan:1992vj}. 
It was at this stage that serious efforts were made to apply PV EFTs to few-nucleon observables.  The shortcomings of the DDH model were identified and addressed in Ref.~\cite{Zhu:2004vw}  and the community continues to move towards a consistent, unifying description of few-body hadronic PV observables.   
An important intermediate stage involves the so-called ``hybrid" approach that combines model and EFT treatments.

In this review we attempt to compile calculations that have been performed since these developments. We hope the reader will obtain a sense of what has been  and what still needs to be done. In particular, where possible we provide a translation so that the results expressed in terms of one parameter set can be understood in terms of parameter sets used in other calculations.  There are  $\sim $ 5-6 (depending upon energy range) independent unknown PV low-energy constants,  and such a basis change might not seem insurmountable.  However, along with different sets of chosen basis operators a number of scale-dependent restrictions make the comparison among some calculations problematic.

In parallel with theoretical advances, new experimental opportunities continue to arise. A naive estimate of the ratio of parity-violating to parity-conserving couplings suggests that the PV components are typically suppressed by a factor of $10^{-6}$ to $10^{-7}$. The detection of the tiny PV asymmetries in few-nucleon systems requires high luminosity, high control of systematics, and very clean systems.  New ideas are needed to discover affordable ways of attaining these conditions. 
The development of high-intensity neutron sources has made it possible to perform experiments  with previously unmatched precision. 
They provide the opportunity to obtain information on hadronic parity violation from few-nucleon systems, for which the relation to underlying nucleon-nucleon (NN) interactions can be more straightforwardly established than in more complex nuclei. 
Two examples are given by the NPDGamma experiment at the Spallation Neutron Source (SNS) at Oak Ridge National Laboratory \cite{Gericke:2011zz} and the measurement of neutron spin rotation in \HeFour at NIST \cite{Snow:2011zza}. 
The further development of high-intensity photon sources presents another opportunity to study parity violation in few-nucleon systems through breakup reactions such as $\vec{\gamma} d \to np$. This possibility is currently being explored at the High Intensity Gamma Source at the Triangle Universities Nuclear Laboratory.

The PV component of nucleon interactions is the manifestation of weak quark-quark interactions on the hadronic level. 
The search for PV nucleon forces \cite{Tanner:1957zz} began shortly after the confirmation of parity violation in the beta decay of $^{60}\text{Co}$ \cite{Wu:1957my} and $\mu$ decay \cite{Garwin:1957hc}. Since at that time the exact form of the weak interactions was not yet determined, hadronic parity violation in nucleons was considered to be an additional test of proposed weak interaction theories. The corresponding experiments are extremely challenging because of the presence of dominant strong and electromagnetic effects, so detailed information on the weak interactions continued to be obtained from leptonic and semi-leptonic processes instead.
While we now have a detailed understanding of the quark-level interactions responsible for PV effects, an interpretation at the nucleonic level is complicated by nonperturbative quantum chromodynamics (QCD).
The motivation to study hadronic parity violation has therefore taken on an additional role.  Not only do we want to make sure that we understand PV in nuclear systems for their own sake, but now we also wish to use PV effects to gain a better understanding of how the strong interactions lead to the observed nonperturbative phenomena at low energies. 
 
We will restrict our discussion to few-body processes. The difficulty of tiny PV effects seen in few-body systems can be circumvented by utilizing complex nuclei, in which near-degenerate energy levels of opposite parity and the admixture of large PC amplitudes can lead to enhancements of the PV effects by several orders of magnitude; see, e.g., Ref.~\cite{Bowman:1989ci}. However, it is theoretically difficult to relate many-nucleon systems to the underlying two-, three-, and few-nucleon interactions in a systematic way without introducing uncontrolled errors. These uncertainties are significantly reduced when restricting the considered processes to systems involving at most five nucleons. While one motivation for this discussion is to hope that 
it can be used as a step in the process of finally understanding hadronic and few-body PV processes in terms of quark degrees of freedom, at the moment that possibility exists in
the future and requires progress in both lattice and analytic efforts.  Instead, we
recognize that achieving the above goal will require a consistent analysis of a suite
of observables and this is what we review here. A quark-level calculation must
be consistent with these results.  In the end, it is likely that understanding PV processes in complex
nuclei will require progress in quark-level (lattice), few-nucleon (effective field
theory), and many-body techniques.

A better understanding of hadronic parity violation might also be able to shed light on another problem involving the interplay of strong and weak interactions. In the strangeness-changing nonleptonic decays of hadrons, amplitudes with $\Delta I=1/2$ are strongly enhanced over those with $\Delta I=3/2$. It is not clear yet whether this enhancement is related to strangeness or whether it is related to some underlying QCD dynamics. Similar isospin patterns found in the strangeness-conserving hadronic weak interaction might point to a better understanding of this phenomenon \cite{RamseyMusolf:2006dz}.

This review is organized as follows:
In Sec.~\ref{sec:obs} we describe the observables that are used to study hadronic parity violation.
The different theoretical methods that have been applied to calculate these observables are discussed in Sec.~\ref{sec:methods}. We also describe how results in the different approaches can in principle be related to one another as well as potential pitfalls in such translations. 
In the following sections we describe available calculations for systems with an increasing number of nucleons:
Section \ref{sec:1N} deals with one-nucleon systems, while two-nucleon observables are discussed in detail in Sec.~\ref{sec:2N}.
We then turn to three-nucleon systems in Sec.~\ref{sec:threeN}, while Sec.~\ref{sec:fourN} contains a discussion of the available calculations involving four and five nucleons. 
A summary and outlook are given in Sec.~\ref{concl}.

%%%%%%%%%%%%%%%%%%%%%%%%%%%%%%%%%%%%%%%%%%%%%%%%%%%%%%%%%%%%%

\section{Observables}\label{sec:obs}

Hadronic parity violation in few-nucleon systems can be detected using a particular class of observables, which we briefly discuss here before considering individual systems. As explained above, PV interactions are typically suppressed by factors of $10^{-6}-10^{-7}$ compared to PC ones in few-nucleon systems. In order to detect their effects, it is necessary to use polarized beams or targets. Most of the observables we will discuss in this review have in common that they are sensitive to correlations between the oriented spin and a momentum, $\vec{\sigma}\cdot \vec{p}$; they are pseudoscalar observables. 
In terms of transition amplitudes, they correspond to the interference terms of PC and PV matrix elements. The pseudoscalar observables include longitudinal asymmetries, angular asymmetries, and spin rotation angles. We will briefly mention here some measurements that have been made, but will present them again when the status of the theory for that measurement is dicussed in later sections. Note that in principle any correlation resulting in a nonzero $\vec \sigma\cdot \vec{p}$ between the available spin and momentum degrees of freedom provides information on the PV interactions. Here we discuss those observables that are relevant for few-nucleon systems and that are experimentally feasible.

The longitudinal analyzing power $A_L$ is used to study the scattering of a polarized beam on an unpolarized target, 
\be\label{def:AL}
A_L = \frac{\sigma_+ - \sigma_-}{\sigma_+ + \sigma_-},
\ee
where $\sigma_+$ ($\sigma_-$) is the total cross section for the scattering of a beam with positive (negative) helicity. The longitudinal analyzing power has been measured at various energies in $\pv p$   \cite{Nagle:1978vn,Eversheim:1991tg,Balzer:1980dn,Balzer:1985au,Kistryn:1987tq,Berdoz:2002sn,Yuan:1986yj} as well as $\pv \alpha$ \cite{Lang:1985jv} scattering, and Ref.[8] provides an upper limit for this asymmetry in $\vec p d$ scattering.
Cross sections are typically not measured over the full solid angle, and the particular angular range has to be taken into account in the comparison between theory and experiment.

Parity-violating interactions can also be studied in angular asymmetries. In the radiative capture of polarized neutrons on proton, deuteron, or \HeThree targets the angular distribution of the outgoing photon with respect to the polarization of the incoming neutron corresponds to a correlation $\vec{\sigma}_n\cdot \kv_\gamma$.  As an example, the PV asymmetry $A_\gamma$ in radiative neutron capture on protons, $\vec{n}p\to d\gamma$, is defined by
\be\label{Agamma}
\frac{1}{\Gamma}\frac{d\Gamma}{d\cos \theta} = 1 +A_\gamma \cos \theta \,,
\ee
with $\Gamma$ the width and $\theta$ the angle between $\vec{\sigma}_n$ and $\kv_\gamma$.
This is the measurement currently underway at the NPDGamma experiment at the SNS \cite{Gericke:2011zz}. 
Similarly, in the charge exchange reaction $\vec{n}\HeThree\to \HThree p$, a PV asymmetry can be related to the correlation $\vec{\sigma}_n\cdot \pv_p$ between the direction of the outgoing proton with respect to the incoming neutron polarization. 

Radiative capture of unpolarized neutron beams can also be used to study PV effects, since the outgoing photons will acquire a circular polarization from the PV interaction, 
\be\label{def:Pgamma}
P_\gamma = \frac{\sigma_{\gamma +} - \sigma_{\gamma -}}{\sigma_{\gamma +} + \sigma_{\gamma -}}\,, 
\ee
with $\sigma_{\gamma\pm}$ the total cross section for photons with $\pm$ helicity. In the two-nucleon system, the circular polarization $P_\gamma$ at threshold (measured from the asymmetry in $np \to d \vec \gamma$) is equal to the helicity asymmetry $A_L^\gamma$ in deuteron breakup with circularly polarized photons, $\vec{\gamma} d \to np$, for exactly reversed kinematics. Since the measurement of the outgoing circular polarization is very challenging, determination of  $A_L^\gamma$ might be more experimentally feasible
than $P_\gamma$.  Note that the observable $A_L^\gamma$ is distinct and independent from the observable $A_\gamma$ in Eq.~\eqref{Agamma}. 

Neutron beams polarized perpendicularly to the beam direction give access to another observable. When these beams traverse an unpolarized target, PV interactions will induce a rotation of the neutron spin around the beam direction, with the rotation angle proportional to the forward scattering amplitude.  As an illustration, consider the case of a beam of neutrons with very low energy interacting with a spin-zero target. Describing the low-energy neutrons with a plane wave, the beam picks up a phase factor as it passes through the target. This phase factor is related to the index of refraction $n$ of the target medium. The accumulated phase for a target thickness $l$ is 
\be
  \varphi=\Re(n-1)kl,
\ee
with $k$ the magnitude of the incoming wave vector. The index of refraction $n$ can be related to the forward scattering amplitude $\calM$, resulting in \cite{Fermi,Griesshammer:2011md}
\be
\varphi= \rho l \frac{\mu}{k} \text{Re}(\calM)
\ee
where $\rho$ is the density of scattering centers in the target and $\mu$ is the reduced mass of the beam-target system. (Note that conventions for the relation between $n$ and the amplitude $\calM$ differ in the literature.)  A perpendicularly polarized beam can be represented as a linear combination of positive- and negative-helicity states. The PV interactions between beam and target result in different phase factors for the different helicity states, $\phi_+$ and $\phi_-$ respectively, causing  the neutron spin to be rotated by $\phi_\text{PV} = \phi_+ - \phi_-$. The spin rotation angle per unit length is given by
\be\label{def:spinrotangle}
\frac{1}{\rho}\,\frac{d \phi_\text{PV}}{d l} = \frac{\mu}{k}\, \Re\left(\calM_+ - \calM_- \right),
\ee
with $\calM_\pm$ the forward scattering amplitude for $\pm$-helicity neutrons. In the case of non-zero target spin, the target presents a statistical mixture of spin orientations to the beam.

A measurement of neutron spin rotation on $\HeFour$ performed at NIST \cite{Snow:2011zza} resulted in an upper bound on the rotation angle consistent with that found in Ref.~\cite{markoff_thesis}. As discussed in Secs.~\ref{sec:npspinrot} and \ref{sec:ndspinrot} spin rotation experiments in proton and deuteron targets would provide important complementary information. They have not been performed to date, but have been considered \cite{markoff_thesis,Markoff:2005,Snow}.
  
An unpolarized beam traversing a target can also pick up a net longitudinal polarization from PV interactions. This polarization $P_n$ is related to
\be
\Im\left(\calM_+ - \calM_- \right).
\ee
Using the optical theorem, this observable is equivalent to the longitudinal asymmetry $A_L$. For a given target length, it also tends to be several orders of magnitude smaller than the spin rotation angle and thus not experimentally accessible.

There is a further quantity that can be used to study PV nucleon interactions. The matrix element of the electromagnetic current evaluated between nuclear and/or nucleon initial and final states can be parameterized in terms of electric and magnetic form factors. Lifting the restriction on parity and time reversal symmetry gives rise to an electric dipole term. Keeping time reversal conservation while allowing parity to be violated yields the so-called anapole form factor \cite{Zeldovich:1957,Flambaum:1980sb,Flambaum:1984}, with the anapole moment its value at zero momentum transfer. Anapole moments can be measured in atomic systems, in which atomic electrons interact with the anapole moment of the nucleus.  In this case atomic systems are used to study properties of nuclei and nucleon-nucleon interactions. Anapole moments could also be observed in PV electron scattering. While the anapole moment itself is not a gauge-invariant quantity, it dominates over other $Z^0$-induced effects in heavy nuclei \cite{Haxton:1989ap}. Indeed, so far only measurements of anapole moments in heavy nuclei have been performed, which complicates the interpretation in terms of nucleon-nucleon interactions.

%%%%%%%%%%%%%%%%%%%%%%%%%%%%%%%%%%%%%%%%%%%%%%%%%%%%%%%%%%%%%

\section{Methods}\label{sec:methods}

\subsection{ Effective field theories (EFTs)}
The story of progress in physics is really a story of effective
theories. Newton's laws are an effective theory of a fully relativistic
and quantum mechanical treatment. Thermodynamics is an effective theory of 
statistical mechanics. The underlying
premise is that one does not need to understand physics at all  scales in order
to make predications of long-distance phenomena \cite{Weinberg:1978kz,Georgi:1994qn,Weinberg:1995mt}.
All that is necessary is to retain the
physics to which the expected measurement will be sensitive.  
For example, the beta decay of a nucleus can be described without
the need to include the dynamics of the top quark.  Instead, this
short-distance (or sometimes simply unknown) physics is encoded in
coefficients of operators that involve the fields that are dynamical
at the energies of interest.  Effective theories without the short-distance dynamics will only be valid
so long as the system possesses a separation of scales.  The
ratio of these disparate scales will form the small expansion
parameter used to cast the predictions of the effective theory 
in a perturbative expansion.  

\begin{figure}
\begin{center}
\includegraphics[width=0.5\textwidth]{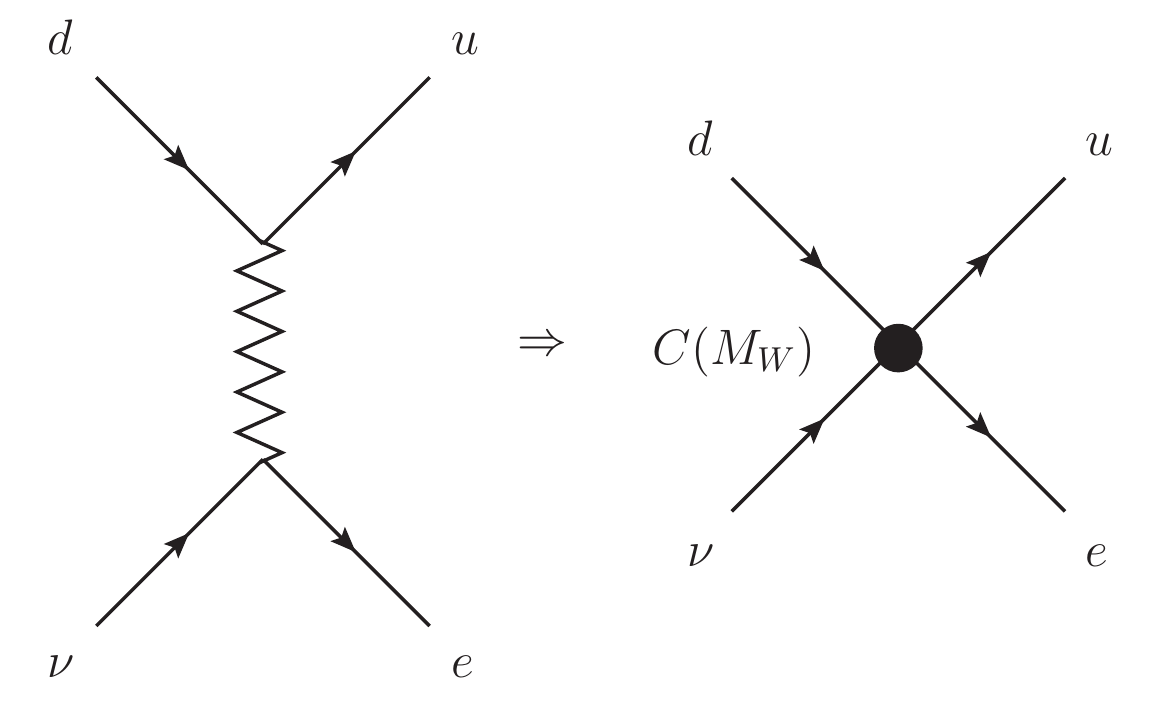}
\end{center}
\caption{Four-Fermion interactions in the underlying and effective theories at the quark level.  \label{fig:udW}}
\end{figure}

Effective {\sl Field} Theories (EFTs) are effective theories that incorporate the advantages of (quantum) field theoretic descriptions, such as gauge invariance and the consistent coupling to external fields. They  are useful for (i) simplifying 
calculations in theories we do know, (ii) making 
predictions from theories we do know but cannot solve and (iii)  probing 
theories we do not know.  
An example  of (i) is Fermi's contact term at the quark-level in Fig.~\ref{fig:udW}.  

We ``know'' the underlying Standard Model (SM) of weak interactions and
can calculate the W-exchange diagram shown on the left side of Fig.~\ref{fig:udW}. However, for something like
low energy $\beta$ decay, the dynamics of the W particle is of no
consequence at moderate precision.  Instead, it can be formally
``integrated out'' of the path integral, leaving a four-Fermi contact
operator and a coefficient, a so-called low-energy constant (LEC), that scales as $1/M_W^2$.  That coefficient
is known because the underyling SM is known and can be matched to
the EFT of Fermi's contact term.  Greater precision is available by
going to higher order in the SM and matching to higher  order operators in the EFT.
In general, the contributions from the SM can be reproduced to a given order by terms of the form $\sum_i  C_i O_i$ where
the $O_i$ are of increasing dimension and decreasing relevance.
Such an EFT formulation, where heavy propagators are  replaced by point couplings, can considerably simplify calculating many-loop QCD corrections, for example.

An example of  an EFT of type (ii) is QCD.  For case (ii) or (iii), it may be possible to model the physics with an assumed
set of particles and interactions, but it may not be possible to estimate
how closely that model mimics reality.  An EFT, on the other hand,
is built following protocols that allow such estimates to be made.
EFTs rely on symmetries and a power counting.  Known (and/or approximate
and/or assumed) symmetries are built into the EFT.  Fortunately, information
on the  symmetries of a theory is often available even in the absence
of a solved or known theory.  Such is the case for QCD.

The power counting is developed by identifying disparate length scales
in the problem.  For Fermi's contact theory, the power counting is
found by noting that $p_{ext}/M_W$ is a small quantity.  Higher-order operators in the EFT are suppressed by ever-higher powers of
$p_{ext}/M_W$ -- a perturbative expansion has been identified.
When $p_{ext}$ becomes so large  that $p_{ext}/M_W$ is no longer
much less than one, the power counting fails and the EFT is no 
longer valid.  Within the realm of the EFT's validity, however, there
is a built-in estimate of corrections to a prediction made to
$(p_{ext}/M_W)^n$: it is on the order of $(p_{ext}/M_W)^{(n+1)}$

\subsection{Pionless EFT: \eftnopi}\label{nopimethod}

For purposes of studying low-energy parity violation among few nucleons,
we need an EFT that includes QCD, electromagnetic, and weak interactions. Since the electromagnetic and weak interactions  can be treated perturbatively in their own coupling constants, the challenge
is, as with parity-conserving observables, to address the complications
of nonperturbative QCD.  Since we do not know how QCD forms hadrons
from quarks and gluons, we instead build an EFT in the language
of the nucleons themselves, and impose the symmetries of QCD
to restrict the form of the EFT Lagrangian.  This EFT will necessarily
be accompanied by coefficients (the analogues of
$C(M_W)$ in Fig.~\ref{fig:udW}), only unlike in Fermi's contact (quark-level) theory, the
coefficients will be unknowns that must be fixed by experiment
or lattice simulations.  However, because this is a field theory,
once a coefficient has been determined from any one observable,
it can be used in the prediction of any other.  

Arguably the simplest theory is one that contains the fewest number of dynamical degrees of freedom.  Gluons and photons are massless, but
the gluons only appear in bound states and so their effect lives in the hadron degrees of freedom. The lowest mass hadrons are the pions; keeping to low enough energies it is possible to treat pions as ``heavy" effective degrees of freedom, and, along with photons, have a complete theory
of pions up to a few MeV \cite{Weinberg:1978kz,Gasser:1983yg,Gasser:1984gg} (see Ref.~\cite{ChPTPrimer} for a pedagogical introduction). 
But the PV observables we wish to describe are those involving nucleons.  At energies well below pion production, one can choose the nucleons and photons as the only dynamical degrees of freedom.   Further, at extremely low energies, the nucleons can be treated as non-relativistic, with corrections to that limit included perturbatively.  It may seem counter-intuitive that we include nucleons as dynamical when they are heavier than the pions that are removed; the concept here is analogous to that of heavy quark effective theory \cite{Isgur:1989vq,Isgur:1989ed}; nucleon-anti-nucleon pairs are not produced in \eftnopi and where the theory is valid all momentum transfers involving nucleon external legs are well below pion excitations.  

\subsubsection{Single-nucleon terms}

\begin{figure}
\begin{center}
\includegraphics[width=0.5\textwidth]{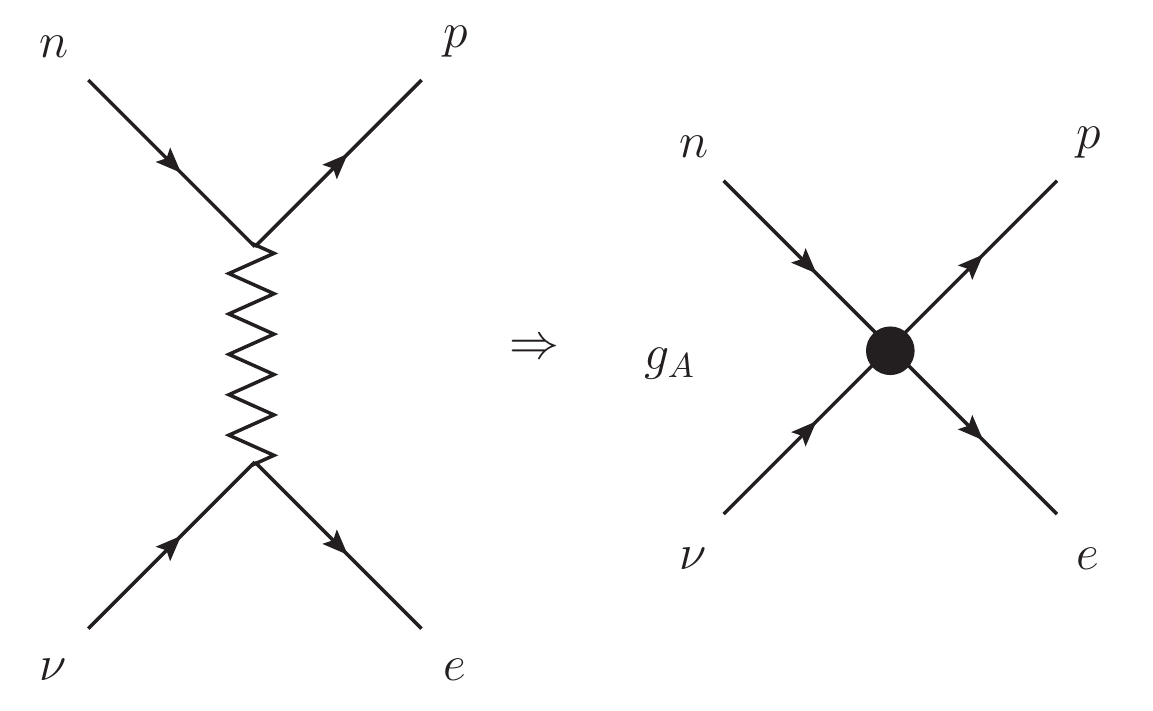}
\end{center}
\caption{Four-Fermion interactions in the underlying and effective theories at the nucleon level.   \label{fig:pnW}}
\end{figure}

For a single-nucleon observable, starting with a kinetic term of the form
\begin{align}\label{rel_kinetic}
i \bar N \slashed{D} N - M (\bar N N) \ \ ,
\end{align}
where $N$ is the SU(2) douplet of nucleons, $D_\mu$ the covariant derivative, and $M$ the nucleon mass,
does not take explicit advantage of
the additional symmetry a very low energy  process provides.  Instead,
with the nucleons non-relativistic, a velocity-dependent phase rotation
is used to explicitly remove the heavy fermion mass:
\begin{align}
N_v=e^{i M (v_\mu x^\mu)} N \ \ , 
\end{align}
where $v_\mu$ is a velocity with $v^2=1$,  in terms of which Eq.~\eqref{rel_kinetic} becomes
 \begin{align}
 i \overline N_v (v \cdot D) N_v \ \ .
 \end{align}
 Now instead of derivatives yielding large momenta $p \sim M v$, they yield
 small ``residual" momenta $k$ such that $k/M$ is a small quantity. 
 The large (but irrelevant for low energy observables) momentum $M v$ has
 been removed from explicitly appearing in the expansion. Higher-order kinetic terms appear as relativistic corrections
 suppressed by powers of $1/M$.  The velocity label will now be suppressed.
One interaction term is just Fermi's contact term, only now in the nucleon rather than quark-level basis (see Fig.~\ref{fig:pnW}).
 For example, with an analogous transformation
\begin{align}
 g_1(q^2) \overline p \gamma_\lambda \gamma_5 n \rightarrow
 g_A \overline p \sigma^i n \ \ , 
\end{align} 
where $g_A = 1.27$ \cite{Beringer:1900zz}  is the nucleon axial-vector coupling constant.

\subsubsection{Parity-conserving two-nucleon terms}

 The most general leading-order (LO)
Lagrangian including two nucleons and imposing parity invariance, but excluding external currents, is
\begin{align}\label{Lag:PCpartial}
\mathcal{L} = & \ND(i \partial_0 + \frac{\vec{\nabla}^2}{2M})N -
\frac{1}{8}{\cal C}_0^{(^1 \! S_0)} (N^T \tau_2 \tau_a \sigma_2 N)^\dagger 
(N^T \tau_2 \tau_a \sigma_2 N) \   \notag\\
&- \frac{1}{8}{\cal C}_0^{(^3 \! S_1)} (N^T \tau_2 \sigma_2 \sigma_i N)^\dagger 
(N^T \tau_2  \sigma_2 \sigma_i N) + \ldots,
\end{align}
where the collection of spin Pauli matrices $\sigma$ and isospin Pauli matrices $\tau$ are
often written as partial wave projection operators $P_a(^1S_0)= \tau_2 \tau_a \sigma_2/\sqrt{8}$ and $P_i(^3S_1)= \tau_2 \sigma_2 \sigma_i/\sqrt{8}$.  
$P_a(^1S_0)$ projects onto the spin-singlet, isospin-triplet \oneS state, while $P_i(^3S_1)$ projects onto the spin-triplet, isospin-singlet \threeS state, where we have used the partial wave notation $^{(2S+1)}L_J$. The $C$ coefficients are unknown in the EFT, and the $\cdots$ stands for terms with more derivatives, such as 
$$
C_2/8 \left[ (N^T P_i N)^\dagger \left(N^T ( P_i \overrightarrow D^2 + \overleftarrow D^2 P_i - 2 \overleftarrow D P_i \overrightarrow D)N\right)\right].
$$ 
For details on the higher-order Lagrangians see, e.g., Refs.~\cite{Ordonez:1993tn,Chen:1999tn,Rupak:1999rk,Beane:2000fx,Bedaque:2002mn} and references therein.
Having removed the pion,  the constraints from the full chiral symmetry of QCD reduce to those of isospin invariance. 
Note that when Coulomb corrections are taken into account they break this isospin symmetry. 

\begin{figure}
\begin{center}
\includegraphics[width=0.8\textwidth]{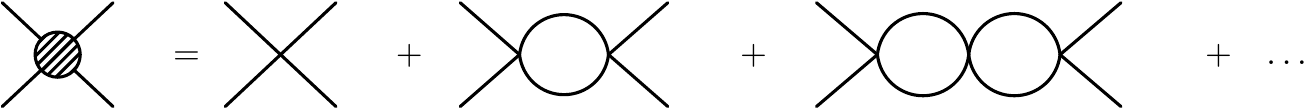}
\end{center}
\caption{Diagrams contributing to PC NN scattering in \eftnopi at LO.\label{fig:2NPCScatter}}
\end{figure}

The smallest ``few-nucleon" system is the deuteron. However, the perturbative expansion of the single-nucleon system does not easily generalize to include a second nucleon.  Instead, because of the presence of shallow bound states, a class of diagrams that are not perturbative must be summed to all orders. 
To see this, consider  the  diagrams in Fig.~\ref{fig:2NPCScatter}. 
 The Feynman rule from 
Eq.~\eqref{Lag:PCpartial}, for either partial wave, yields 
\begin{align}\label{eq:bubblesum}
iA = -iC-C^2 \frac{M p}{4 \pi} +i C^3 \left(\frac{M p}{4 \pi}\right)^2 + \cdots  \ \ ,
\end{align}
where  $C$ is
one of the partial wave coupling constants and $p = \vert\vec{p}\,\vert$ is the magnitude of the nucleon momentum in the center-of-mass frame. If this is understood as a perturbative series  in momentum, one could compare it
to the effective range expansion
\begin{align}
iA=-i\frac{4 \pi / M}{1/a - (r_0/2) p^2 + \cdots + ip}=-i\frac{4 \pi}{M} \left(a-i a^2 p - (a^3 - a^2 r_0/2) p^2 + \cdots \right) \ \ ,
\end{align}
where $a$ is a scattering length, $r_0$ an effective range,
and matching would yield ${\cal C}=\frac{4 \pi a}{M}$, etc.
But the large value of the scattering
length in either channel: $a^{(^1S_0)}= - \frac{1}{8\, \MeV}$ and
$a^{(^3S_1)}= \frac{1}{36\, \MeV}$, would require that $p \ll 1/a$ for such an expansion to  be useful. To address momenta higher than this the series should be summed, as expected for
a bound state. 
The result is 
\begin{align}
-iA=\frac{4 \pi i}{ M} \frac{1}{1/a+ip} \ \ ,
\end{align}
plus perturbative corrections in the effective range and other higher-order
shape parameters that reproduce the scattering characteristics of nucleons.
If the Power Divergence Subtraction (PDS) scheme \cite{Kaplan:1998tg} is used to renormalize loop integrals, the relation between the coupling and the scattering parameters is given by
\be
C = \frac{4\pi}{M}\frac{1}{1/a-\mu}\ ,
\ee
where $\mu$ is the renormalization scale, typically taken to be on the order of $p$.

The EFT not only reproduces the results of effective range theory, but goes beyond it.
It can accommodate external
currents so that we have a theory not only of QCD, but of QED and weak
interactions as well.  Leading-order QED effects are obtained by gauging
the derivatives: $D_\mu N= \partial_\mu N+ i \frac{e}{2} (1 + \tau_3) A_\mu N$,
where $A_\mu$ is the EM field, and by including further contact terms order by
order, such as the magnetic term
\begin{align}
\frac{e}{2M} N^\dagger (\kappa_0+\kappa_1 \tau_3) \sigma \cdot B N \  ,
\end{align}
where $\kappa_0$ and $\kappa_1$ are the isoscalar and isovector anomalous magnetic moments.
Calculations to very high precision (e.g., \cite{Rupak:1999rk}) have been accomplished
in this formalism.

There exists a different formulation of  \eftnopi that provides a number of calculational advantages, in particular for three-nucleon systems. In this so-called ``dibaryon'' formalism two additional dynamical fields are introduced. These fields $d_t$ and $d_s$ have the quantum numbers of the real and virtual $NN$ bound states in the $\threeS$ and $\oneS$ channels, respectively%} 
\cite{Kaplan:1996nv,Bedaque:1997qi,Beane:2000fi}. In this formalism, four-point nucleon-nucleon contact interactions are replaced by couplings of two nucleon fields to a dibaryon field.  The corresponding LO Lagrangian reads
 \begin{align}
  \label{eq:PCLag}
  \mathcal{L}_{PC}^d =& \ND\left(i D_0 +
    \frac{\vec{D}^2}{2M}\right)N -y_t\left[ d_t^{i\dagger} (N^T P^i_t N)
    +\mathrm{h.c.}\right]-y_s\left[ d_s^{a \dagger} (N^T P^a_s N)
    +\mathrm{h.c.}\right] \notag\\
  & +d_t^{i\dagger}\left[\Delta_t
    -c_{0t}\left(i D_0+\frac{\vec{D}^2}{4M}+\frac{\gamma_t^2}{M}\right)
  \right] d_t^i +d_s^{A\dagger}\left[\Delta_s -c_{0s}\left(i D_0+
      \frac{\vec{D}^2}{4M}+\frac{\gamma_s^2}{M}\right)\right] d_s^A
  \notag \\   & + \ldots,
\end{align}
where the $d$ superscript in $\calL^d$ indicates the dibaryon formalism, 
 $\gamma_t$ ($\gamma_s$) is the binding momentum of the real (virtual) bound state in the \threeS (\oneS) channel, $y_{s/t}$, $\Delta_{s/t}$ and $c_{0s/t}$ are low-energy couplings, and $P_t^i=P_i(^3S_1)$, $P_s^a=P_a(^1S_0)$. 
The dibaryon fields are auxiliary fields; however, the field $d_t$ can be used as the deuteron interpolating field since it has the same quantum numbers.
The auxiliary nature of the dibaryon fields is also apparent in the negative sign of the dibaryon kinetic energy terms in Eq.~\eqref{eq:PCLag}. The corresponding bare propagators are dressed with an infinite series of nucleon ``bubble'' diagrams that lead to to a ``dressed'' dibaryon propagator at LO. The explicit form of the dressed propagator depends on certain conventions, which are discussed below.  

The low-energy couplings $y_{s/t}$, $\Delta_{s/t}$ and $c_{0s/t}$ are adjusted to reproduce physical observables. Different conventions exist for their relation to the effective range parameters. One of the advantages of the dibaryon formalism in the two-nucleon sector is that it can be used to resum all contributions proportional to the effective ranges even at LO \cite{Beane:2000fi}. 
However, this convention can cause technical problems in 3N calculations. 
As one aim of this review is to present a variety of calculations in a unified framework, we will not use this convention in presenting results, and effective range contributions will be treated perturbatively and of NLO. 
There still exists additional freedom in how to fix the LECs to parameters extracted from observables. 
While these various prescriptions in principle only differ by higher-order contributions, the convergence of the perturbative expansion can be improved by a convenient choice.
Following the so-called $Z$-parameterization prescription of Refs.~\cite{Phillips:1999hh,Griesshammer:2004pe}, $y_s=y_t=y=\sqrt{\frac{4 \pi}{M}}$,
the $\Delta_{s/t}$ are fit to the poles of
the $NN$ \wave{}{S}{}-wave scattering amplitudes at momenta $i\gamma_{s/t}$,
and the $c_{0s/t}$ include effective range corrections. 
In the \threeS channel, this corresponds to fitting the couplings to the effective range expansion around the deuteron pole, and not around zero momentum. This ensures that the deuteron pole is correctly reproduced at LO instead of perturbatively, which in turn speeds up convergence \cite{Phillips:1999hh}. In the \oneS channel the difference between the two different approaches of fixing the LECs is far smaller \cite{Griesshammer:2004pe}, but we also use the $Z$-parameterization in this channel. 
To present results in a consistent formalism, in this review we use the conventions 
\begin{equation}\label{yDeltaConv}
y^2=\frac{4\pi}{M},\quad \Delta_{s/t}= \left( \gamma_{s/t}-\mu\right),
\end{equation}
where $y=y_t=y_s$. The scale $\mu$ appears in loop integrals; the LECs depend on this scale such that observables are $\mu$-independent.

\subsubsection{Parity-violating two-nucleon terms}

Lifting the requirement of invariance under parity yields five additional operators at leading order.  
This can be understood
as follows:  A PV interaction, at lowest order and at lowest energy,
connects an S-wave and a P-wave.  The only possible S-waves are
$\threeS$ and $\oneS$.  Conserving
total angular momentum, $\threeS$ can only connect to $\oneP$ and $\threePone$;
while $\oneS$ can connect only to $\threePzero$.  Isospin provides additional constraints.
For the isospin zero $\threeS$ state
there is only one way to get to either $\oneP$ or $\threePone$.
On the other hand, $\oneS$  is isospin 1, as is $\threePzero$.  Since
$1 \otimes 1 = 0 \oplus 1 \oplus 2$, there are three  isospin combinations for \oneS to \threePzero transitions, yielding the five independent operators.

The corresponding Lagrangian can be written as \cite{Phillips:2008hn,Schindler:2009wd} 
\begin{align}\label{Lag:PV}
\mathcal{L}_{PV}=  -  & \left[ \CA \left(N^T\sigma_2 \ \VS \tau_2 N \right)^\dagger
\cdot  \left(N^T \sigma_2  \tau_2 i\LRd N\right) \right. \notag\\
& +\CB \left(N^T\sigma_2 \tau_2 \VT N\right)^\dagger
\left(N^T\sigma_2 \ \VS \cdot \tau_2 \VT i\LRd  N\right) \notag\\
& +\CC \ \epsilon^{3ab} \left(N^T\sigma_2 \tau_2 \tau^a N\right)^\dagger
\left(N^T \sigma_2  \ \VS\cdot \tau_2 \tau^b \LRd N\right) \notag\\
& +\CD \ \mathcal{I}^{ab} \left(N^T\sigma_2 \tau_2 \tau^a N\right)^\dagger
\left(N^T \sigma_2 \ \VS\cdot \tau_2 \tau^b i \LRd N\right) \notag\\
& +\left. \CE \ \epsilon^{ijk} \left(N^T\sigma_2 \sigma^i \tau_2 N\right)^\dagger
\left(N^T \sigma_2 \sigma^k \tau_2 \tau_3 \LRd{}^{\!j} N\right) \right] + h.c.,
\end{align}
where $a\, \mathcal{O}\LRd b = a\,\mathcal{O}\vec D b - (\vec D a)\mathcal{O} b\
$ with $\mathcal{O}$ some spin-isospin-operator, and $\mathcal{I}=\text{diag}(1,1,-2)$. 
The coefficients ${\cal C}$ contain the short-distance details of the SM interactions and are not fixed by the EFT. A theoretical determination would require the calculation of nonperturbative QCD effects. Instead, the values of the constants can be fit to data.
There are five  PV coefficients at leading order, corresponding to the
five independent allowed combinations of structures with only one derivative; the
five independent S-P wave combinations. These are  the five Danilov amplitudes \cite{Danilov:1965,Danilov:1971fh} (see Sec.~\ref{subsec:models}) 
cast in a field theory formalism, to be included systematically along
with the QCD \eftnopi. 

This partial-wave representation is only one of the ways to represent the LO Lagrangian. A different representation is the one of Refs.~\cite{Zhu:2004vw,Girlanda:2008ts},  which in the minimal version presented by Girlanda \cite{Girlanda:2008ts} reads\footnote{While it was pointed out that only five independent operators exist at leading order, the original version of the Lagrangian of Ref.~\cite{Zhu:2004vw} contained ten structures. Reference \cite{Girlanda:2008ts} showed how these can be reduced to five. } 
\begin{align}\label{STversion}
\mathcal{L}_{PV}^{Gir}=&\left\{\calG_1 (N^\dagger \vec{\sigma} N \cdot N^\dagger i \stackrel{\leftrightarrow}{\nabla} N -N^\dagger N N^\dagger i \stackrel{\leftrightarrow}{\nabla}\!\cdot\;\VS    N) \right. \notag\\
& -\tilde \calG_1 \epsilon_{ijk} N^\dagger \sigma_i N \nabla_j(N^\dagger \sigma_k N) \notag\\
& -\calG_2 \epsilon_{ijk}\left[ N^\dagger \tau_3\sigma_i N \nabla_j(N^\dagger \sigma_k N) + N^\dagger \sigma_i N \nabla_j (N^\dagger \tau_3\sigma_k N) \right] \notag\\
& -\tilde \calG_5 \mathcal{I}_{ab}\epsilon_{ijk}N^\dagger \tau_a\sigma_i N \nabla_j(N^\dagger \tau_b \sigma_k N) \notag\\
& + \left. \calG_6 \epsilon_{ab3}\vec{\nabla}(N^\dagger \tau_a N)\cdot N^\dagger \tau_b \vec{\sigma} N \right\},
\end{align} 
where the $\calG_i$ are related to the $\calC_i$ of Ref.~\cite{Girlanda:2008ts} by $\calG_i = \calC_i/\Lambda_\chi^3$, with $\Lambda_\chi^3$ the scale of chiral symmetry breaking. 
The representation of Eq.~\eqref{STversion} and the partial-wave representation are equivalent and can be related using Fierz transformations \cite{Schindler:2009wd,Vanasse:2011nd}: %
\begin{equation}\label{eq:wave2Gir}
\begin{split}
\CA&=\frac{1}{4}(\calG_1-\tilde \calG_1)  \ , \\
\CB&=\frac{1}{4} (\calG_1+\tilde \calG_1) \ , \\
\CC&=\frac{1}{2} \calG_2  \ , \\
\CD&=-\frac{1}{2} \tilde \calG_5 \ , \\
\CE&=\frac{1}{4} \calG_6 \  .
\end{split}
\end{equation}

Again, it is often more convenient to use the dibaryon formalism. The PV Lagrangian then takes the form \cite{Schindler:2009wd}
\begin{align}
  \label{eq:PVLag}
  \mathcal{L}_{PV}^d =& - \left[ \gA d_t^{i\dagger} \left(N^T
      \sigma_2 \tau_2\,i\LRd_i N\right) \right. \notag\\
  &\quad\quad +\gB d_s^{A\dagger}
  \left(N^T\sigma_2 \ \VS \cdot \tau_2 \tau_A \,i\LRd  N\right) \notag\\
  &\quad\quad +\gC \ \epsilon^{3AB} \, d_s^{A\dagger}
  \left(N^T \sigma_2  \ \VS\cdot \tau_2 \tau^B \LRd N\right) \notag\\
  &\quad\quad +\gD \ \mathcal{I}^{AB} \, d_s^{A\dagger}
  \left(N^T \sigma_2 \ \VS\cdot \tau_2 \tau^B \,i \LRd N\right) \notag\\
  &\quad\quad \left. +\gE \ \epsilon^{ijk} \, d_t^{i\dagger} \left(N^T \sigma_2
      \sigma^k \tau_2 \tau_3 \LRd{}^{j} N\right) \right] +\mathrm{h.c.}
  +\ldots,
\end{align}
Parts of the PV Lagrangian in the dibaryon formalism are also given in Ref.~\cite{Shin:2009hi}.
As in the PC sector, the parameters in the dibaryon formalism can be related to the ones without dibaryons by integrating out the dibaryon fields in the path integrals \cite{Schindler:2009wd}:
\be\label{eq:dib2wave}
g^{(X-Y)} = \sqrt{8} \, \frac{\Delta_X}{y_X}\, \calC^{(X-Y)},
\ee
where $\Delta$ and $y$ are the dibaryon couplings in the PC sector. 
These PC couplings, as well as the PC couplings of Eq.~\eqref{Lag:PCpartial} and the PV ones of Eqs.~\eqref{Lag:PV} and \eqref{STversion}, are dependent on the renormalization scale $\mu$. As shown in Refs.~\cite{Phillips:2008hn,Schindler:2009wd}, the renormalization scale dependence of the PV $NN$ couplings is dictated by that of the PC $NN$ LECs. However, for the PV dibaryon couplings of Eq.~\eqref{eq:PVLag} the scale dependence of the terms in Eq.~\eqref{eq:dib2wave} cancels such that the PV couplings $g^{(X-Y)}$ are in fact scale independent at least up to next-to-leading order (NLO). A more in-depth discussion of parameter relationships in different formalisms can be found in Sec.~\ref{sec:translate}.

\subsubsection{Parity-conserving three-nucleon terms}

Since the EFT Lagrangian must have the most general form allowed by symmetries, it also contains interaction terms involving more than two nucleons. In addition, the power counting in principle predicts the relative importance of the various terms in the Lagrangian. Applying a ``naive'' power counting based on dimensional analysis suggests that in \eftnopi three-nucleon interactions are suppressed compared to two-nucleon interactions. 
However, an analysis of neutron-deuteron scattering in the spin doublet channel ($S=1/2$) showed \cite{Bedaque:1999ve} that without a three-nucleon interaction, the three-nucleon scattering amplitude at leading order depends strongly on a cutoff $\Lambda$ that is introduced to regularize the corresponding integral equation (discussed below). This undesired (and unphysical) cutoff dependence can be absorbed by introducing a cutoff-dependent three-nucleon term at leading order. 
This is an example of how a renormalization group analysis
can uncover a power counting that differs from naive power counting. With the inclusion of the three-nucleon term at LO, the strong cutoff dependence of the scattering amplitude is removed and the result is properly renormalized.
In the dibaryon formalism, the corresponding Lagrangian has the form \cite{Bedaque:1999ve}
\begin{equation}
  \label{eq:3NLag}
  \mathcal{L}_{3N}=\frac{y^2M\,H_0(\Lambda)}{3\Lambda^2}
  \left[d^i_t(\sigma_iN)-d_s^A(\tau_AN)\right]^\dagger
  \left[d^i_t(\sigma_iN)-d_s^A(\tau_AN)\right],
\end{equation}
where $y$ is defined as in the two-nucleon Lagrangian of Eq.~\eqref{eq:PCLag}. $H_0(\Lambda)$ is the cutoff-dependent three-body coupling, which can be determined from scattering data or bound state properties.
For more details, see, e.g., the reviews in Refs.~\cite{Beane:2000fx,Bedaque:2002mn,Platter:2009gz}.

\begin{figure}
\begin{center}
\includegraphics[width=0.8\textwidth]{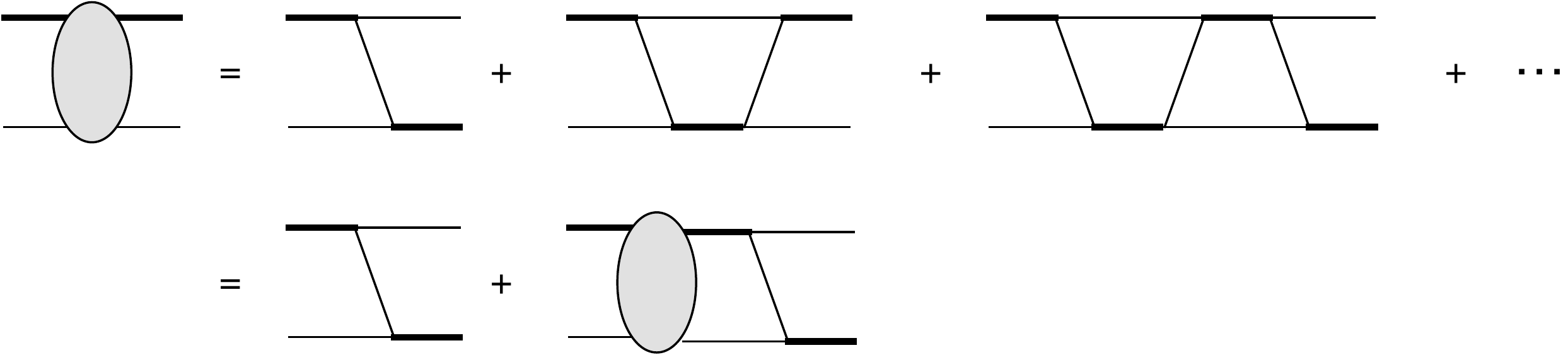}
\end{center}
\caption{Diagrams contributing to neutron-deuteron scattering in \eftnopi at LO in the spin quartet $(S=3/2)$ channel. A thin solid line denotes a nucleon, while a thick line stands for a dibaryon. The grey oval denotes the $nd$ scattering amplitude.\label{fig:3Nwithout}}
\end{figure}

The main features of a three-body calculation in \eftnopi are illustrated by neutron-deuteron scattering below the breakup threshold.  It is again useful to employ  the dibaryon formalism, introduced above, with one auxiliary field in each of the two S-wave channels. According to the power counting of \eftnopi, an infinite number of diagrams contribute at leading order, see Fig.~\ref{fig:3Nwithout}.
But a summation of these diagrams can be performed by considering an integral equation for the scattering amplitude $T$ \cite{Bedaque:1997qi,Bedaque:1998mb}. The amplitude in the spin quartet channel ($S=3/2$), projected onto orbital angular momentum $L$, is the solution to the equation \be
\label{3N-4Seq}
T^{(L)}(E;k,p) = -4\pi \calK^{(L)}(E;k,p)+ \frac{2}{\pi} \int_0^\Lambda dq\; q^2 \calK^{(L)}(E;q,p) D_t(E-\frac{{\qv}^{\, 2}}{2M},\qv)T^{(L)}(E;k,q),
\ee
with total nonrelativistic energy $E$ and incoming (outgoing) momentum $k$ ($p$). $\calK$ denotes the nucleon-exchange kernel and $D_t$ is the \threeS dibaryon propagator.
At LO and in the conventions used in the PV calculations of Refs.~\cite{Griesshammer:2010nd,Griesshammer:2011md} (discussed below), the dibaryon propagator is
\be
\label{Dt-prop}
D_t(q_0,\qv) = \frac{1}{\gamma_t -\sqrt{\frac{\qv^{\, 2}}{4}-Mq_0-i\epsilon}},
\ee
and the projected kernel $\calK^{(L)}$ reads
\be
\calK^{(L)} (E;q,p) = \half \int_{-1}^1 d\cos\theta\,\frac{P_L(\cos\theta)}{p^2+q^2-ME+pq\cos\theta}\ ,
\ee
where $P_L(\cos\theta)$ is the $L$th Legendre polynomial of the first kind.
The cutoff $\Lambda$ serves as a regulator for the integral equation. For $\Lambda \to \infty$, Eq.~\eqref{3N-4Seq} can be viewed as the generalization of the Skorniakov--Ter-Martirosian equation \cite{STM:1958}. As discussed above, power counting predicts that the first 3N interaction term appears at \NtwoLO for S-wave scattering, which means that three-body observables  should be described to about $10\%$ using only 2N interactions. Good agreement with experiment is found without need for a 3N contact interaction \cite{Bedaque:1997qi,Bedaque:1998mb} in the spin quartet  channel. For large enough values  of the cutoff $\Lambda$, the scattering amplitude is independent of $\Lambda$ and is properly renormalized. 

\begin{figure}
\begin{center}
\includegraphics[width=0.9\textwidth]{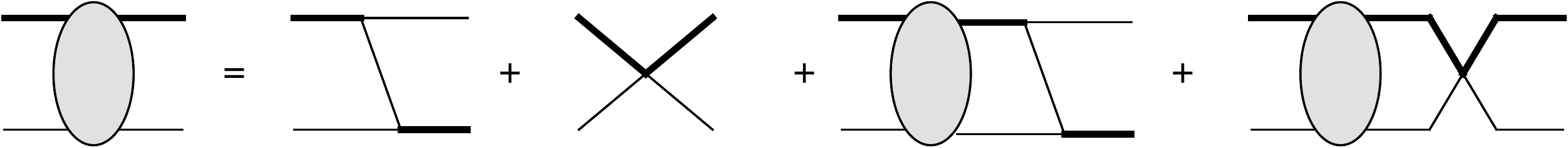}
\end{center}
\caption{Diagrams contributing to neutron-deuteron scattering in \eftnopi at LO in the spin doublet $(S=1/2)$ channel. Symbols as in Fig.~\ref{fig:3Nwithout} above. \label{fig:3N-H0}}
\end{figure}

 In the spin doublet channel ($S=1/2$) both S-wave dibaryon fields can contribute to the scattering amplitude, and Eq.~\eqref{3N-4Seq}  has to be replaced by a system of coupled differential equations \cite{Bedaque:1997qi,Bedaque:1999ve}. However, unlike in the  quartet channel, the solution of the integral equations for angular momentum $L=0$ does not approach a unique limit for $\Lambda \to \infty$ \cite{Danilov:1961,Bedaque:1999ve}. Instead, the solution exhibits a strong dependence of the scattering amplitude on the cutoff $\Lambda$. Introducing a 3N interaction at LO, Eq.~\eqref{eq:3NLag}, with an appropriate dependence on $\Lambda$ removes the cutoff dependence of the solution. The strength $H_0$ of the 3N interaction is determined from a single 3-body observable and then used in all other calculations in the 3N sector \cite{Bedaque:1998kg,Bedaque:1999ve}. The resulting integral equation is shown schematically in Fig.~\ref{fig:3N-H0}.

Unlike in the two-nucleon sector, where dimensional regularization can be applied and closed-form expressions obtained for NLO results, the three-body diagrams are complicated (and nested) enough that it is not yet
known how to solve them other than numerically. This also implies that cutoff independence and the expected size of higher-order corrections have to be checked numerically.

\subsubsection{Parity-violating three-nucleon terms}

To extend the discussion of low-energy parity violation to include three nucleons up to and
including NLO requires
only modifications in the strong sector.  As shown in Ref.~\cite{Griesshammer:2010nd}, through that order there are no new PV terms necessary; the PV physics is all contained within the two-nucleon PV operators.  This is  important for a comprehensive analysis of PV observables because it means that low-energy PV experiments including two and/or three nucleons will all give constraints on the five leading PV LECs and only those,
at least up to about 10 percent corrections.

The power counting of \eftnopi predicts that a PV 3N contact term first contributes at \NtwoLO \cite{Griesshammer:2010nd}. But the experience with the 3N operator in the PC sector above requires that this be verified by a renormalization analysis. Since the operators in the effective Lagrangian are constructed according to their symmetry properties, it is not clear whether the PC 3N contact term is sufficient to completely renormalize the PV 3N sector. Reference~\cite{Griesshammer:2010nd} addressed this question for S-P wave transitions by considering the UV behavior of the PV $Nd$ scattering amplitude. The outcome is that the only possibly divergent contribution at LO vanishes because of an angular integration that is identically zero. At NLO, Ref.~\cite{Griesshammer:2010nd} showed that the available PV 3N operators have a different  spin and isospin structure than any potential divergence. Therefore they cannot renormalize the expressions contributing at NLO. No PV 3N operator is thus required at LO and NLO in the $Nd$ system, which means that at least up to corrections expected to be of order 10\%  only 2N contact terms have to be considered in the PV Lagrangian.

At NLO, which is as far as we can go without possibly requiring additional
PV  operators, there are a variety of choices about
how to proceed in calculating three-nucleon observables. At this time only one application of \eftnopi to PV three-nucleon observables at NLO has appeared \cite{Griesshammer:2011md}. This uses the partially-resummed
formalism of Ref.~\cite{Bedaque:2002yg}. 
The phrase ``partially-resummed" refers to the
fact that only some of the range corrections are included, rather than all of them,
as can be done using the dibaryon formalism in the two-body sector. 
Instead,  the kernel is expanded perturbatively and then
used to solve the three-body integral equations.  
The effect is to include some higher-order diagrams, but because each is of naive-power-counting size, the precision of the NLO prediction is preserved.  

%%%%%%%%%%%%%%%%%%%%%%%%%%%%%%%%%%%%%%%%%%%%%%%%%%%%%%%%%%%%%

\subsection{Chiral EFT}\label{sec:chiralEFT}

In this section we review the basics of chiral  EFT.  For more in-depth reviews see, e.g., Refs.~\cite{vanKolck:1999mw,Beane:2000fx,Bedaque:2002mn,Epelbaum:2008ga,Machleidt:2011zz} which also offer a guide to further literature.
At energies beyond about 20 MeV, pions become dynamical degrees of freedom and have to be included as such in the EFT.  With pions present, the (approximate) chiral symmetry of QCD becomes relevant. Chiral symmetry refers to the invariance of the QCD Lagrangian in the limit of massless up and down quarks under $SU(2)_L \times SU(2)_R \times U(1)_V$ transformations.  While chiral symmetry is a symmetry of the QCD Lagrangian, the QCD ground state only exhibits approximate invariance under $SU(2)_V \times U(1)_V$, i.e., chiral symmetry is spontaneously broken. The  pions, which are much lighter than all other known hadrons, are identified as the corresponding Goldstone bosons. Chiral symmetry and its breaking impose restrictions on the possible interactions, and these constraints are used in the construction of the EFT Lagrangian involving pions.

Chiral symmetry was first used to construct an EFT for interactions of pions with themselves and external fields  \cite{Weinberg:1978kz,Gasser:1983yg,Gasser:1984gg}, called Chiral Perturbation Theory (\chiPT). 
At leading-order tree level, the results of \chiPT are equivalent to those of current algebra, but the EFT approach allows a systematic extension to higher-order loop diagrams. 
Chiral Perturbation Theory has matured into an important tool in the study of low-energy QCD phenomena; two loop calculations are now available. 
For recent reviews, see, e.g., Refs.~\cite{Bernard:2006gx,Bijnens:2006zp,Birse:2007zz}. Subsequently, the interactions of nucleons have also been considered. The case of a single nucleon interacting with pions and external fields is referred to as baryon \chiPT (B\chiPT). Starting from a manifestly Lorentz-invariant form, the corresponding Lagrangian is expanded in powers of derivatives and quark masses,
\be
\calL_{\pi N} = \calL_{\pi N}^{(1)} + \calL_{\pi N}^{(2)}  + \calL_{\pi N}^{(3)} +\cdots ,
\ee 
where the superscript denotes the order in the chiral power counting.
As pointed out in Ref.~\cite{Gasser:1987rb}, the application of dimensional regularization in combination with a minimal subtraction scheme as in the purely mesonic sector does not result in a consistent power counting. One solution to this problem is to apply the nonrelativistic reduction described in Sec.~\ref{nopimethod} to the complete baryonic Lagrangian. The resulting so-called Heavy Baryon \chiPT (HB\chiPT) Lagrangian \cite{Jenkins:1990jv,Bernard:1992qa} corresponds to an expansion in not only powers of  derivatives and quark masses, but also inverse powers of the nucleon mass.  For other solutions to the power counting problem of baryon \chiPT see, e.g., \cite{Becher:1999he,Gegelia:1999gf,Fuchs:2003qc}.

The pion fields are collected in an exponential matrix 
\be
\xi = \exp\left( i\frac{\Pi(x)}{F} \right),
\ee
where 
\be
\Pi(x)=\frac{1}{\sqrt{2}}\begin{pmatrix} \frac{\pi^0}{\sqrt{2}} &  \pi^+ \\ \pi^- & - \frac{\pi^0}{\sqrt{2}} \end{pmatrix} = \frac{1}{2}\pi^a\tau^a\ ,
\ee
the $\tau^a$ are Pauli matrices in isospin space, and $F$=92.4 MeV. The $x$
dependence in the $\Pi$ matrix of fields will be suppressed in what follows. The transformation properties under combined left-handed transformations $L$ and right-handed transformations $R$  of 
$SU(2)_L \times SU(2)_R$ are 
\be \label{xi}
\xi \rightarrow L \xi U^\dagger = U \xi R^\dagger \ \ ,
\ee
where $U$ is a function of the pion fields. Two currents useful for constructing operators are
\be
{\cal A}_\mu = \frac{i}{2} \left( \xi D_\mu \xi^\dagger  - \xi^\dagger D_\mu \xi \right) \ , \ \  {\cal V}_\mu = \frac{1}{2} \left( \xi D_\mu \xi^\dagger  + \xi^\dagger D_\mu \xi \right) \ \ ,
\ee
where $D_\mu \pi = \partial_\mu \pi +i e A_\mu [{\rm Q},\pi]$  includes the electromagnetic
field $A_\mu$ and ${\rm Q}={\rm diag} (2/3,-1/3)$.\footnote{The covariant derivative can be generalized to include other external fields.}

The leading Lagrangian involving only pions is 
\be\label{eq:PionLag}
\calL_\pi=\frac{F^2}{4} \text{Tr} \left[D_\mu \Sigma D^\mu \Sigma^\dagger\right] + \frac{F^2}{2} \lambda \text{Tr} \left[m_q \left (\Sigma + \Sigma^\dagger\right)\right] + \cdots \ \ ,
\ee
where $\Sigma=\xi^2$, $\lambda$ is such that the second term yields the pion mass to leading order and $m_q={\rm diag}(m_u,m_d)$.  Including a single nucleon field results in the following leading order terms  
\be\label{eq:NucLag}
\calL_{\pi N}=N^\dagger \left( i {\cal D}_0 + \frac{\vec {\cal D}^2}{2M} \right) N + \frac{g_A}{2F} N^\dagger \vec \sigma \cdot \vec {\partial} \pi^a \tau^a N + \cdots \ \ ,
\ee
where $g_A=1.25$ is the axial coupling and 
\be
{\cal D}_\mu N = \partial_\mu N + [{\cal V}_\mu,N] \ \ .
\ee

The complete single-nucleon Lagrangian up to fourth order in the power counting in both the manifestly Lorentz invariant and the heavy-baryon forms can be found in \cite{Fettes:2000gb}. Baryon \chiPT and HB\chiPT have been applied to a large number of observables, including nucleon form factors, Compton scattering, pion-nucleon scattering, etc. For reviews, see e.g., \cite{Bernard:1995dp,Bernard:2006gx,Bernard:2007zu,Scherer:2009bt}.
Once the restriction on conservation of parity is lifted, further terms are allowed.
Defining currents \cite{Kaplan:1992vj}
\be X^a_L=\xi^\dagger \tau^a \xi \ , \ \ X^a_R=\xi \tau^a \xi^\dagger \ \ ,
\ee
where the chiral transformation is $X \rightarrow UXU^\dagger$, the $U$
matrix is the same as in Eq.~\eqref{xi}, and $a$ is an isospin index, yields \cite{Kaplan:1992vj,Savage:1998rx}
\bea\label{NPV}
 \calL^{pv} =h^{0}_V N^\dagger {\cal A}_0 N 
-h_{\pi NN}^{1} \frac{F}{2\sqrt{2}} N^\dagger \left(X^3_L-X^3_R\right) N+ \frac{h_V^{1}}{2} N^\dagger N \text{Tr} \left[{\cal A}_0\left(X^3_L+X^3_R\right)\right] - \nonumber \\ \nonumber \frac{h_A^{1}}{2} N^\dagger \sigma^a N \text{Tr} \left[{\cal A}_a\left(X^3_L-X^3_R\right)\right] + h_V^{2} \mathcal{I}_{ab} N^\dagger \left(X^a_R {\cal A}_0 X^b_R + X^a_L {\cal A}_0 X^b_L \right) N - \\ h_A^{2}  \mathcal{I}_{ab} N^\dagger \left(X^a_R \vec \sigma \cdot \vec {\cal A} X^b_R + X^a_L \sigma \cdot \vec {\cal A} X^b_L \right) N + \cdots \ \ ,
\eea
where the $\cdots$ contain higher order terms.  The superscripts on the couplings
indicate the isospin structure of the associated operator.   $h^{0}_V$ is the PV parameter for the $\Delta I=0$ operator; $h_{\pi NN}^{1}$,  $h_V^{1}$, and $h_A^{1}$ are PV parameters for the $\Delta I=1$ operators; and $h_V^{2} $ and  $h_A^{2}$ are PV parameters for the $\Delta I=2$ operators. The subscript $V$ or $A$ indicates whether the operator contains a vector or axial vector nuclear current, and 
$\mathcal{I}_{ab}$ is the same matrix that appears in the \eftnopi expressions, see Eq.~\eqref{Lag:PV}.
As usual, the new parameters are not constrained by the EFT and must
be  determined independently.   
The leading-order term describing the coupling of a single pion to a nucleon stems from the second term in Eq.~\eqref{NPV} and is given by
\be
\calL^{pv}_{\pi NN} = -i h^1_{\pi NN} (\bar p n \pi^+-\bar n p \pi^-) + \cdots \ \ .
\ee

With the exception of $h_{\pi NN}^{1}$, each of the above includes photon couplings through gauging the embedded derivatives.  For example, a term that contributes to the isovector anapole moment of the nucleon comes from expanding Eq.~\eqref{NPV},
\be 
\calL^{pv}_{\gamma \pi NN}=-\frac{i e}{ \sqrt{2} F} (h_V^0+\frac{4}{3} h_V^2) \pi^+ \bar p A_\mu \gamma^\mu n\ .
\ee
In addition, explicit electromagnetic operators are \cite{Zhu:2000gn}
\be\label{ZhuCs}
\calL^{pv}_{\gamma NN}=\frac{c_1}{\Lambda_\chi} \bar N \sigma^{\mu \nu} \left[F^+_{\mu \nu},X^3_L-X^3_R\right]_+ N + \frac{c_2}{\Lambda_\chi} \bar N \sigma^{\mu \nu} F^-_{\mu \nu} N + \frac{c_3}{\Lambda_\chi} \bar N \sigma^{\mu \nu} \left[F^-_{\mu \nu},X^3_L+X^3_R\right]_+ N
\ee 
where 
\be F^\pm_{\mu \nu}=\frac{1}{2} (\partial_\mu A_\nu - \partial_\nu A_\mu)(\xi Q^\prime \xi^\dagger \pm \xi^\dagger Q^\prime \xi) 
\ee
and $Q^\prime = {\rm diag}(1,0)$

The LECs contained in Eq.~\eqref{NPV} were used in several calculations involving single-nucleon physics.   For terms involving one
$\gamma$, and/or one pion, and including leading two-pion terms involving $h_A^i$, 
the only combinations of LECs that appear are $h^1_{\pi NN}$, $h_V^0+\frac{4}{3} h_V^2$, $h_A^1\pm 2h_A^2$, and $-4c_1+c_2$ \cite{Zhu:2000gn}.\footnote{We have modified the couplings of \cite{Zhu:2000gn}   to conform with the convention of Eq.\eqref{NPV}.} However, Ref.~\cite{Chen:2001rc} showed that a field redefinition removes the $h_V$ from observables, at least through this order.
Analogous terms involving the $\Delta$ as an explicit degree of freedom are provided, e.g., in Appendix A of Ref.~\cite{Zhu:2001re}.

The EFT LEC $h^1_{\pi NN}$ \ corresponds to the coupling $h_\pi^1$ in the DDH model (see Sec.~\ref{subsec:models}).
 Both represent the PV coupling of a pion to two nucleons.  However, different
conventions for factors of 2, $\sqrt{2}$, and signs (because of $\gamma_5$ definitions,
phase choices for chiral transformations, etc.) prevail, so care should be taken
when comparing calculations using different conventions. In Ref.~\cite{Zhu:2000fc} the behavior of the coupling $h_\pi^1$  was studied using  chiral EFT. In 
particular, the authors noted that the estimate 
$$h^1_{\pi } ({\rm best}) = 7 g_\pi \ \ {\rm and}  \ \ h^1_{\pi }({\rm range}) = (0-30) g_\pi $$ 
(where  $g_\pi = 3.8 \times 10^{-8}$) does not include chiral symmetry breaking corrections.   They calculated these corrections using HB$\chi$PT and find that not only does the coupling $h^1_\pi$ itself become renormalized at one loop, but that corrections from additional hadronic terms cannot be neglected.  The motivation for this analysis was the observation that the cleanest experimental measurement of $h^1_\pi$,
from $^{18}\text{F}$ \cite{Page:1987ak,Bini:1988zz}, where the analog $^{18}\text{Ne}$ state can be used to remove uncertainties about
nuclear details \cite{Haxton:1981sf},
$$h^1_{\pi} ({\rm exp}) = (0.73 \pm 2.3) g_\pi \ \  , $$
suggests a deviation from the DDH estimate.  Estimated chiral corrections indicate that what is
actually measured is \cite{Zhu:2000fc}
\be
h_\pi^{EFF} = 0.5 h^1_\pi + 0.25 h^1_A - 0.24 h_\Delta + 0.079 h^\Delta_A \ \ ,
\ee
where $h^1_A$ (which starts at $NN\pi \pi$) is the coefficient found in the $\Delta I=1$ PV term in Eq.~\eqref{NPV}, and the last two parameters arise from including the $\Delta$ as a degree of freedom; $h_\Delta$ is the coefficient of the PV Yukawa $\Delta \Delta \pi$ term and $h^\Delta_A$ is the coefficient in front of PV 
$N \Delta \pi \pi$ terms.
Other estimates of $h^1_\pi $ include a two-flavor
Skyrme model analysis \cite{Kaiser:1989fd} yielding a magnitude of $\approx 4 \times 10^{-8}$ 
and a three-flavor Skyrme estimate  \cite{Meissner:1998pu}  of 
$h^1_\pi \approx (1.1 - 1.8 ) \times 10^{-7}$.
QCD sum rules give \cite{Henley:1995ad,Henley:1995dw,Henley:1998xh}
$h_\pi^1 \approx 3 \times 10^{-7}$ .

By identifying the dimension-six four-quark operators and running them to
the low energy scale, Ref.~\cite{Kaplan:1992vj} matches to the chiral EFT PV
operators to make the following estimates, using dimensional analysis:
\begin{align}
[h^1_{\pi NN}]_{\rm nda} &\approx 5 \times 10^{-7} \approx 10 g_\pi \ \  , \\
[h^1_A]_{\rm nda} \approx [h_V^{0,1}]_{\rm nda} &\approx 5 \times 10^{-8} \approx g_\pi \ \  , \\
(h^2_V)_{\rm nda} \approx [h_A^2]_{\rm nda} &\approx 1 \times 10^{-8} \approx 0.3 g_\pi \ \  ,
\end{align}
where the first coupling is largest because it is lower order in the chiral
expansion. This  suggests that the higher order $h^1_A$ is  
not important for a 10 percent estimate of $h_\pi^{EFF}$ above.  On the  other hand,
appealing to SU(3), or even the analog behavior of the $\Delta S=1$ version of
the $\Delta S=0$ PV terms argues that $h_A^1$ may be within a factor of two
of $h^1_{\pi NN}$ \cite{Kaplan:1992vj}. Meanwhile, Ref.~\cite{Feldman:1991tj} estimates $h_\Delta$ to
be $\approx - 20 g_\pi$. A factorization argument yields \cite{Zhu:2000fc} 
$h_A^\Delta \approx (0-{\rm few}) g_\pi$.

Ref.~\cite{Zhu:2009nj} uses a large-$N_c$ analysis to estimate PV LECs, finding
\begin{align}
h_\pi &\sim \calO\left(\frac{1}{\sqrt{N_c}}\right) \ \  ,\nonumber \\ 
h_V^i &\sim \calO(1) \ \  , \nonumber\\
h_A^1 &\le \calO(1) \ \  ,\\ \nonumber 
h_A^2 &\le \calO(N_c^{-1}) \ \  ,\\ \nonumber
h_\Delta &= -\frac{3}{\sqrt{5}} h_\pi \ \  ,
\end{align}
as well as additional relationships involving $\Delta$ couplings.

\subsubsection{Two-nucleon sector}

For the case of two and more nucleons, the interactions are no longer perturbative. 
Instead, a potential is  defined, which consists of all connected, two-nucleon irreducible diagrams \cite{Weinberg:1990rz,Weinberg:1991um}. It is derived from the \chiPT Lagrangians involving pions and nucleons, supplemented by two-nucleon contact terms that also take into account the constraints from chiral symmetry. The potential is then inserted into a Schr\"odinger or Lippmann-Schwinger equation. The corresponding EFT is called chiral EFT. In chiral EFT, the power counting is applied to the potential, which is expanded in powers $\nu$ of a small parameter $Q$. Each connected, irreducible diagram is assigned a chiral index $\nu$ which is determined by (see, e.g. Ref.~\cite{Epelbaum:2008ga})
\be
\nu = -4 +2N +2L + \sum_i V_i \Delta_i  \  ,
\ee
where $N$ is the number of nucleons, $L$ is the number of pion loops, and $V_i$ is the number of vertices derived from a Lagrangian of type $\Delta_i$, where
\be
\Delta_i = d+\frac{f}{2}-2  \ ,
\ee
with $d$ the number  of derivatives and $f$ the number of fermion fields.

\begin{figure}
\begin{center}
\includegraphics[width=0.6\textwidth]{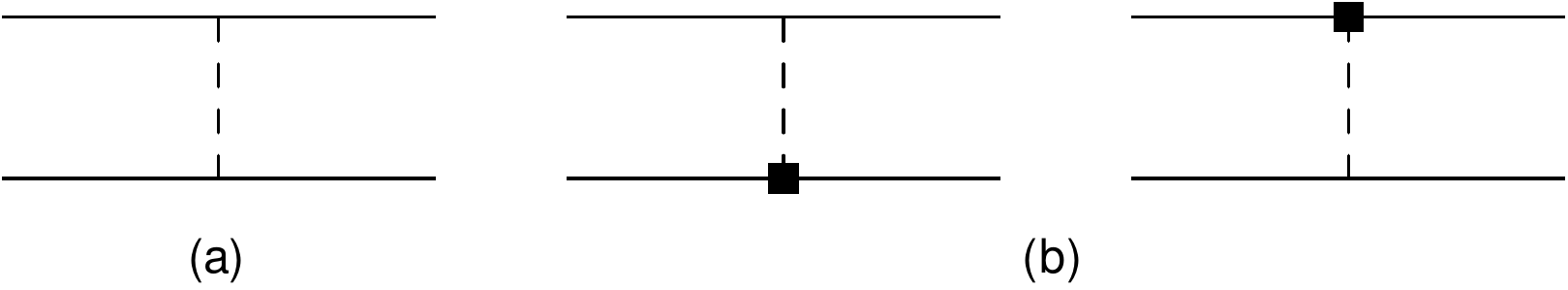}
\end{center}
\caption{One-pion-exchange  contributions to the 2N potential. (a): Parity-conserving potential, (b):  parity-violating potential. Solid lines denote nucleons, the dashed line is a pion. The solid square stands for a PV vertex.  \label{fig:OPE}}
\end{figure}

 In the PC sector, the LO potential has $\nu=0$ and consists of contributions from the S-wave contact operators analogous to those of Eq.~\eqref{Lag:PCpartial} as well as a one-pion exchange (OPE) contribution, see Fig.~\ref{fig:OPE}(a).
This is given by
\be\label{VPCLO}
V_{\nu=0}^\text{PC} = -\frac{g_A^2}{4F^2}\,\frac{(\vec{\sigma}_1\cdot \vec{q})(\vec{\sigma}_2\cdot \vec{q})}{\vec{q}^{\;2}+m_\pi^2}\,(\vec{\tau}_1 \cdot \vec{\tau}_2) + C_S + C_T\, (\vec{\sigma}_1\cdot \vec{\sigma}_2)\ ,
\ee
where $\sigma_i$ ($\tau_i$) denotes the spin (isospin) of the nucleon $i$, $\vec{q} = \vec{p} - \vec{p}^{\;\prime}$  is the momentum of the exchanged pion, and the LECs $C_S$ and $C_T$ correspond to the two LO contact operators.
The chiral EFT can be consistently extended to higher orders. 
For example, two-pion exchange contributes at NLO. It is also possible to take into account the coupling to external fields in the same framework that is used for the derivation of the potential, see, e.g., Refs.~\cite{Pastore:2011ip,Kolling:2011mt} for the latest results of the electromagnetic currents. In fact, by employing the EFT formalism, it is possible to establish a direct connection from the purely mesonic sector to interactions between two and more nucleons \cite{Weinberg:1992yk}. Two-nucleon potentials were first constructed in Refs.~\cite{Ordonez:1992xp,Ordonez:1993tn,Ordonez:1995rz} and have now been derived in EFT up to \NthreeLO \cite{Entem:2003ft,Epelbaum:2004fk}.  However, the issue of proper renormalization in the two-nucleon sector has been a topic of intense discussion, see, e.g., Refs.~\cite{Kaplan:1996xu,Kaplan:1998tg,Kaplan:1998we,Beane:2001bc,PavonValderrama:2004nb,Nogga:2005hy,Birse:2005um,Epelbaum:2006pt,Gegelia:2004pz,Yang:2009kx,Epelbaum:2009sd,Epelbaum:2012ua} and references therein. 

For the PV sector, the LO potential scales as $\nu=-1$.  This is obtained from Fig.~\ref{fig:OPE} (a) by replacing one of the PC vertices with the PV $h^{1}_{\pi NN}$ vertex, see Fig.~\ref{fig:OPE} (b).  The PC vertex scales as one power of momentum, as can be seen from Eq.~\eqref{VPCLO}, while the PV vertex does not have a derivative. So the scaling of this term in the potential is
\be 
\sim \frac{\vec q}{\vec{q}^{\;2} + m_\pi^2} \ .
\ee
All the other terms in Eq.~\eqref{NPV} have an additional derivative accompanying
the PV pion vertex and so occur at least one order higher. 
The LO potential containing pion exchange takes the form \cite{Zhu:2004vw}
\be\label{VPVLO}
V_{\nu=-1}^\text{PV} = - i \frac{g_A h_{\pi NN}^1}{2\sqrt{2}F}\,\frac{(\vec{\sigma}_1+ \vec{\sigma}_2) \cdot \vec{q}}{\vec{q}^{\;2}+m_\pi^2}\,(\vec{\tau}_1 \times \vec{\tau}_2)_z.
\ee
This in fact constitutes the complete potential at LO.
As with the PC operators, the two-nucleon contact PV operators for the chiral EFT have
the same form as those for \eftnopi, but the coefficients are different
in the two theories.  
Unlike in the PC case  of Eq.~\eqref{VPCLO}, however, their contributions to the potential are of higher order than the  one-pion exchange potential.   

The first subleading contributions to the PV potential occur for $\nu=1$. They originate from the contact terms, two-pion exchange contributions proportional to $h_{\pi NN}^{1}$, as well as  one-pion exchange contributions proportional to higher-order $\pi N$ couplings. However, as discussed in Refs.~\cite{Zhu:2004vw,RamseyMusolf:2006dz,Liu:2006dm} these latter contributions can be absorbed by a redefinition of lower-order LECs. They thus do not contribute new, independent structures in the potential.
The potential at subleading order therefore contains two types of contributions: a short-range part from the PV contact operators and a medium-range part from two-pion exchange that is proportional to $h_{\pi NN}^{1}$,
\be
V_{\nu=1}^\text{PV} = V^\text{PV}_\text{contact} + V^\text{PV}_{2\pi}.
\ee
Explicit expressions and a detailed discussion of the derivation can be found in Ref.~\cite{Zhu:2004vw}. 
Detailed studies of the two-pion exchange potential, also including $\Delta$ degrees of freedom, are presented in Refs.~\cite{Kaiser:2007zzb,Desplanques:2008jx}.

In an EFT, currents can be derived consistently in the same framework as the potential. Some of the same LECs that contribute to the PV potential also contribute to the PV current after gauging of derivatives. There is an additional operator containing a $\gamma\pi NN$ coupling proportional to an independent LEC $\bar C_\pi$ which is a linear combination of the couplings in Eq.~\eqref{ZhuCs}. This operator contributes to the current, but not to the potential, and is therefore only  relevant for processes involving photons. 

\subsubsection{Three-nucleon sector}

In chiral EFT the first contributions to the 3N potential naively appear at NLO and stem from pion-exchange diagrams. However, it can be shown that one of the three topologies is shifted to higher orders due to an additional suppression by $1/M$, where $M$ is the nucleon mass. The remaining contributions can be treated in two different ways: in an energy-independent formalism they are again suppressed by factors of $1/M$ \cite{Epelbaum:2008ga}, while in an energy-dependent formalism based on time-ordered perturbation theory these contributions cancel with recoil corrections of the iterated 2N interaction \cite{Weinberg:1990rz,Weinberg:1991um,Ordonez:1992xp,vanKolck:1994yi,Friar:1998zt}. Thus the first nonzero contribution to the 3N potential appears at \NtwoLO and is due to both pion-exchange terms as well as a 3N contact operator. The complete 3N potential at \NtwoLO has been worked out in Ref.~\cite{vanKolck:1994yi,Epelbaum:2002vt}, and the contributions at \NthreeLO  can be found in Refs.~\cite{Ishikawa:2007zz,Bernard:2007sp,Bernard:2011zr}. The derivation of the contributions at \NXLO{4} is ongoing \cite{Krebs:2012yv}.

Parity-violating 3N interactions formally appear at NLO. The contributing diagrams are tree-level pion-exchange diagrams with one PV $\pi N$ vertex.  Analogously to the PC sector, however, they again cancel against contributions from the iterated 2N potential and can thus be neglected in an energy-independent formalism \cite{Zhu:2004vw}. 
As in the pionless case, 3N interactions are suppressed and can only start to contribute at \NtwoLO.

%%%%%%%%%%%%%%%%%%%%%%%%%%%%%%%%%%%%%%%%%%%%%%%%%%%%%%%%%%%%%

\subsection{Effective field theories beyond three nucleons}\label{sec:fewbody}

EFTs have also been used in few-body calculations with more than three nucleons.
Pionless EFT calculations have recently been extended to systems with up to $A=6$ nucleons  \cite{Platter:2004zs,Stetcu:2006ey,Stetcu:2009ic,Kirscher:2009aj,Kirscher:2011uc},  
using no-core shell model (NCSM) \cite{Stetcu:2006ey,Stetcu:2009ic}  (see Ref.~\cite{Stetcu:2012ie,Barrett:2013nh} for recent reviews) or resonating group method (RGM) methods \cite{Kirscher:2009aj,Kirscher:2011uc} to solve the few-body problem.
While at larger $A$ prevailing energies are such that pion exchange has to be taken into account explicitly, the calculations of Refs.~\cite{Stetcu:2006ey,Stetcu:2009ic,Kirscher:2009aj,Kirscher:2011uc} show that calculations up to $A=6$ are within reach of \eftnopi. 
The main focus though has been in the application of chiral EFT interactions, again in the NCSM framework (see, e.g., Refs.~\cite{Nogga:2005hp,Navratil:2007we,Maris:2011as,Maris:2012bt}), a combination of NCSM and RGM methods (see, e.g., Refs.~\cite{Navratil:2010jn,Navratil:2011zs}), and in the hyperspherical harmonics approach (see, e.g., Refs.~\cite{Kievsky:2008es,Viviani:2010qt,Viviani:2012vg}).

Another very promising recent development in the application of effective field theories to few-nucleon systems is the combination of lattice methods with  EFTs (see, e.g., Ref.~\cite{Lee:2008fa} for a recent review). In contrast to lattice QCD \add{,} which uses quarks and gluons, the  dynamical degrees of freedom in lattice EFT are the nucleon (and potentially pion) fields. The EFT Lagrangian is discretized on a space-time lattice, and Monte Carlo methods can be employed to evaluate path integrals to obtain observables in two-, three-, and few-nucleon systems. An application using \eftnopi in the three-nucleon sector can be found in Ref.~\cite{Borasoy:2005yc}, while the main focus has recently been on chiral EFT applications in the few-body sector (see Ref.~\cite{Lee:2008fa} and references therein).

All these approaches show that EFT methods can be extended to systems beyond $A = 3$. Applying these methods to an analysis of PV observables in $A \ge 3$ systems would contribute significantly to an improved understanding of hadronic parity violation .

%%%%%%%%%%%%%%%%%%%%%%%%%%%%%%%%%%%%%%%%%%%%%%%%%%%%%%%%%%%%%

\subsection{Hybrid calculations}\label{subsec:hybrid}

 Fully consistent EFT calculations are still in their infancy for applications to four or more nucleons. For these systems, a combination of traditional phenomenological models and EFTs, the so-called ``hybrid'' approach, has been an important tool. In hybrid calculations, matrix elements are calculated by evaluating operators derived in an EFT framework between model wave functions,
\be
\langle \Psi_\text{model} \vert O_\text{EFT} \vert \Psi_\text{model} \rangle.
\ee
The motivation behind the hybrid approach is to take advantage of the existing expertise in applying modern phenomenological models. Not only do these models provide good parameterizations of NN phase shift data, electroweak properties, and three-nucleon systems, they have also been used in the development and application of calculational tools for few-body  systems and reactions. In the hybrid approach, wave functions for three- and few-nucleon systems do not have to be recalculated when combined with various operators. 
Hybrid calculations have played an important role in the acceptance of EFT methods. For example, Ref.~\cite{Park:1994sr} showed that the cross section for the reaction $n+p\to d+\gamma$ with thermal neutron energies is very accurately reproduced when using the electromagnetic current derived in chiral EFT up to \NtwoLO in combination with Argonne $v_{18}$ (AV18) \cite{Wiringa:1994wb}  wavefunctions. Similarly, Ref.~\cite{Park:2002yp} employed electroweak currents up to \NthreeLO to calculate astrophysical $S$ factors for $p+p\to d + e^+ + \nu_e$ and $p+\HeThree\to\HeFour + e^+ + \nu_e$ to previously unmatched precision. 

Numerical differences between  chiral EFT  and hybrid calculations are expected to be small. The various phenomenological potentials describe low-energy data well, and therefore agree with the long-range part of the EFT interactions. The difference lies in the parameterizations of short-distance details. While in an EFT these details are subsumed in the couplings that accompany the low-energy operators of increasing dimensions, models make specific assumptions about the high-energy components of the interaction, e.g., by the inclusion of heavy degrees of freedom. If low-energy observables are independent of these short-distance details, it could be argued that the difference between hybrid and consistent EFT calculations represents higher-order effects in the EFT expansion and should therefore be small. However, without a consistent power counting in place for both wave functions and operators, there is no mechanism for predicting the hybrid method's accuracy. A number of comparisons using different phenomenological potentials in combination with EFT operators have been performed to assess the impact of short-distant physics on low-energy observables (see, e.g., Refs.~\cite{Park:1995ku,Beane:1997iv,Beane:1999uq,Phillips:2003jz,Kim:2005kd,Song:2008zf}). These studies found that in general hybrid methods work well numerically.  Some  applications of the hybrid method in few-body systems include muon capture on deuterium and \HeThree \cite{Ando:2001es,Gazit:2008vm,Marcucci:2010ts}, neutrino reactions on \HeFour \cite{Gazit:2007jt}, and recently the calculation of magnetic moments and M1 transitions in nuclei with up to $A=9$ nucleons \cite{Pastore:2012rp}. Hybrid calculations have also been employed for hadronic parity violation, see, e.g., Refs.~\cite{Hyun:2001yg,Liu:2006dm,Hyun:2006cb,Schiavilla:2008ic,Song:2010sz,Viviani:2010qt,Song:2012yx}.

Hybrid calculations  have been performed for few-body systems where EFT calculations are currently unavailable. However, there are a number of potential pitfalls in hybrid calculations. For example, the models used to derive the wave functions and the EFTs that form the basis of the operators in most cases contain different degrees of freedom. This could be interpreted as a mismatch in the short-distance resolution. It is also not clear how consistent the different treatment of wave functions and operators is. For example, it may be that certain contributions are incorrectly included when operators derived in one approach are evaluated between wave functions derived in a different framework. It is common to use different regularizations in different parts of the calculations, which requires special care in the renormalization procedure to ensure that results are not regularization dependent. It is particularly unlikely that the combination of pionless EFT operators with model wavefunctions will yield sensible results.  Fortunately, there appears to be no barrier to proceeding with many-body techniques applied to chiral potentials and chiral wavefunctions, and in fact the community is moving in this direction.

%%%%%%%%%%%%%%%%%%%%%%%%%%%%%%%%%%%%%%%%%%%%%%%%%%%%%%%%%%%%%

\subsection{Models}\label{subsec:models}

Following the discovery of parity violation in beta decay, Feynman and Gell-Mann proposed a ``universal'' four-fermion interaction \cite{Feynman:1958ty} that accounts for beta and muon decay as well as the decays of other mesons and hyperons. The theory also predicts the existence of a parity-violating component in the interaction of two nucleons that is first order in the weak coupling. The proposed current-current form of the weak interaction was then combined with general symmetry arguments to derive a first PV nucleon-nucleon potential \cite{BlinStoyle:1960zz}. 

Subsequently, going back to work by Michel \cite{Michel:1964zz}, PV nucleon potentials were described as arising from meson exchanges, with one of the vertices describing a PV meson-nucleon coupling. At low energies the potential should be dominated by the exchange of light mesons. It is customary to include charged pions as well as $\rho$ and $\omega$ mesons. The exchange of a neutral pion is not considered, as a PV coupling of a neutral scalar or pseudoscalar meson to an on-shell nucleon would also violate CP \cite{Barton:1961eg}. The general form of the parity-violating but time-reversal conserving potential in the nonrelativistic limit is then given by (see, e.g., \cite{Henley:1977qn,Desplanques:1979hn}) 
\begin{align}\label{eq:ddhpotential}
V_\text{PV} = & i \frac{h_\pi^1 g_A M}{\sqrt{2} F_\pi} \left( \frac{\VT_1 \times \VT_2}{2} \right)_z (\VS_1 + \VS_2)\cdot \left[\frac{\pv_1-\pv_2}{2M},w_\pi(r) \right] \notag \\
& -g_\rho \left( h_\rho^0 \, \VT_1\cdot \VT_2 + h_\rho^1  \left( \frac{\VT_1+ \VT_2}{2} \right)_z +h_\rho^2 \, \frac{3\tau_1^z \tau_2^z - \VT_1\cdot \VT_2}{2\sqrt{6}} \right) \notag\\
& \times \left( \left(\VS_1-\VS_2  \right) \cdot \left\{ \frac{\pv_1-\pv_2}{2M},w_\rho(r) \right\} +i \left(1+\chi_\rho\right)  \left(\VS_1 \times \VS_2\right) \cdot \left[\frac{\pv_1-\pv_2}{2M},w_\rho(r) \right] \right) \notag\\
& - g_\omega  \left( h_\omega^0 + h_\omega^1  \left( \frac{\VT_1+ \VT_2}{2} \right)_z \right) \\
& \times \left( \left(\VS_1-\VS_2  \right) \cdot \left\{ \frac{\pv_1-\pv_2}{2M},w_\omega(r) \right\} +i \left(1+\chi_\omega\right)  \left(\VS_1 \times \VS_2\right) \cdot \left[\frac{\pv_1-\pv_2}{2M},w_\omega(r) \right] \right) \notag \\
& - \left(g_\omega h_\omega^1 - g_\rho h_\rho^1 \right) \left( \frac{\VT_1- \VT_2}{2} \right)_z \left(\VS_1+\VS_2 \right) \cdot  \left\{ \frac{\pv_1-\pv_2}{2M},w_\rho(r) \right\} \notag\\
& -g_\rho h_\rho^{\prime 1} i \left( \frac{\VT_1 \times \VT_2}{2} \right)_z  \left(\VS_1+\VS_2 \right) \cdot  \left[ \frac{\pv_1-\pv_2}{2M},w_\rho(r) \right]. \notag
\end{align}
Using the conventions of Refs.~\cite{Zhu:2004vw,RamseyMusolf:2006dz}, the coupling constants $h_M^i$ denote the PV interactions of meson $M$ with a nucleon, the superscript $i$ indicates the isospin structure of the corresponding operators, and the $g_M$ stand for PC couplings. The pion-nucleon coupling has been expressed in terms of the axial-vector coupling $g_A$ and the pion-decay constant $F_\pi$ via the Goldberger-Treiman relation \cite{Goldberger:1958tr,Nambu:1960xd}. In this convention $F_\pi = 92.4\, \MeV$.\footnote{While it is common to write the PV $\pi NN$ coupling as $f_\pi$, this is not done here to avoid confusion with the pion-decay constant in SU(2) chiral perturbation theory.}  The $\chi_M$ are the ratio of Pauli and Dirac couplings of meson $M$.  In addition, $\pv_i = -i\nabla_i$  is the momentum operator for nucleon $i$, $r = \vert \vec{x}_1-\vec{x}_2 \vert$, and the $w_M(r)$ stand for Yukawa functions, 
\begin{equation}
w_M(r) = \frac{\exp(-m_M r)}{4\pi r}\ \ .
\end{equation}

The PV $MNN$ couplings correspond to the matrix element of the weak Hamiltonian between one-nucleon and nucleon-meson states,
\begin{equation}\label{eq:weakmatrixelement}
\langle N \vert \mathcal{H}_\text{weak} \vert NM \rangle.
\end{equation}
They incorporate the short-distance details not captured in the description of the potential in terms of nucleons and mesons. This includes the exchanges of weak gauge bosons, but also the strong interaction effects that lead to the formation of nucleon and meson states. A direct calculation of these effects in terms of QCD has not been achieved. Various approaches and models have been proposed to calculate the matrix elements in Eq.~\eqref{eq:weakmatrixelement}. The most widely used calculation was performed by Desplanques, Donoghue, and Holstein (DDH) \cite{Desplanques:1979hn}, who combined quark model, current algebra, and symmetry arguments to estimate the PV meson-nucleon couplings. Accounting for uncertainties, they found a broad ``reasonable range''  for each of the couplings $h_M^i$ with the exception of ${h_\rho^1}^\prime$. The authors of Ref.~\cite{Desplanques:1979hn} also list ``best'' values for the couplings, although they caution that these ``may seem little more than guesses.'' The results are shown in Tab.~\ref{tab:ddhvalues}.
\begin{table}
\begin{center}
\begin{tabular}{|ccc|}
\hline
Coupling & Reasonable range & ``Best'' value \\ 
&  (each ${}\times 3.8\times 10^{-8}$) & (each ${}\times 3.8\times 10^{-8}$) \\ \hline
$h_\pi^1$             & $0 \to 30$         & 12 \\
$h_\rho^0$          & $30 \to -81$   & -30 \\
$h_\rho^1$          & $-1 \to 0$      & -0.5 \\
$h_\rho^2$          & $-20 \to -29$ & -25 \\
$h_\omega^0$   & $15 \to -27$ & -5 \\
$h_\omega^1$   & $-5 \to -2$ &-3 \\ \hline
\end{tabular}
\end{center}
\caption{DDH \cite{Desplanques:1979hn} ``reasonable'' ranges and ``best'' values for the PV couplings $h_M^i$.}\label{tab:ddhvalues}
\end{table}
Note in particular the value $h_\pi^1 = 4.6 \times 10^{-7}$ for the PV pion-nucleon coupling; its measurement continues to be a primary experimental objective. The coupling ${h_\rho^1}^\prime$ was estimated separately and is expected to only give an insignificant contribution to observables \cite{Holstein:1981cg}. It is therefore commonly neglected. A variety of other calculations to estimate the PV couplings, in particular of $h_\pi^1$,  exist in the literature, e.g., based on chiral soliton and Skyrme models \cite{Kaiser:1988bt,Kaiser:1989fd,Kaiser:1989ah,Meissner:1998pu,Lee:2012fx},
sum rules \cite{Khatsimovsky:1986xp,Henley:1995ad}, and holographic QCD \cite{Gazit:2008gz}. While other sets of coupling values have been proposed, see, e.g., Refs.~\cite{Dubovik:1986pj,Feldman:1991tj}, the DDH set still remains the most widely used determination of these couplings. A different approach was taken by Bowman \cite{Bowman:int07}, who used available experimental information to extract values for some of the meson-nucleon couplings. As not enough reliable results were available to fit all parameters, additional constraints were applied. Since the couplings $h_\rho^1$ and $h_\omega^1$ only enter observables with small numerical coefficients, they were fixed at their DDH ``best values,'' with the reasonable ranges taken into account in the experimental errors. The remaining constants were then extracted from a fit to 10 experimental results ranging from low-energy $pp$ scattering to asymmetries in $^{181}$Ta and the anapole moment of $^{133}$Cs. The resulting values are listed in Tab.~\ref{tab:bowman}, and not all are consistent with the DDH estimated ranges. However, it should be noted that the fit used a large range of different systems and relied on use of a PV nuclear one-body potential. Further, as Bowman points out, uncontrolled
errors may arise in trying to extract the DDH couplings, which are coefficients of a two-nucleon Lagrangian, from the many-body systems in which  measurements are made.
\begin{table}
\begin{center}
\begin{tabular}{|ccc|}
\hline
Coupling & Value & Error \\ 
&  (each ${}\times 3.8\times 10^{-8}$) & (each ${}\times 3.8\times 10^{-8}$) \\ \hline
$h_\pi^1$             & $-1.20$   &  2.40 \\
$h_\rho^0$          & $-114$ & 23.1    \\
$h_\rho^2$          & $97.6$ & 33.8 \\
$h_\omega^0$   & $36.0$ & 24.7  \\ \hline
\end{tabular}
\end{center}
\caption{Values for the PV couplings $h_M^i$ from fit to experimental results \cite{Bowman:int07}. $h_\rho^1$ and $h_\omega^1$ fixed at DDH best values.}\label{tab:bowman}
\end{table}

The potential of Eq.~\eqref{eq:ddhpotential} includes single-meson exchanges and nucleon, $\pi$, $\rho$, and $\omega$ degrees of freedom. 
While the DDH model was an important step towards
unifying the description of PV in nuclear systems, more recent understanding of relevant distance scales requires us to re-interpret its components.  
Low-energy parity violation depends upon long-distance physics and should be
independent of the description of short-distance effects.   The $\rho$ and $\omega$,
for example,  ``particles" in the DDH model are  used to model this short-distance physics. However, this corresponds to certain assumptions about these short-distance details, which may not always be appropriate. In fact, given that a PV potential with a specific short-distance form is often combined with a variety of PC potentials that differ in their short-distance description, it is possible that a mismatch of the short-distance details of the PC and PV frameworks leads to different results \cite{Desplanques:1976mt}.
Several extensions of the potential have been proposed to also include two-meson-exchange contributions (see, e.g., Refs.~\cite{Pirner:1973wy,Chemtob:1973uj,Niskanen:2007hf} and references therein) and additional degrees of freedom such as the $\Delta$ resonance \cite{Silbar:1989gk,Feldman:1991tj,Iqbal:1992xm}. Given the model dependence and the difficulties in determining reliable values for the various couplings, the importance of these additional terms is not clear.

The one-meson-exchange potential has been used to calculate PV observables in a wide range of physical systems, from proton-proton scattering to anapole moments in nucleon-rich nuclei. If the experimental results are used to extract values for the PV couplings, individual results tend to lie within the DDH ranges. However, the agreement between different experiments is not as clearly established.  For example, the value of the isovector combination of meson-nucleon couplings as extracted from $^{133}\text{Cs}$ seems to differ from values based on extractions in other systems. Whether this discrepancy is due to approximations in the shell-model calculations required to describe the $^{133}\text{Cs}$ nucleus or due to inconsistencies in the description of the PV couplings is currently unclear \cite{Haxton:2001mi}.

To avoid some of these obstacles, one can forego any model assumptions about the short-distance details and instead use PV transition amplitudes without reference to any underlying mechanism. At very low energies, S-P wave transitions should dominate and the corresponding amplitudes were delineated in Ref.~\cite{Danilov:1965,Danilov:1971qg,Danilov:1971fh}. Danilov suggested that the energy dependence of the weak amplitudes at low energies is dominated by strong interaction effects, and  that the PV amplitude can therefore be written as a set of constants $c$, $\lambda_t$, etc.~parameterizing the PV interactions multiplying the appropriate PC scattering amplitudes. For the case of neutron proton interactions the PV amplitude is parameterized as \cite{Danilov:1965} (with slightly different notation)
\be
\begin{split}
A_{PV}^{np}(\kv,\kv^\prime) & =  c\, a_t(k) (\VS_p+\VS_n) \cdot (\kv + \kv^\prime) + \half \lambda_t a_t(k) \left[(\VS_p-\VS_n)\cdot (\kv + \kv^\prime)-i (\VS_p\times\VS_n)\cdot (\kv^\prime-\kv)\right]\\
& \quad +\half \lambda_s^{np} a_s(k) \left[(\VS_p-\VS_n)\cdot (\kv + \kv^\prime)+i (\VS_p\times\VS_n)\cdot (\kv^\prime-\kv)\right],
\end{split}
\ee
where $a_s(k)$ and $a_t(k)$ are the PC scattering amplitudes in the \oneS and \threeS channels, respectively, and $\kv,\kv^\prime$ are the initial and final momenta in the center of mass. The three constants $c$, $\lambda_t$, and $\lambda_s^{np}$ encode parity violation for the $\threeS - \threePone$, $\threeS - \oneP$, and $\oneS - \threePzero$ channels. This analysis was extended in Refs.~\cite{Moskalev:1968a,Missimer:1975wb,Desplanques:1976mt} to also describe $nn$ and $pp$ amplitudes with the introduction of two additional parameters $\lambda_s^{nn}$ and $\lambda_s^{pp}$ in the $\oneS - \threePzero$ channel, bringing the total to 5 independent terms. The three $\oneS - \threePzero$ parameters can also be expressed in terms of the total isospin $\Delta I = 0,1,2$, resulting in a parameterization in terms of the constants
\be
c,\lambda_t,\lambda_s^0, \lambda_s^1, \lambda_s^2.
\ee

While this approach relates the energy dependence of the weak amplitudes to the much better determined strong ones, it does not make any predictions for the PV constants. In order to avoid any model assumptions it was suggested that the PV constants be determined from a number of experiments in the two-nucleon sector \cite{Missimer:1975wb}. Unfortunately, due to a lack of feasible measurements so far this has not been accomplished. The philosophy behind this approach is very similar to the ideas of effective field theory, in particular \eftnopi as discussed in Sec.~\ref{nopimethod}.

In order to extend this model-independent approach to higher energies, Desplanques and Missimer combined the five low-energy amplitudes with a one-pion exchange contribution \cite{Desplanques:1976mt}. The reasoning behind this approach is that pions start to become dynamical at energies above about 20 MeV and need to be included in the description of the long-range component of the PV interaction. At the same time, this approach maintains the advantage of being largely independent of the details of short-range interactions, as it avoids using heavier meson exchanges. PV interactions are thus described in terms of six parameters: the five constants corresponding to the Danilov parameters, plus a PV pion-nucleon coupling. Note that this corresponds to separating the Danilov \threeS-\threePone coupling into two parts: a short-range contribution (which is now different from the one in the Danilov approach) plus one-pion exchange. This approach can be regarded as the precursor of the chiral PV EFT approach described in Sec.~\ref{sec:chiralEFT}.

%%%%%%%%%%%%%%%%%%%%%%%%%%%%%%%%%%%%%%%%%%%%%%%%%%%%%%%%%%%%%

\subsection{Relations between different formalisms}\label{sec:translate}

One goal of this review is to help make it possible to interpret few-body hadronic PV experiments in terms of a unifying theoretical description.  
Different authors have used different underlying assumptions, different calculational strategies, or simply different variable choices for parameters.  
Since the 1980's, most PV calculations have been expressed in terms of the DDH parameters.  
More recently EFT descriptions, both with and without explicit pion degrees of freedom, have been adopted to ensure consistency between PC and PV interactions and currents.  
Finally, instead of using the Lagrangian directly, hybrid calculations use a potential derived from the EFT Lagrangian combined with models for the PC interactions. In this section we attempt to create a dictionary, to the extent possible, that translates from one language to the next, e.g., from the DDH parameters to (various conventions for) the LECs.  
There are some inherent uncertainties involved, particularly when cutoffs and subtraction points in one scheme are not compatible with another, so some of these translations cannot be considered exact and should be interpreted carefully.\footnote{We thank J.~Vanasse for his work on connecting different formalisms and for many helpful discussions on this topic.}

The PV \eftnopi potential of Ref.~\cite{Schiavilla:2008ic}, which is also used in the hybrid calculations of Refs.~\cite{Song:2010sz,Song:2012yx}, reads
\be
\begin{split}\label{eq:GirPot}
V^{Gir}_{12} & = \frac{1}{\Lambda_\chi^3}\left [ 2C_1  +  (C_2+C_4) (\tau_1 + \tau_2)_z  + 2 C_5 \mathcal{I}_{ab} \tau_1^a \tau_2^b \right] (\vec{\sigma}_1 - \vec{\sigma}_2) \cdot \vec{X}_{12,+}  \\
& \quad +\frac{1}{\Lambda_\chi^3} \left [ 2\tilde{C}_1 (\vec{\sigma}_1 \times \vec{\sigma}_2) +2 C_6 \epsilon_{3ab} \tau_1^a \tau_2^b (\vec{\sigma}_1 + \vec{\sigma}_2) \right] \cdot \vec{X}_{12,-} \ , 
\end{split}
\ee
where 
\be
\vec{X}_{12,+} = \left\{ -i\vec{\nabla} ,\frac{\mu_P^2}{4\pi r} e^{-\mu_P r} \right\},\quad \vec{X}_{12,-} = i \left[ -i\vec{\nabla},\frac{\mu_P^2}{4\pi r} e^{-\mu_P r} \right].
\ee

The authors of Ref.~\cite{Schiavilla:2008ic} derived this form of the potential from the Lagrangian of Eq.~\eqref{STversion}. 
(In an earlier version by one of the authors of Ref.~\cite{Schiavilla:2008ic}, the LECs $(C_2+C_4)$ are replaced simply by $C_2$, which corresponds to the notation in Eq.~\eqref{STversion} with $\calG_i = C_i/\Lambda_\chi^3$).
However, because of the need to regularize the potential and due to minor misprints, the LECs in the potential do not exactly correspond to those of the Lagrangian \cite{Vanasse:2011nd,Vanasse:Priv}. 
A straightforward derivation of the potential with only the terms of the LO Lagrangian would result in $\delta$-functions instead of the regulator functions
\be\label{eq:GirReg}
f_{\mu_P}(r) = \frac{\mu_P^2}{4\pi r} e^{-\mu_P r}
\ee
of Eq.~\eqref{eq:GirPot}.
While 
$$
\lim_{\mu_P \to \infty}  f_{\mu_P}(r)= \delta^3(\vec{r}),
$$
in hybrid calculations the regulator $\mu_P$ is conventionally chosen in the region $138\,\MeV \le \mu_P \le 1\,\text{GeV}$. 
This introduces some regulator dependence in the translation from the Lagrangian to the potential parameters. 
(Alternatively, the regulator function should in principle correspond to including the summation of an infinite number of higher-order terms in the Lagrangian.)
With this caveat, the relation between the two sets of parameters of Eqs.~\eqref{STversion} and \eqref{eq:GirPot} is given by
\be
\calG_1 =\frac{1}{\Lambda_\chi^3} C_1, \quad \tilde \calG_1  = \frac{1}{\Lambda_\chi^3}\tilde C_1, \quad \calG_2  = \frac{1}{2}\frac{1}{\Lambda_\chi^3}(C_2+C_4), \quad \tilde \calG_5  = \frac{1}{\Lambda_\chi^3} C_5, \quad \calG_6  = -2 \frac{1}{\Lambda_\chi^3}C_6.
\ee
In applications, results using the potential of Ref.~\cite{Schiavilla:2008ic} are conventionally presented in terms of coefficients $c_n^\slashed{\pi}$, which are given by
\begin{align}
c_1^\slashed{\pi} & = 2\frac{\mu_P^2}{\Lambda_\chi^3} C_6, &
c_4^\slashed{\pi} & = \frac{\mu_P^2}{\Lambda_\chi^3} (C_2+C_4), &
c_6^\slashed{\pi} & = -2\frac{\mu_P^2}{\Lambda_\chi^3} C_5, \notag \\ 
c_8^\slashed{\pi} & = 2\frac{\mu_P^2}{\Lambda_\chi^3} C_1, &
c_9^\slashed{\pi} & = 2\frac{\mu_P^2}{\Lambda_\chi^3} \tilde C_1.
\end{align}
Combined with Eqs.~\eqref{eq:wave2Gir} and \eqref{eq:dib2wave}, these expressions also give the connection between the LECs in the partial wave basis and the 
potential parameters, see Tab.~\ref{tab:couplings}. 
\begin{table}
{    \renewcommand{\arraystretch}{2.2}
\begin{tabular}{|c|c|c|c|c|}  
\hline
$\calL_{PV}^d$ [Eq.~\eqref{eq:PVLag}] & $\calL_{PV}$ [Eq.~\eqref{Lag:PV}] & $\calL_{PV}^{Gir}$ [Eq.~\eqref{STversion}] & Hybrid \cite{Schiavilla:2008ic} & Danilov    \\[0.5ex] \hline
$\gA$ 
	& $\sqrt{8}\dfrac{\Delta_t}{y_t}\CA(\mu)$ 
	& $ \dfrac{\sqrt{8}}{4}\dfrac{\Delta_t}{y_t}\left( \calG_1(\mu) - \tilde\calG_1(\mu) \right)$ 
	& $ \dfrac{\sqrt{8}}{8}\dfrac{\Delta_t}{y_t}\dfrac{1}{\mu_P^2}\left( c^\slashed{\pi}_8 - c^\slashed{\pi}_9 \right)$ 
	& $ - \dfrac{y_t}{\sqrt{8}} \lambda_t$  \\[1ex] \hline
$\gB$ 
	& $\sqrt{8}\dfrac{\Delta_s}{y_s}\CB(\mu)$ 
	& $\dfrac{\sqrt{8}}{4}\dfrac{\Delta_s}{y_s}\left( \calG_1(\mu) + \tilde\calG_1(\mu) \right)$ 
	& $\dfrac{\sqrt{8}}{8}\dfrac{\Delta_s}{y_s}\dfrac{1}{\mu_P^2}\left( c^\slashed{\pi}_8 + c^\slashed{\pi}_9 \right)$ 
	& $ - \dfrac{y_s}{\sqrt{8}}\lambda_s^0$  \\[1ex] \hline
$\gC$ 
	& $\sqrt{8}\dfrac{\Delta_s}{y_s}\CC(\mu)$ 
	& $\dfrac{\sqrt{8}}{2}\dfrac{\Delta_s}{y_s}  \calG_2(\mu) $ 
	& $\dfrac{\sqrt{8}}{4}\dfrac{\Delta_s}{y_s}\dfrac{1}{\mu_P^2} c^\slashed{\pi}_4 $ 
	& $ - \dfrac{y_s}{\sqrt{8}} \lambda_s^1$  \\[1ex] \hline
$\gD$ 
	& $\sqrt{8}\dfrac{\Delta_s}{y_s}\CD(\mu)$ 
	& $-\dfrac{\sqrt{8}}{2}\dfrac{\Delta_s}{y_s}  \tilde\calG_5(\mu) $ 
	& $\dfrac{\sqrt{8}}{4}\dfrac{\Delta_s}{y_s}\dfrac{1}{\mu_P^2}  c^\slashed{\pi}_6 $ 
	& $ - \dfrac{y_s}{4\sqrt{3}} \lambda_s^2$  \\[1ex] \hline
$\gE$ 
	& $\sqrt{8}\dfrac{\Delta_t}{y_t}\CE(\mu)$ 
	& $\dfrac{\sqrt{8}}{4}\dfrac{\Delta_t}{y_t} \calG_6(\mu) $ 
	& $ - \dfrac{\sqrt{8}}{4}\dfrac{\Delta_t}{y_t}\dfrac{1}{\mu_P^2}  c^\slashed{\pi}_1 $ 
	& $ \dfrac{y_t}{\sqrt{8}} \rho_t$  \\[1ex] \hline
\end{tabular}
}	
\caption{Naive translation among PV couplings. In this review the conventions $y_s=y_t=y=\sqrt{\frac{4\pi}{M}}$, $\Delta_{s/t} = \gamma_{s/t} -\mu$ are used. \label{tab:couplings}}
\end{table}

In principle, these relations provide a comparison between results obtained in EFT and hybrid calculations. 
However, there are some additional caveats.
The authors of Refs.~\cite{Schiavilla:2008ic,Song:2010sz,Song:2012yx} advocate the value $\mu_P = 138\,\MeV$ as appropriate for a pionless potential.
While observables in EFT are independent of the regularization scale $\mu$, the values of the couplings are in general scale-\emph{dependent}. 
Again, $\mu = 138\,\MeV$ is often considered an appropriate scale, but it is important to keep in mind that there is no straightforward relation between the 
mass parameter $\mu_P$ and the scale $\mu$ in dimensional regularization.
This intrinsic scale-dependence, coupled with the congenital scale dependence in deriving the relations between the Lagrangian and potential parameters, makes it difficult to perform reliable comparisons between EFT and hybrid calculations.

\begin{table}
\begin{center}
\begin{tabular}{|c|c|c|c|c|c|c|}
\hline
n & \pbox{5cm}{Hybrid \cite{Schiavilla:2008ic} \\$\mu_P = 138\,\MeV$} & $\mu = 100\, \MeV$ & $\mu = 125 \,\MeV$ & $\mu = 138\, \MeV$ & $\mu = 170\, \MeV$ & $\mu = 200\, \MeV$  \\ \hline
1 & 63.2 & 85.0 & 118.4 & 136 & 178 & 219 \\
4 & 57.8 & 42.4 & 52.2 & 57.3 & 69.8 & 81.5 \\
8 & -75.2 & -53.2 & -68.9 & -77.1 & -97.3 & -116 \\
9 & -6.11 & -10.4 & -9.33 & -8.77 & -7.40 & -6.12 \\ \hline
\end{tabular}
\caption{Values for $I_n$ of Eq.~\eqref{eq:ndCompare}. Second column: hybrid results, following columns: LO \eftnopi results. All values in fm. \label{tab:RegDepCompare}}
\end{center}
\end{table}

To demonstrate the regulator dependence in the matching, consider as an example neutron-deuteron spin rotation (see Sec.~\ref{sec:ndspinrot} for details of the different calculations). 
The rotation angle per unit length can be written as 
\begin{equation}\label{eq:ndCompare}
\frac{1}{\rho}\,\frac{d\phi}{dl} =   \sum_n c_n I_n,
\end{equation}
where $\rho$ is the target density.
Table~\ref{tab:RegDepCompare}  contains the results of the hybrid calculation of Ref.~\cite{Schiavilla:2008ic} using the AV18 and UIX potentials in combination with the potential of Eq.~\eqref{eq:GirPot} compared with those of a LO \eftnopi calculation \cite{Griesshammer:2011md,Vanasse:2011nd}. 
For this comparison, $\mu_P=138\,\MeV$ and in the \eftnopi calculations the cutoff in the solution of the 3N equations is chosen as $\Lambda = 1500\,\MeV$.
While the original \eftnopi results are entirely independent of $\mu$, there is clear regulator dependence in the translated results.
In addition, the hybrid results for  $\mu_P=1\,\text{GeV}$ differ by those with $\mu_P=138\,\MeV$ by up to an order of magnitude \cite{Schiavilla:2008ic}, demonstrating further why a translation between the different formalisms is ambiguous at best.

For completeness the relations to the zero-range amplitudes in the Danilov formalism are also included in Tab.~\ref{tab:couplings}. Reference~\cite{Vanasse:2011nd} also includes a translation from the LECs to the DDH parameters, which is given by
\begin{align}
\frac{\gA}{y_t} & = -\frac{M\left( \frac{1}{a_t} - \mu_P \right)}{8\sqrt{2}\pi} \left[ \frac{g_\omega\chi_\omega}{Mm_\omega^2}h_\omega^0 - \frac{3g_\rho \chi_\rho}{Mm_\rho^2}h_\rho^0\right]  \ ,\\
\frac{\gB}{y_s} & = \frac{M\left( \gamma_s - \mu_P \right)}{8\sqrt{2}\pi} \left[ \frac{g_\omega(2+\chi_\omega)}{Mm_\omega^2}h_\omega^0 + \frac{g_\rho (2+\chi_\rho)}{Mm_\rho^2}h_\rho^0\right]  \ ,\\
\frac{\gC}{y_s} & = \frac{M\left( \gamma_s - \mu_P \right)}{8\sqrt{2}\pi} \left[ \frac{g_\omega(2+\chi_\omega)}{Mm_\omega^2}h_\omega^1 + \frac{g_\rho (2+\chi_\rho)}{Mm_\rho^2}h_\rho^1\right]  \ ,\\
\frac{\gD}{y_s} & = \frac{M\left( \gamma_s - \mu_P \right)}{8\sqrt{2}\pi} \left[ \frac{g_\rho (2+\chi_\rho)}{\sqrt{6}Mm_\rho^2}h_\rho^2\right]  \ ,\\
\frac{\gE}{y_t} & = \frac{M\left( \frac{1}{a_t} - \mu_P \right)}{8\sqrt{2}\pi} \left[ \frac{g_{\pi NN}}{\sqrt{2}Mm_\pi^2}h_\pi^1 + \frac{g_\rho}{M m_\rho^2}h_\rho^1 - \frac{g_\omega}{M m_\omega^2}h_\omega^1\right]  \ .
\end{align}
Here, $a_t$ and $a_s = 1/\gamma_s$ are the \threeS and \oneS scattering lengths, respectively. The same caveats as in the comparison between hybrid and EFT results applies to these relations, as they rely on the matching of different forms of the potential and include implicit scale dependence. They should therefore only be viewed as order of magnitude estimates.

%%%%%%%%%%%%%%%%%%%%%%%%%%%%%%%%%%%%%%%%%%%%%%%%%%%%%%%%%%%%%

\subsection{Lattice QCD}

Hadronic parity violation is governed by Standard Model dynamics; in principle the PV nucleon couplings can be determined from the strong and weak interactions at the quark level. 
In practice, the nonperturbative nature of QCD at low energies complicates this approach significantly. 
However, strong interaction dynamics at low energies can be calculated nonperturbatively in lattice QCD; see, e.g., Refs.~\cite{Rothe:1992nt,Gupta:1997nd,Gattringer:2010zz} and references therein. 
The application of lattice QCD to nuclear physics problems continues to be a very active area of research; see, e.g., Ref.~\cite{Beane:2010em} and references therein.
Because of high computational costs, though, direct calculations of few-nucleon reactions on the lattice are still far in the future.
An alternative approach that appears more feasible on shorter time scales is to determine LECs from lattice QCD, which are then used in EFT calculations of hadronic observables. 
In the PC sector, scattering parameters such as the scattering lengths and effective ranges can be determined on the lattice. These in turn are related to the EFT LECs. However, the relation between the LECs and the scattering parameters is regularization (and renormalization) scheme dependent. While in some cases, such as in \eftnopi, the regulator dependence of a number of LECs is known analytically, in other cases this dependence can so far only be studied numerically. For some of the LECs in chiral EFT the running of the LEC with the regulator is significant. Therefore, special care has to be taken in relating the LECs to lattice results. Similar complications are expected in the PV sector. 
See, e.g., Ref.~\cite{Beane:2013br} for a recent determination of 2N scattering parameters.
The combination of lattice and EFT methods can provide a systematic link from the underlying Standard Model dynamics to, e.g., few-nucleon systems.
While a model-independent determination of LECs can also be achieved by extraction from experiments, the approach based on lattice  is of particular interest where experimental results are difficult to obtain, such as in the case of hadronic parity violation.
The application of lattice QCD to hadronic parity violation was first addressed in Ref.~\cite{Beane:2002ca}.  
To prepare for a future lattice simulation with unphysical quark masses, the authors provided expressions for  the PV $\pi N$ coupling $h_{\pi NN}^{(1)}$  and for the anapole form factor of the proton using partially-quenched QCD. 
A first result for $h_{\pi NN}^{(1)}$ was recently presented in Ref.~\cite{Wasem:2011tp}.

Since lattice QCD generally only considers the three light quark flavors as dynamical degrees of freedom, the Standard Model PV quark operators to be evaluated have to be evolved to the hadronic scale. 
This is achieved by first integrating out the weak boson, followed by the heavy-quark flavors by renormalization group running \cite{Dai:1991bx,Kaplan:1992vj,Tiburzi:2012hx,Tiburzi:2012xx}. 
The resulting four-quark operators containing only $u$, $d$, and $s$ quarks can then be used in lattice calculations. 
In order to extract hadronic coupling constants, correlation functions of these PV operators are evaluated in combination with a suitable choice of interpolating operators for the hadronic states needed.

\begin{figure}
\begin{center}
\includegraphics[width=0.7\textwidth]{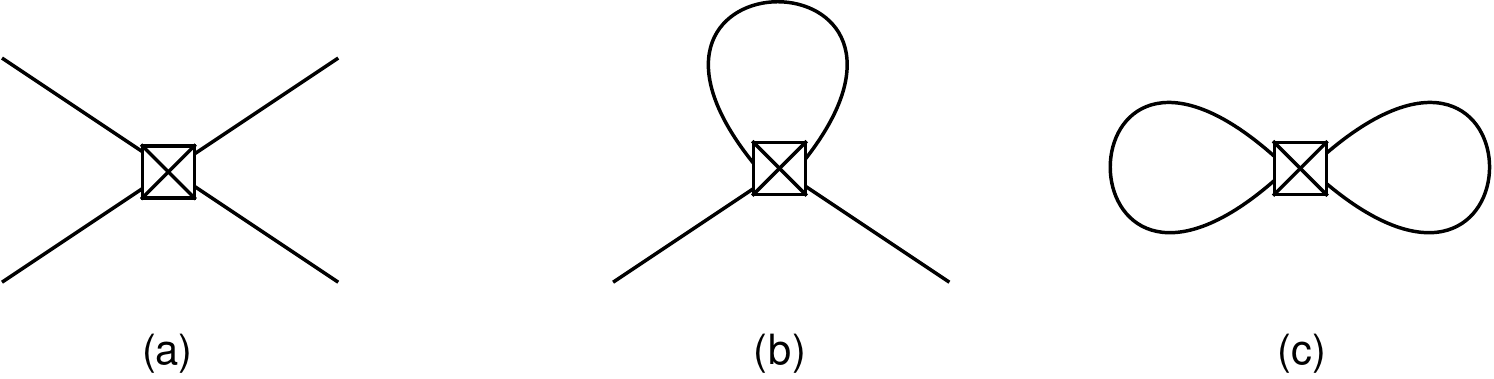}
\end{center}
\caption{PV quark-level diagrams. The crossed square represents the PV four-quark vertex. (a) Connected, (b) quark-loop, (c) disconnected. \label{fig:lattice}}
\end{figure}

The first lattice calculation of hadronic parity violation was performed in Ref.~\cite{Wasem:2011tp}, which determined the PV $\pi N$ coupling $h_{\pi NN}^{(1)}$. 
 The Lagrangian  is written as 
\be
\calL_{\pi NN}^\text{PV} = h_{\pi NN}^{(1)} (\bar{p}\pi^+ n -\bar{n} \pi^- p).
\ee
The coupling can be extracted from  three-point correlation functions
\be
C^{ij}_{A\to B}(t,t^\prime) = \langle 0| \calO_{B,j}(t) \calO^{\Delta I = 1}_\text{PV}(t^\prime) \calO_{A,i}(0) |0 \rangle,
\ee
where $\calO^{\Delta I = 1}_\text{PV}(t^\prime)$ represents a PV four-quark operator and the interpolating operators $\calO_{A,i}$ and $\calO_{B,j}$ create and destroy states with quantum numbers corresponding to a proton or a neutron and pion, respectively. Reference~\cite{Wasem:2011tp} uses three-quark operators in both cases, which reduces calculational costs. Three general types of diagrams appear due to the contraction of the available quark fields, see Fig.~\ref{fig:lattice}. 
The first type (``connected'' diagrams) connects each of the four quarks of the PV operator with quark fields of the interpolating operators. In the second type (``quark-loop'' diagrams), only two of the four quarks are connected to the interpolating operators while the remaining two are contracted with each other to form a quark loop. In the last type (``disconnected'' diagrams) all four quarks are contracted with each other, without any contraction with the interpolating quark operators. Since Ref.~\cite{Wasem:2011tp} works in the isospin limit of equal up- and down-quark masses, the disconnected diagrams cancel and do not contribute.  The calculation of the quark-loop diagrams shows large noise and no signal is extracted. Future improvements are expected to improve this situation. Therefore, the results of Ref.~\cite{Wasem:2011tp}  stems solely from the connected diagrams. With the calculation performed at a single lattice spacing and at a nonphysical pion mass of $\sim 389\,\MeV$, the PV pion-nucleon coupling is determined to be
\be
h_{\pi NN}^{(1),\text{con}} = \left( 1.099 \pm 0.505^{+0.058}_{-0.064}  \right) \times 10^{-7},
\ee
with the first uncertainty being statistical and the second one systematic. This value is consistent with the DDH range for $h_{\pi NN}^{(1)}$ and the values determined in other model calculations, as well as the current experimental bounds \cite{Wasem:2011tp}. 

The work of Ref.~\cite{Wasem:2011tp} presents an important first step in the model-independent determination of the PV couplings. With future improvements in computing power and algorithms, it is important to also consider the quark-loop diagrams and to extend the calculations to multiple sets of lattice parameters and pion masses. The $\Delta I = 2$ LECs present another opportunity for lattice QCD to have great impact on the better understanding of hadronic parity violation. Since observables in general depend on linear combinations of couplings with different values of $\Delta I$, removing uncertainty on the $\Delta I=2$ couplings would further constrain the remaining LECs. Since the $\Delta I=2$ couplings are very difficult to extract from experiments, a lattice determination can provide important constraints. A step towards preparing the SM operators for this effort is given in Ref.~\cite{Tiburzi:2012xx}. While quark-loop diagrams do not contribute,  the $\Delta I = 0$ sector is the least accessible in the lattice approach, as in principle it requires the calculation of disconnected diagrams which are computationally very expensive. 

The calculations discussed above show the potential impact that lattice QCD can have in the systematic study of hadronic parity violation. While computationally involved, with advancements in methods and computing power these and similar efforts can prove to be complementary to other ongoing theoretical and experimental efforts. The importance of lattice QCD to an improved understanding of hadronic parity violation was also recently acknowledged at a workshop on ``Forefront Questions in Nuclear Science and the Role of Computing at the Extreme Scale''\cite{GrandChallenge}. 
Crucially, the field of low energy hadronic parity violation affords the opportunity for lattice techniques to provide predictions in the absence of experimental data.

%%%%%%%%%%%%%%%%%%%%%%%%%%%%%%%%%%%%%%%%%%%%%%%%%%%%%%%%%%%%%

\section{Single-Nucleon Systems}\label{sec:1N}

The discovery of parity violation was made on a very heavy nucleus \cite{Wu:1957my}, yet we
now know that the underlying mechanism was a simple single-nucleon process,
the beta decay of a neutron: $n \rightarrow p e \nu^-$.  Similarly, for hadronic parity violation, naive power counting suggests that the dominant effect occurs via
few-nucleon PV interactions.  In particular, if pion dynamics are important then the
most important term is a single nucleon interaction.   This is the physics
origin of the famous DDH $h^1_\pi$ coupling  that has been the focus of considerable theoretical and experimental
effort. 
The leading-order term in the chiral Lagrangian is
\be
\calL=-{h^1_{\pi NN}  F \over 2 \sqrt{2}} \bar N X^3_- N = -i h^1_{\pi NN} (\bar p n \pi^+-\bar n p \pi^-) + \cdots \ \ , 
\ee
 where $X^3_-=\xi^\dagger \tau^3 \xi - \xi \tau^3 \xi^\dagger$.
Estimates of the value of $h^1_{\pi NN}$ and possible corrections to that value are discussed in Sec.~\ref{sec:chiralEFT}. 
The DDH $h_\pi^1$ and the EFT $h_{\pi NN}^1$ are related.  
They are both the coupling of the pion field to two nucleons.  In what follows, we attempt to
reproduce results found in the literature in terms of the EFT $h_{\pi NN}^1$ as defined in this review.  However,
it is not always possible to do this because not all authors  specify
their sign conventions.  In those cases we echo the result in
the literature and note any change in notation that is necessary to avoid
confusion. 

\subsection{Anapole Form Factor} \label{1anapole}

One way of extracting $h^1_{\pi NN}$  from a single nucleon observable is to consider the anapole moment \cite{Zeldovich:1957} of the nucleon.    The structure of the corresponding Lagrangian is \cite{Zhu:2000gn} 
\be
\calL^{AM}={e \over \Lambda_\chi^2} \overline N (a_s + a_v \tau_3) \gamma_\mu \gamma_5 N \partial_\nu F^{\nu \mu} \ \ , 
\ee
where $F^{\mu \nu}$ is the EM field strength tensor and $a_s$ and $a_v$ are the 
isoscalar and isotensor terms in the anapole operator.  
Refs.~\cite{Maekawa:2000qz,Maekawa:2000bd} obtain the nucleon anapole form factor via (for earlier work see Refs.~\cite{Haxton:1989ap,Savage:1999cm})
\be
J^\mu_{AM}={2 \over M^2} (a_0 F_A^{(0)}(-q^2)+ a_1 F_A^{(1)}(-q^2)\tau_3)(S^\mu q^2 - \vec S \cdot \vec q q^\mu) \ \ ,
\ee
 where $a_0$ and $a_1$ are the isoscaler and isotensor terms and the $F_A^{0,1}$ are the associated form factors as functions of the nucleon momentum transfer $q$.

The leading contribution  to the isoscalar term (adjusted to the conventions of Eq.~\eqref{NPV}) is \cite{Kaplan:1992vj}
\be 
a_0= - {e g_A h^1_{\pi NN} \over 48 \sqrt{2} \pi} {M^2 \over m_\pi F} \sim -0.43 \ h^1_{\pi NN} \ \ ,
\ee
while the isovector term $a_1=0.$

Ref.~\cite{Zhu:2000gn} uses HB$\chi$PT to calculate loop corrections and estimate counterterms to 
$\frac{1}{\Lambda_\chi^2}$. At subleading order additional operators are needed.  In particular there are 
now contact terms $\tilde a_0$ and $\tilde a_1$ \cite{Maekawa:2000bd}, 
\be 
\calL = {2 \over m_N^2} \bar N (\tilde a_0 + \tilde a_1 \tau_3) S_\mu N \partial_\nu F^{\mu \nu} \ \ , 
\ee
as well as dependence upon the LEC $h_A^{1}$ from Eq.~\eqref{NPV} in Sec.~\ref{sec:chiralEFT}.
The isovector anapole moment receives its first nonzero contribution at this
order  \cite{Maekawa:2000bd}, 
\be
a_1^{NLO} = {e M^2 \over 6 (4 \pi F)^2} \left[ 2 h_A^2 + g_A(h_V^0+{4\over 3} h_V^2)\right] \left[{\rm ln} \left( {\mu^2 \over m_\pi^2} \right) + {1 \over \epsilon} + 1 -\gamma -{2 \over 3} + {\rm ln} 4 \pi\right]  + \tilde a_1  \ \ ,
\ee
where the second bracket contains the usual dimensional regularization terms. The counterterm absorbs both the dependence on the  subtraction point $\mu$ and the pole
as $\epsilon \rightarrow 0$.  The PV LECs $h_A^2$, $h_V^0$, and $h_V^2$ are found in Eq.~\eqref{NPV} of Sec.~\ref{sec:chiralEFT}.

\subsection{Compton scattering}

The parameter $h^1_{\pi NN}$ is also the dominant contribution to PV asymmetries from Compton scattering on the proton.  Ref.~\cite{Bedaque:1999dh} considered the interference between PV and PC amplitudes that comes from polarizing the proton in
$\gamma \vec p$ scattering, while Ref.~\cite{Chen:2000mb} argued that the
faster switching of $\gamma$ polarization compared to proton polarization would
reduce systematic uncertainties for that observable.  
Adjusted to the conventions of Eq.~\eqref{NPV}, the asymmetry from proton polarization is estimated at forward angles from a one-loop calculation to be \cite{Bedaque:1999dh}
\be -1.5 \times 10^{-9} \left( {h^1_{\pi NN} \over 5 \times 10^{-7}} \right) \left({\omega \over 20 \ {\rm MeV}} \right)^2 \left( 1 + \calO \left({\omega^2 \over m_\pi^2}\right) \right) \ \  ,
\ee
where $\omega$ is the energy of the incoming photon.
For incoming photon polarization the asymmetry is
\be
A_{\gamma \gamma}(\omega, \theta) = { \left. {d \sigma \over d \Omega}\right\vert_{+} -\left. {d \sigma \over d \Omega}\right\vert_{-} \over \left. {d \sigma \over d \Omega}\right\vert_{+} +\left. {d \sigma \over d \Omega}\right\vert_{-}} \ \ ,
\ee
where  the $\pm $ subscripts indicate the incoming photon helicity. 
Ref.~\cite{Chen:2000mb} estimates a general size for low energies by considering $\theta= {\pi \over 2}$ and finds
\be
\left\vert A_{\gamma \gamma}( \omega \ll m_\pi,{\pi \over 2}) \right \vert \approx
8.8 \times 10^{-9} \left( {h^1_{\pi NN} \over 5 \times 10^{-7}} \right) \left({\omega \over 70 \ {\rm MeV}} \right)^3 \ \ 
\ee
up to corrections of about 25\%.

\subsection{Pion production}

A possibly more promising route to finding $h^1_{\pi NN }$ is to look at the process
$\vec \gamma p \rightarrow \pi^+ n$.  It appears to be clean theoretically, at least at leading order, and has the advantage of a large and well understood PC cross section.  Early meson exchange analyses were performed in Refs.~\cite{Woloshyn:1978qk,Li:1982fd}. Ref.~\cite{Woloshyn:1978qk} found a maximum asymmetry 
\be
A_\gamma={\sigma_+-\sigma_- \over  \sigma_+ + \sigma_-} \ \ ,
\ee
in terms of total cross sections for circularly polarized photons of $\pm$ helicity,
of $4 \times 10^{-7}$. 
 
Ref.~\cite{Chen:2000hb}  used HB$\chi$PT (including $\Delta$ degrees of freedom) to study the process.  The parity-conserving terms in the T matrix are
\be
T^{PC}=N^\dagger \left[ i \calA_1 \vec \sigma \cdot \vec \epsilon + i \calA_2 \vec \sigma \cdot \hat q \vec \epsilon \cdot \hat k + i \calA_3 \vec \sigma \cdot \hat k \vec \epsilon \cdot \hat k + \calA_4 \vec \epsilon \cdot \hat q \times \hat k \right] N  \ \ ,
\ee
where $\hat q$ and $\hat k$ are unit vectors in the direction of the incoming photon momentum and the outgoing pion momentum, respectively.  At leading order
only $A_{1-3}$ are nonzero but they reproduce available data to about 10\% up to photon energies of 200 MeV (see  Fig.~1 in Ref.~\cite{Chen:2000hb}).  The 
PV violating terms in the T matrix are \cite{Chen:2000hb,Zhu:2000hf}
\be
T^{PV}=N^\dagger \left[ \calF_1\hat k \cdot \hat  \epsilon + i \calF_2 \vec \sigma \cdot \hat \epsilon \times \hat q + i \calF_3 \vec \sigma \cdot \hat \epsilon \times \hat k \right] N \ \ ,
\ee
where only $\calF_{1-2}$ are nonzero to subleading order:
\be
\calF_1=-{e h^{(1)}_{\pi NN} |\vec k| \over q \cdot k} \  , \ \ \ \calF_2={e h^{(1)}_{\pi NN} \over 2 M} \left[ \mu_p - {\omega \over \omega_\pi} \mu_n \right] \ \ ,
\ee
where $h^{(1)}_{\pi NN}$ is the PV coupling as defined in Ref.~\cite{Chen:2000hb}. Here, $\omega$ is the CM photon energy, $\omega_\pi$ is the pion energy, and $\mu_{p,n}$ are the nucleon magnetic moments, yielding an asymmetry
\be
A_\gamma(\omega, \theta) = { \left. {d \sigma \over d \Omega}\right\vert_{+} -\left. {d \sigma \over d \Omega}\right\vert_{-} \over \left. {d \sigma \over d \Omega}\right\vert_{+} +\left. {d \sigma \over d \Omega}\right\vert_{-}} \ \ .
\ee
 At threshold this becomes \cite{Chen:2000hb}
\be
A_\gamma(\omega_{th}, \theta) = {\sqrt{2} F(\mu_p-\mu_n) \over g_A M} h^{(1)}_{\pi NN} \approx 0.52 h^{(1)}_{\pi NN} \ \ ,
\ee
which remains stable at forward and backward angles even for energies that are 10s of MeV beyond threshold.  
The authors of Ref.~\cite{Chen:2000hb} argue that measuring for a few months at a facility that produces $10^{37}$  photons/(cm$^2$ s)  will see this effect to $\calO (10^{-7}$). Higher-order corrections were explored in Refs.~\cite{Zhu:2000hf,Zhu:2001br,Chen:2001rc}.  They involve dependence on more LECs; $h_V=h_V^0+{4 \over 3} h_V^2$ (see Eq.~\eqref{NPV} in Sec.~\ref{sec:chiralEFT}) and the coefficient $C=-2\sqrt{2} c_1 + c_2/\sqrt{2}$  from the operator (Eq.~\eqref{ZhuCs} from Sec.~\ref{sec:chiralEFT})
\be
-i e {C \over \Lambda_\chi F} \bar p \sigma^{\mu \nu} F_{\mu \nu} n \pi^+\ .
\ee
While the leading-order prediction of the above observables depends on $h^{(1)}_{\pi NN}$ and only $h^{(1)}_{\pi NN}$, the experimental challenge is formidable.  Electroproduction at threshold $\vec e + p \rightarrow e^\prime + \pi^+ + n$ is 
another possibility, but Ref.~\cite{Chen:2000km} argues that even running for $10^7$ sec. at a luminosity of $10^{38}$  photons/(cm$^2$ s) will only provide an accuracy of $2 \times 10^{-7}$ for $h^{(1)}_{\pi NN}$.

%%%%%%%%%%%%%%%%%%%%%%%%%%%%%%%%%%%%%%%%%%%%%%%%%%%%%%%%%%%%%

\section{Two-nucleon systems}\label{sec:2N}

Because of the relative simplicity of the two-nucleon systems, their PV asymmetries have been calculated in many different ways. The prospect of additional experiments on 2N asymmetries has caused the community to revisit these observables using more modern techniques. 
For example, the circular polarization of the outgoing photon in unpolarized neutron capture was the starting point for the zero-range formalism by Danilov \cite{Danilov:1965}. More recently it has been evaluated using DDH models,  hybrids hybrid methods, and finally EFTs. Similarly, the  photon angular asymmetry in polarized neutron capture has  been the subject of various one-meson-exchange approaches \cite{Tadic:1969xx,Desplanques1975423,Lassey:1975kv,Gari:1975qd,Morioka:1986cp} and then revisited using more modern models, hybrid calculations and EFTs.

Below we describe two-nucleon observables, including those that involve
photons.  Because our goal is to review the status of hadronic parity violation in few-nucleon systems,  we do not consider events involving neutrinos or electron scattering unless they are sensitive to the same PV parameters as those involved in the hadronic processes. We will discuss low-energy phenomena and in general restrict our discussion to energies below pion-production threshold.

\subsection{Longitudinal asymmetry in $\vec{N}N$ scattering}

\subsubsection{\eftnopi results}

\begin{figure}
\begin{center}
\includegraphics[width=0.4\textwidth]{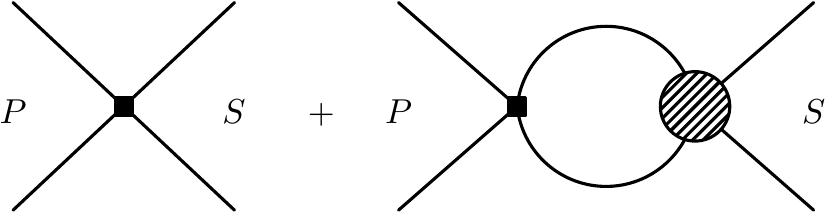}\caption{Diagrams contributing to PV NN scattering in \eftnopi at LO. The square denotes the vertex from the PV Lagrangian, while the shaded disc represents the LO PC scattering amplitude. The labels indicate the angular momentum of the 2N state.\label{fig:2NPVScatter}}
\end{center}
\end{figure}

The LO scattering amplitudes in the PC sector can be calculated analytically in \eftnopi. The corresponding diagrams are shown in Fig.~\ref{fig:2NPCScatter}. 
As discussed in Sec.~\ref{nopimethod}, this corresponds to S-wave scattering. 
Introducing parity violation leads to an S-P-wave transition. P-wave scattering is perturbative in the \eftnopi power counting, and does not need to be resummed. Figure~\ref{fig:2NPVScatter} shows the diagrams contributing to the PV
part of the interference process.

Neglecting Coulomb interactions in the $pp$ case, the results for the asymmetries are given by \cite{Zhu:2004vw,Holstein:2006bv,Phillips:2008hn}
\begin{align}
A^{\vec nn}_L &= - 32 p  \, \frac{\CB -  \CC +  \CD}{\CSing} \ ,\\ 
A^{\vec pp}_L &= - 32p \, \frac{\CB + \CC + \CD}{\CSing} \ , \\
A^{\vec np}_L &= - 32 p \,\frac{\frac{d\sigma^{^1\!S_0}}{d\Omega}}
{\frac{d\sigma^{^1\!S_0}}{d\Omega}+3\frac{d\sigma^{^3\!S_1}}{d\Omega}}\, 
\frac{\CB - 2 \CD}{\CSing}  \nonumber \\ \label{np}
&\quad  - 32 p \, \frac{\frac{d\sigma^{^3\!S_1}}{d\Omega}}
{\frac{d\sigma^{^1\!S_0}}{d\Omega}+3\frac{d\sigma^{^3\!S_1}}{d\Omega}}\, 
\frac{\CA +2 \CE}{\CTrip} \ \ ,
\end{align}
 where to leading order
\begin{equation}
\frac{d \sigma}{d \Omega}=\left[\frac{1}{a^2}
+ p^2\right]^{-1}\ ,
\end{equation}
with $a$ the scattering length in the corresponding channel.

Coulomb corrections can also be calculated in the \eftnopi formalism \cite{Zhu:2004vw,Holstein:2006bv,Phillips:2008hn}. Taking into account a finite angular range $\theta_1 \le \theta \le \theta_2$, which is typical for experimental measurements, Ref.~\cite{Phillips:2008hn} finds that the $pp$ asymmetry including Coulomb effects is approximately
\be\label{eq:ppCoul}
A_L^{pp}\approx  - 8 p \frac{{\cal A}_{pp}}{{\cal C}^{^1S_0}_0}\left[1 + \eta \left(\frac{1}{a_S(\mu) p} \right) \frac{1}{\cos \theta_1 - \cos \theta_2} \ln\left(\frac{1- \cos \theta_1}{1- \cos \theta_2}\right) + {\cal O}(\eta)^2 \right],
\ee
 where
\be
{\cal A}_{pp} = 4 \left( \CB + \CC + \CD \right), 
\ee
where $\eta\equiv\frac{M\alpha}{2p}$ is the Coulomb parameter and $a_S(\mu)$ is the strong $pp$ scattering length. Since $\eta \ll 1$ for the energies of interest experimentally, the expansion in Eq.~\eqref{eq:ppCoul} should be valid. In fact, at the lowest energy so far considered in a measurement, Coulomb corrections amount to about 3\%, which is much smaller than the uncertainties in the experiment and in a LO \eftnopi calculation.

One of the motivations for this review is to present the various results of few-nucleon reactions in a unified framework. 
Since calculations in the three-body sector are significantly simplified when using the dibaryon formalism, we also present the longitudinal asymmetries in this approach. 
 Using the conventions of Eq.~\eqref{yDeltaConv} in Sec.~\ref{nopimethod} for consistency,
the asymmetries are given by 
\begin{align}
A^{nn}_L &= - \sqrt{\frac{32 M}{\pi}}\, p  \left( \gB -  \gC +  \gD \right)  \ , \\ 
A^{pp}_L &= - \sqrt{\frac{32 M}{\pi}}\, p  \left(\gB + \gC + \gD\right) \ , \\
A^{np}_L &= - \sqrt{\frac{32 M}{\pi}}\, p \frac{\frac{d\sigma^{^1\!S_0}}{d\Omega}}
{\frac{d\sigma^{^1\!S_0}}{d\Omega}+3\frac{d\sigma^{^3\!S_1}}{d\Omega}}\, 
\left(\gB - 2 \gD\right)  \nonumber \\ \label{npdib}
&\quad - \sqrt{\frac{32 M}{\pi}}\, p \, \frac{\frac{d\sigma^{^3\!S_1}}{d\Omega}}
{\frac{d\sigma^{^1\!S_0}}{d\Omega}+3\frac{d\sigma^{^3\!S_1}}{d\Omega}}\, 
\left(\gA +2 \gE\right) \ .
\end{align}

\subsubsection{Hybrid method results}

The hybrid calculation in Ref.~\cite{Liu:2006dm} includes the combination of the AV18 PC potential with a PV potential derived in \eftnopi. Results are given for two energies, 13.6 MeV and 45 MeV, corresponding to two existing measurements \cite{Eversheim:1991tg,Kistryn:1987tq}. The author of Ref.~\cite{Liu:2006dm} notes that the ratio of the asymmetry at 45 MeV to that at 13.6 MeV of the ``pionless hybrid'' result is $\approx 1.0$, much smaller than the naive experimental ratio of $\approx 1.7$ (ignoring errors).
However, the energy of 45 MeV is outside the domain of validity of a pionless description, therefore the higher-energy prediction of an {\eftnopi}+hybrid formalism cannot be trusted. 
Note, however, that in the consistent pionless EFT calculation the ratio is given simply by the ratio of center of mass momenta, since the PV parameters are energy independent at leading order. This yields $\approx 1.8$, and thus much closer to the experimental ratio. 
The ratio of the results using AV18 and a PV chiral EFT potential is $1.8$, in rough agreement with the experimental ratio. Ref.~\cite{Liu:2006dm} also contains expressions for the asymmetry at 221 MeV. At this energy \eftnopi is not applicable, and a reliable calculation in chiral EFT would require contributions beyond the leading order considered in  Ref.~\cite{Liu:2006dm}.

\subsubsection{Model results}

In terms of model calculations, we will discuss the calculation of Ref.~\cite{Carlson:2001ma}. This work can be viewed as an update and extension of the earlier detailed study of Ref.~\cite{Driscoll:1988hg}. Reference~\cite{Carlson:2001ma} calculates the longitudinal asymmetry using the AV18, Bonn-2000, and Nijmegen-I potentials in combination with the DDH PV potential. Since neutral pion exchange would violate CP invariance \cite{Barton:1961eg}, this amounts to a PV potential solely in terms of $\rho$ and $\omega$ meson exchanges. The results show very little dependence on the choice of PC potential used. They show, however, that a description of the three experimental data points at 13.6 MeV, 45 MeV, and 221 MeV cannot be achieved with the DDH best values. The authors of Ref.~\cite{Carlson:2001ma} adjust two combinations of PV $\rho$ and $\omega$ couplings by performing a fit to the data. The resulting parameter set will be referred to as ``DDH-adj.'' These values are still compatible with the ``reasonable ranges'' in Ref.~\cite{Desplanques:1979hn}, even though some of them lie towards the limits of these ranges.

However, since the resulting DDH-adj values are not close to most existing model estimates of the couplings, Ref.~\cite{Liu:2005sn} studies the dependence of the fitted $\rho$ and $\omega$ couplings on a number of modifications, such as the variation of strong interaction parameters or the introduction of form factors for the PV meson-nucleon couplings. The authors show that while the fit results are quite sensitive to these modifications, they do not improve the agreement with the existing model estimates.

The contributions from including the $\Delta$ resonance and two-pion exchange are studied in Refs.~\cite{Iqbal:1992xm,Niskanen:2007hf,Partanen:2012qw} in combination with the Reid93 \cite{Stoks:1994wp} and DDH models. These contributions turn out to be significant at 221 MeV, the energy of one of the existing experimental results.  However, the calculations rely on a number of model assumptions, and the consistency of the various ingredients is unclear. 

\subsubsection{Experimental status}

Of the $\vec NN$ PV elastic scattering observables listed above,
only $\vec p p$ is available.  That experiment was performed at 13.6 MeV,
yielding \cite{Eversheim:1991tg} 
$$
A_L^{\vec p p}(13.6\,\MeV) = (-0.93 \pm 0.21) \times 10^{-7} \ \ .
$$  
This provides one of the five measurements needed to fix the 5 PV LECs of \eftnopi.  
The measurements by Ref.~\cite{Kistryn:1987tq} 
were made at 45 MeV; outside the realm of applicability of \eftnopi\
but important for a chiral treatment and to constrain PV observables
among  higher energies. The result for the asymmetry at this energy is
$$
A_L^{\vec p p}(45\,\MeV) = (-1.50 \pm 0.22) \times 10^{-7} 
$$
for the measured angular range of $23^\circ < \theta_{lab} < 52^\circ$. For the asymmetry using total cross sections, the result is given as
$$
A_L^{\vec p p, tot}(45\,\MeV) = (-1.57 \pm 0.23) \times 10^{-7} \ \ .
$$

\subsection{Neutron-proton spin rotation}\label{sec:npspinrot}

\subsubsection{ \eftnopi results}

\begin{figure}
\begin{center}
\includegraphics[width=0.75\textwidth]{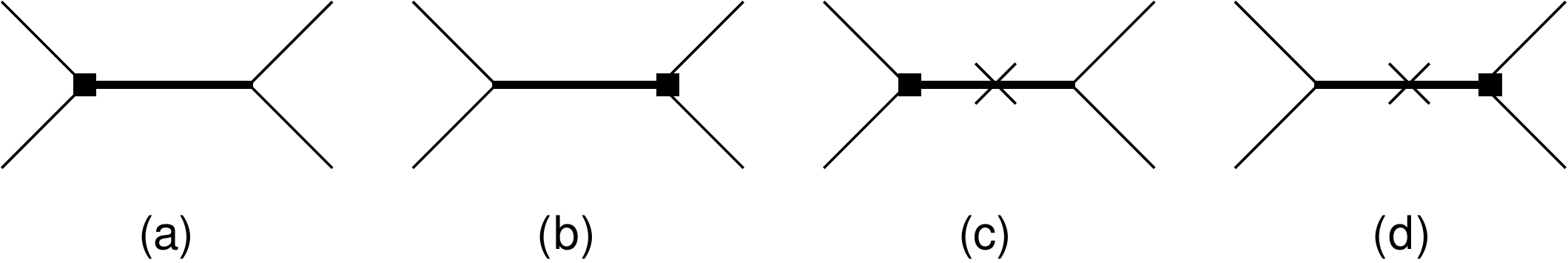}\caption{Diagrams contributing to PV NN scattering. (a) and (b): LO; (c) and (d) NLO. Thin line: nucleon; thick line: dibaryon; square: PV vertex; cross: insertion of effective range correction.\label{fig:npspinrot}}
\end{center}
\end{figure}

As discussed in Sec.~\ref{sec:obs}, the spin rotation angle is related to the forward scattering amplitude. Using the dibaryon formalism, Ref.~\cite{Griesshammer:2011md} calculates the amplitudes and the rotation angle up to NLO. The diagrams contributing up to this order are shown in Fig.~\ref{fig:npspinrot}. Note that in the dibaryon formalism there are no new PV operators at NLO; the contributions at this order stem solely from an insertion of the effective-range correction to the dibaryon propagator. The result is
\begin{align}
  \label{eq:npspinrotdecomposed}
  \frac{1}{\rho}\;\left.\frac{\dd \phi_\text{PV}^{np}}{\dd l}\right|_{\text{LO+NLO}}
  = & 4 \sqrt{2 \pi M} \left( \frac{2\gE+\gA}{\gamma_t} \frac{Z_t+1}{2}
+ \frac{ \gB-2 \gD}{\gamma_s} \frac{Z_s+1}{2}\right) 
\end{align}  
where $Z_{s/t} = \frac{1}{1- \gamma_{s/t} \rho_{s/t}}$, 
$\rho$ is the relevant effective range and $\gamma$ the poles in
the NN scattering amplitude.  
Note that this corrects an error of a factor of two in the result of Ref.~\cite{Griesshammer:2011md}.
An order-of-magnitude estimate yields a rotation angle
\be\label{npspinrotest}
\left| \frac{\dd \phi}{\dd l} \right| \approx \left[10^{-7}\ldots10^{-6} \right] \frac{\text{rad}}{\text{m}}\ .
\ee

\subsubsection{Hybrid method results}

Reference \cite{Liu:2006dm} performs a hybrid calculation of the np spin rotation angle using the AV18 potential combined with PV interactions derived from both pionless and chiral EFT. A comparison of the pionless and chiral results shows good agreement, as expected for the very low energy considered in this process.  By mapping the DDH best values to the LECs and using a target density of $0.24\times 10^{23}/\text{cm}^3$, Ref.~\cite{Liu:2006dm} finds a rotation angle of 
\be
\frac{d\phi}{dl} = 6.7 \times 10^{-7} \text{rad}/\text{m}
\ee
or 
\be
\frac{d\phi}{dl} = 5.1 \times 10^{-7} \text{rad}/\text{m},
\ee
depending on the regulator functions in the PV potential.

\subsubsection{Model results}

The hybrid results also agree with the DDH model calculation of Ref.~\cite{Schiavilla:2004wn} when the same values for the couplings are used. This model calculation compares results obtained with the AV18, Bonn, and Nijmegen potentials to study model dependence. The authors of Ref.~\cite{Schiavilla:2004wn} also consider two different sets of values for the DDH couplings: one corresponds to the DDH best values; the other is the ``DDH-adj'' described above. For the DDH best values the rotation angle is found to be in the range $[7.19,7.64]\times 10^{-7}\text{rad/m}$, while for the ``DDH-adj'' parameter set the rotation angle is  somewhat smaller at $[4.63,5.09]\times 10^{-7}\text{rad/m}$. The ranges correspond to the different PC potentials used.

Neutron proton spin rotation has also been used to study contributions not present in the standard DDH approach. In addition to the DDH potential, Ref.~\cite{Partanen:2012ny} employs two different PV two-pion exchange potentials, one without \cite{Desplanques:2008jx} and one with \cite{Kaiser:2007zzb} explicit $\Delta$ degrees of freedom. Note however that for the latter case the $\Delta$ is not explicitly included in the one-meson-exchange part of the potential. These potentials are combined with the PC Reid93 potential. In contrast to all previous calculations the author of Ref.~\cite{Partanen:2012ny} argues that the molecular density of $\text{H}_2$ and not the proton density should be used in the determination of the spin rotation angle. This introduces a systematic factor of two difference in the results, which should be taken into account when comparing with previous calculations. Adjusting for the difference in densities, the result with only the one-meson exchange potentials using the DDH best values agrees with the results of Refs.~\cite{Schiavilla:2004wn,Liu:2006dm}. The two-pion contributions are of the order of 20\% and in general of opposite sign compared to the one-meson exchange parts.

\subsubsection{Experimental status}

If measured, this reaction would yield one constraint which depends,
using the partial wave basis, on four of the PV LECs  in \eftnopi.
At present there are no experimental limits available on this reaction, 
and indeed no firm plans to measure it.  
The possibility of measuring neutron spin rotation on a variety
of targets has been discussed at both Oak Ridge's SNS \cite{Snow}
and NIST \cite{Markoff:2005}  but no experiments are currently planned.

\subsection{Angular asymmetry $A_\gamma$ in $\vec{n} p \to d\gamma$}\label{sec:npAgamma}

The most general amplitude for the process $np \leftrightarrow d \gamma$ combines the  photon momentum $\vec q$,  the deuteron polarization  $\vec{\epsilon}_d$, and the photon polarization $\vec{\epsilon}_d$ to yield
\begin{align}\label{Res:AmpPara}
\mathcal{M} =& e X N^T \tau_2 \sigma_2 \left[ \vec{\sigma}\cdot\vec{q}\ \vec{\epsilon}_d^{\; *}\cdot\vec{\epsilon}_\gamma^{\; *} - \vec{\sigma}\cdot\vec{\epsilon}_\gamma^{\; *} \ \vec{q}\cdot\vec{\epsilon}_d^{\; *}\right]N 
+ ie Y \epsilon^{ijk} \epsilon_d^{* i} q^j \epsilon_\gamma^{k*} \left(N^T \tau_2\tau_3 \sigma_2 N\right) \notag\\
& + ie W \epsilon^{ijk} \epsilon_d^{* i}\epsilon_\gamma^{k*} \left(N^T \tau_2 \sigma_2 \sigma^j N\right) 
+ e V \vec{\epsilon}_d^{\; *}\cdot \vec{\epsilon}_\gamma^{\; *} \left(N^T \tau_2 \tau_3\sigma_2 N \right) + h.c. +\ldots \ \,
\end{align}
where the $X$ and $Y$ terms are the PC parts of the amplitude, and the $V$ and $W$ terms are PV. The ellipses include terms that are suppressed or of higher order. In the 
$np \leftrightarrow d \gamma$ processes  we look for an interference between PC and PV terms that is linear in the PV interaction.  
One such observable comes from
polarized neutron capture $\vec n p \rightarrow d \gamma$.  The resulting photon asymmetry $A_\gamma$ is the subject of intense interest at the moment by the NPDGamma 
experiment \cite{Gericke:2011zz} at the Spallation 
Neutron Source at Oak Ridge.   The asymmetry is defined through
\begin{equation}\label{Res:AsymDef}
\frac{1}{\Gamma}\,\frac{d\Gamma}{d\cos \theta}=1+A_\gamma \cos \theta,
\end{equation}
where $\theta$ is the angle between the emitted photon and the axis defined by the incoming neutron polarization, as discussed in Sec.~\ref{sec:obs}.

\subsubsection{\eftnopi results}

At leading order the $\vec n p \rightarrow d \gamma$ asymmetry comes from the interference between the $W$ and $Y$ amplitudes in Eq.~\eqref{Res:AmpPara} 
 and is given by
\be
A_\gamma=-2\frac{M}{\gamma^2}\,\frac{\mbox{Re}[Y^* W]}{|Y|^2} \ \ ,
\ee
where $\gamma=\sqrt{MB}$ is the deuteron momentum and $B$ is the deuteron binding energy.
This quantity has been calculated  in Refs.~\cite{Savage:2000iv,Schindler:2009wd}  but we will give the results in terms of the notation used in Ref.~\cite{Schindler:2009wd}, and again in the dibaryon language of Eqs.~\eqref{eq:PCLag} and \eqref{eq:PVLag} of Sec.~\ref{nopimethod}.  The latter is done so that, as explained above, we have a consistent convention for the two- and three-nucleon \eftnopi results presented in this review.

Using the Lagrangians of Eqs.~\eqref{Lag:PCpartial} and \eqref{Lag:PV} yields \cite{Schindler:2009wd}
\be
A_\gamma=\frac{32}{3}\,\frac{M}{\kappa_1 \left(1-\gamma \aSing \right) }\,\frac{\CE}{\CTrip}.
\ee
In terms of the dibaryon formalism of Eqs.~\eqref{eq:PCLag} and \eqref{eq:PVLag} of Sec.~\ref{nopimethod}, the asymmetry becomes\footnote{The dibaryon conventions used here differ from those in Ref.~\cite{Schindler:2009wd}, which used the dibaryon formalism to resum contributions proportional to the effective range.} 
\be
A_\gamma = \frac{4}{3} \sqrt{\frac{2}{\pi}} \,  \frac{M^\frac{3}{2}}{\kappa_1 \left(1-\gamma \aSing \right) }\,\gE.
\ee
Reference~\cite{Savage:2000iv} considers an additional contribution in the PC amplitude proportional to the coupling $L_1$ of a four-nucleon-photon contact interaction, which is formally of higher order in our convention.\footnote{After adjusting for different conventions  there appears to be  an overall sign difference between Refs.~\cite{Schindler:2009wd} and  \cite{Savage:2000iv}.}.

\subsubsection{Chiral EFT results}

The PV asymmetry $A_\gamma$ provides one of the few hadronic parity violating observables that has been calculated using chiral EFT methods for both the PC and PV contributions \cite{Kaplan:1998xi}.  The leading-order PV interaction is given by a pion-nucleon coupling, while NN contact terms are formally of higher order (see Sec.~\ref{sec:chiralEFT}). Reference \cite{Kaplan:1998xi} uses the so-called KSW power counting \cite{Kaplan:1998tg,Kaplan:1998we}, in which pion exchange is treated perturbatively. The obtained asymmetry 
\be \label{KSSWresult}
A_\gamma = 0.17 h_{\pi NN}^{1}
\ee
is somewhat larger than those found in phenomenological models, to be discussed below. 
However,  the KSW perturbative treatment of pions suffers from convergence problems for some partial waves \cite{Gegelia:1998ee,Fleming:1999ee}, so the above result may receive large corrections.

\subsubsection{Hybrid method results}

Reference~\cite{Hyun:2006cb} considers the PV EFT potential including the leading one-pion exchange plus subleading two-pion-exchange and contact contributions. This potential is combined with wave functions obtained from the AV18 potential and electromagnetic currents are taken into account through Siegert's theorem \cite{Siegert:1937yt}. Note that this does not take into account all PV contributions to the current operators. Different regularizations are used for different parts of the potential, and the individual pieces as well as the total result show considerable dependence on the values of the regulation parameters. The authors estimate their final result to be in the range
\be\label{AgammaHybrid1}
A_\gamma = -[0.08,0.11] h_{\pi}^{1},
\ee
with the two-pion-exchange contribution to the asymmetry to be about 10\% of the dominant one-pion exchange.

Similarly, and also relying on the Siegert theorem, Ref.~\cite{Liu:2006dm} combines the PC AV18 potential with PV potentials obtained from pionless and chiral EFT. The result in the latter case is in agreement with previous calculations. 
An earlier hybrid calculation avoids use of the Siegert theorem by combining AV18 wave functions with PC and PV current operators derived in chiral EFT up to NLO \cite{Hyun:2001yg}. The result of this calculation is
\be
A_\gamma = - 0.10 h_\pi^1,
\ee
and in agreement with Eq.~\eqref{AgammaHybrid1} and the model results discussed below.

\subsubsection{Model results}

The model calculation of Ref.~\cite{Hyun:2004xp} uses the DDH potential combined with the Argonne $v_{18}$ (AV18) \cite{Wiringa:1994wb}, Bonn \cite{Machleidt:1987hj}, and Paris \cite{Lacombe:1980dr} potentials. Exchange currents for the electromagnetic transitions are taken into account via Siegert's theorem. Between the AV18 and Bonn potentials, Ref.~\cite{Hyun:2004xp} finds good agreement for the asymmetry $A_\gamma$,
\be
\begin{split}
A_\gamma & = -0.117 h_\pi^1 -0.001h_\rho^1 +0.002h_\omega^1  \quad (\text{AV18}),\\
A_\gamma & = -0.117 h_\pi^1 -0.001h_\rho^1 +0.003h_\omega^1\quad (\text{Bonn}),
\end{split}
\ee
while the pion contribution for the Paris potential is larger by a factor of 1.27. The authors trace this discrepancy to the fact that the Paris potential results in a \oneS scattering length $\aSing=-17.6\,\text{fm}$, which in turn also yields a PC $np\to d \gamma$ $M1$ amplitude that is smaller compared to the AV18 result by the same factor of 1.27.

Another recent model calculation \cite{Schiavilla:2004wn} with the DDH model in combination with the Argonne $v_{18}$, CD-Bonn \cite{Machleidt:2000ge}, and Nijmegen I \cite{Stoks:1994wp} potentials obtains similar results. Using only the pion part  of the DDH potential, Ref.~\cite{Schiavilla:2004wn} finds 
\be
A_\gamma = - 0.11 h_\pi^1,
\ee
with the difference between the various PC potentials being less than 3\%. The pion-exchange part is expected to dominate the photon asymmetry. Using the ``DDH-adj'' parameter, the asymmetry changes by less than 2\% for each strong potential compared to the case with a PV pion-exchange part only. 
These results are in agreement with a number of earlier model calculations. An analysis of these \cite{Desplanques:2000ej} concludes that, after adjusting for various conventions, they all obtain a value close to 
\be
A_\gamma = - 0.11 h_\pi^1.
\ee

Reference \cite{Partanen:2010yt} includes the contributions due to the $\Delta$ in a coupled-channel approach by introducing PV meson-$N\Delta$ couplings in addition to the DDH potential. Strong interactions are taken into account via either the Reid93 or AV18 potentials.  
The $\Delta$ contributions turn out to be small since different effects that are individually of the order of 5\% cancel in the final asymmetry.

Reference~\cite{Partanen:2012ny} considers two-pion exchange contributions in a model approach. As in the case for neutron-proton spin rotation, in addition to the Reid93 potential for the PC and the DDH model for the main part of the PV interactions, two different versions of PV two-pion-exchange potentials are considered, one without \cite{Desplanques:2008jx} and one with \cite{Kaiser:2007zzb} explicit $\Delta$ degrees of freedom. The purely one-meson-exchange part of the calculation agrees with previous results. The two-pion-exchange contribution reduces the photon asymmetry by up to 20\% depending on the choice of potential as well as regularization procedure. Note that in one case, though, the two-pion-exchange contribution increases $A_\gamma$ by about 4\%. 

For a discussion on the apparent sign difference between the chiral EFT result of Eq.~\eqref{KSSWresult} and the hybrid and model results see Ref.~\cite{Savage:2000iv}.

\subsubsection{Experimental status}

For the PV observables from $np \leftrightarrow d \gamma$, only
limits are available.  For $A_\gamma$ the limits are
$(-0.6 \pm 0.21) \times 10^{-7}$\cite{Cavaignac:1977uk}
and $(-0.15 \pm 0.47) \times 10^{-7}$\cite{Alberi:1988fd}.
The long-standing
NPDGamma \cite{Lauss:2006es,Gericke:2011zz} experiment at LANSCE at LANL has moved to the 
higher
neutron fluxes available at Oak Ridge's Spallation Neutron Source and is
now collecting data.

\subsection{Photon circular polarization $P_\gamma$ in $np \to d \vec{\gamma}$ }\label{sec:npPgamma}

An observable that isolates  a $V^*Y$ interference term (see Eq.~\eqref{Res:AmpPara}) can be obtained from  $np\to d \protect\overset{\circlearrowleft}\gamma$, which yields the circular polarization 
\begin{equation}\label{PolDef}
P_\gamma=\frac{\sigma_+-\sigma_-}{\sigma_++\sigma_-} \ \ ,
\end{equation}
where $\sigma_{+/-}$ is the total cross section for  outgoing photons with positive/negative helicity.

\subsubsection{\eftnopi results}
At LO in \eftnopi this asymmetry can be rewritten as
\begin{equation}\label{Res:PolDef}
P_\gamma = 2\frac{M}{\gamma^2} \frac{\mbox{Re}[Y^*V]}{|Y|^2}\ \ ,
\end{equation}
where $Y$ and $V$ are the amplitudes as defined in Eq.~\eqref{Res:AmpPara}.
In the formalism using the Lagrangians of Eqs.~\eqref{Lag:PCpartial} and \eqref{Lag:PV}, the asymmetry is \cite{Schindler:2009wd}
\begin{equation}\label{Res:PolRes}
P_\gamma=-16\frac{M}{\kappa_1\left(1-\gamma\aSing\right)}\,\left[ \left(1-\frac{2}{3}\gamma\aSing\right)\frac{\CA}{\CTrip} +\frac{\gamma\aSing}{3}\frac{\CB-2\CD}{\CSing} \right].
\end{equation}
In terms of the dibaryon formalism of Eqs.~\eqref{eq:PCLag} and \eqref{eq:PVLag} of Sec.~\ref{nopimethod} this becomes
\begin{align}
P_\gamma = -2\sqrt{\frac{2}{\pi}}\,\frac{M^\frac{3}{2}}{\kappa_1 \left( 1 - \gamma\aSing \right)}\,\left[ \left(1-\frac{2}{3}\gamma\aSing \right) \gA 
+\frac{\gamma\aSing}{3} \left( \gB - 2\gD \right)\right]\ ,
\end{align}
where again, for consistency with other results presented in this review, we have used conventions that differ from the dibaryon ones used in Ref.~\cite{Schindler:2009wd}.
These expressions for $P_\gamma$ agree, adjusting for different conventions, with that of Ref.~\cite{Shin:2009hi} if the formally higher-order term $L_1$ is neglected.

\subsubsection{Hybrid method results}

Reference~\cite{Liu:2006dm} calculates the circular polarization $P_\gamma$ in a hybrid approach, combining the AV18 potential in the PC sector with potentials from pionless and chiral EFT for the PV interactions. Siegert's theorem is used for the E1 transition that corresponds to the PV part of the calculation. In this work, this transition involves not only the capture from the \threePzero wave to the even orbital momentum part of the deuteron, but also a \oneP admixture in the deuteron wave function.  For the pionless case, the author of Ref.~\cite{Liu:2006dm} finds correction terms that are large and put into question the hybrid analysis. For the chiral PV potential, these corrections are much smaller. Using a matching to the DDH best values, the circular polarization found is
\be\label{eq:LiuPgamma}
P_\gamma = 2.5\times 10^{-8}.
\ee

\subsubsection{Model results}
Comparison of the above result with earlier model calculations is not straightforward. Since in model calculations the internal consistency between PC and PV interactions is not immediately clear, Ref.~\cite{Hyun:2004xp} uses a variety of PC potentials (Paris, several versions of Bonn, AV18) in combination with the DDH interactions to study theoretical uncertainties. Unlike in $\vec{n}p\to d\gamma$, the result for $P_\gamma$ depends strongly on the employed PC potential, even after correcting for differences in the PC $np$ scattering length. While the Paris and AV18 potentials give similar results and are in agreement with Eq.~\eqref{eq:LiuPgamma}, the results using the Bonn and Bonn-B potentials are larger by factors of 5 and 2, respectively. This discrepancy is explained by the dominance of PV $\rho$ and $\omega$ exchanges, which represent short-range contributions. While the various potentials all result in good overall descriptions of scattering phase shifts, the short-distance details vary significantly. These differences clearly show up in the results for $P_\gamma$.

The circular polarization $P_\gamma$ was also the subject of the paper that introduced the so-called Danilov parameters \cite{Danilov:1965}, see the discussion in Sec.~\ref{subsec:models}. Reference \cite{Danilov:1965} showed that for very low energies this observable  does not receive any contributions from the isovector part. 

\subsubsection{Experimental status}

As with the observable $A_\gamma$ above, only a limit of $P_\gamma = (1.8 \pm 1.8)\times 10^{-7}$ exists \cite{Knyazkov:1984zz}. 

\subsection{Asymmetry $A_L^\gamma$ in $\vec{\gamma} d \to np$}\label{sec:npAgammaL}

By time-reversal symmetry, this observable should be identical to $P_\gamma$ for exactly reversed kinematics. The \eftnopi calculations of $P_\gamma$ \cite{Schindler:2009wd,Shin:2009hi} therefore do not contain separate expressions for the observable $A_L^\gamma$

\subsubsection{Model results}

The model calculation of Ref.~\cite{Liu:2004zm}, which uses the Argonne $v_{18}$ potential in the PC sector and the DDH formalism for PV interactions, goes beyond the threshold and considers the asymmetry up to an energy 10 MeV above threshold. Siegert's theorem is used for the $E1$ transition, while two-body currents are explicitly considered in the $M1$ case. 
At a photon energy of $\omega = 2.235\,\MeV$, i.e., close to threshold, the DDH best values give an asymmetry of 
\be
A^\gamma_L = 2.53 \times 10^{-8}.
\ee
The asymmetry decreases for larger energies and changes sign around a photon energy of $\approx 4\,\MeV$.
The calculation shows that  the pion-exchange contribution stays suppressed throughout the considered energy range.
Reference \cite{Schiavilla:2004wn} finds very similar results for the same potentials and parameter sets. However, it also shows that the results are highly sensitive to the choice of PV couplings; e.g., $P_\gamma$ varies by about a factor of two between the DDH-adj (see above) and DDH best value parameter sets. The authors of Ref.~\cite{Schiavilla:2004wn} also consider the PC Bonn potential, which results in asymmetries larger by almost a factor of 2 compared to the AV18 potential when the same PV parameters are used.
 The energy dependence and the suppression of pion-exchange contributions in these calculcations confirm  the results of Ref.~\cite{Khriplovich:2000mb}, which uses a zero-range interaction for the PC sector. These findings contradict an earlier calculation \cite{Oka:1983sp}, which found an enhancement of the asymmetry even a few MeV above threshold as well as substantial one-pion-exchange contributions in the same energy region. However, as shown in Refs.~\cite{Khriplovich:2000mb,Liu:2004zm}, this is due to an insufficient treatment of parity admixed states in the continuous spectrum.

Analogous to the considerations for $A_\gamma$, Ref.~\cite{Partanen:2010yt} investigates the role of the $\Delta$ resonance in the process $\vec\gamma d \to np$ in a coupled-channel approach up to energies of 10 MeV. The $\Delta$ contributions to $A_L^\gamma$ are again small in the considered energy region. 

\subsubsection{Experimental status}

The experiment of Ref.~\cite{Earle:1988fc} did not have the required sensitivity for an observation of $A_L^\gamma$.  However, with improvements in high-intensity photon sources, a determination of this observable may be feasible.
This reaction is of particular interest to the High Intensity Gamma Source (HIGS) facility at Duke University's Triangle Universities Nuclear Laboratory (TUNL) and is the flagship
experiment in a proposed upgrade of the facility to``HIGS2."  The upgrade would produce high intensity circularly polarized photons at the few MeV energies required to obtain this observable.

\subsection{Deuteron anapole moment and form factor}\label{sec:deutana}

The anapole moment of the deuteron comes from its spin-dependent
interaction with an electromagnetic field $F^{\mu \nu}$.  The operator is
\be
\calL=iA_d \frac{1}{M_N^2} \epsilon_{abc} d^{a \dagger} d^b \partial_\mu F^{\mu c} \ \ ,
\ee
where $A_d$ is to be found by the methods listed below, and $d^a$ is
the deuteron interpolating field.
The deuteron anapole moment consists of contributions from the individual anapole moments of the proton and neutron as well as ``non-nucleon'' contributions that involve the coupling of the photon to more than a single nucleon.
It can potentially be measured by electron scattering off the deuteron.

\subsubsection{\eftnopi result}

The \eftnopi calculation was done in Ref~\cite{Savage:2000iv}. 
The contact operator involved is
\be
\calL=i {h_{33}^{(1)} \over \sqrt{8 M \rTrip }} \epsilon^{ijk} t_i ^\dagger N^T \sigma_2 \sigma_j \tau_2 \tau_3 {1 \over 2} \left( i \overrightarrow \bigtriangledown- i \overleftarrow \bigtriangledown \right)_k N + h.c. + \cdots \ \ ,
\ee
where $h_{33}^{(1)}$ is an unknown parameter and $t_i$ is the dibaryon
field for the deuteron. The contribution proportional to $h_{33}^{(1)}$ is 
\be
A_d=-{ e h_{33}^{(1)} M^{3/2} \over 8 \sqrt{2 \pi}}{1 \over 1-\gamma\rTrip} \left( \kappa_1 - {1 \over 6}\right)\ \ .
\ee
 Ref~\cite{Savage:2000iv} uses the dibaryon formalism to resum contributions proportional to the effective range. Without resummation of the effective range contributions and expressed in the conventions of Eq.~\eqref{eq:PVLag} of Sec.~\ref{sec:methods} the non-nucleonic part of the anapole moment is 
\be
A_d = -e \gE {M^{3/2} \over \sqrt{2 \pi} } \left( \kappa_1 - {1 \over 6}\right) \ \ .
\ee

\subsubsection{Chiral EFT result}
The leading contribution comes from single pion exchange; like $A_\gamma$ this
depends upon the $\Delta I=1$ parameter $h_{\pi NN}^{1}$  \cite{Savage:1998rx}:
 \be
 A_d=-{e g_A h^{1}_{\pi NN} M^2 \over 12 \pi \sqrt{2} F } \left[\kappa_1 {m_\pi + \gamma \over (m_\pi + 2 \gamma)^2} + {2 m_\pi + 9 \gamma \over 6 (m_\pi + 2 \gamma)^2} \right] \ \ ,
\ee
where $\gamma=\sqrt{MB}$ is the deuteron binding momentum. 
Putting this in the language of Ref.~\cite{Hyun:2002in} (see below) by removing a factor of ${M^2 \over 4 \pi}$ and multiplying by (197  $\MeV\, \text{fm}$)$^2$ to obtain units of fm$^2$ yields $\vec A_d \sim -1.3 h^1_{\pi NN} e \vec I$,  where $\vec I = {1 \over 2} (\vec \sigma_p + \vec \sigma_n)$.

\subsubsection{Hybrid method result}
Reference~\cite{Hyun:2002in} employs a hybrid formalism: deuteron wave functions are obtained with the AV18 potential and a parity-odd admixture is calculated from a PV pion-exchange potential. The currents are derived in HB\chiPT, while the results for the nucleon anapole moments are taken from earlier works. The result for the deuteron anapole moment is
\be
\vec{A}_d = -0.909 e \vec{I} h_{\pi NN}^{1} \ \ ,
\ee
 which is smaller then the LO EFT and the zero-range approximation results. Given that the EFT calculations are performed at LO, and that it is not clear how consistent with each other the various parts of the hybrid calculation are, no clear conclusions can be drawn from this discrepancy.

\subsubsection{Model result}

As an extension of this result, Ref.~\cite{Liu:2003au} considers not only the pion-exchange contribution to the PV potential, but the complete DDH potential also including $\rho$ and $\omega$ exchanges. The authors also improve upon the previous calculation by considering current conservation in more detail. The final result for the deuteron anapole moment is
\be
\vec{A}_d = \left(-0.756 h_\pi^1 - 0.387 h_\rho^1 + 0.010 h_\omega^1 + 0.007 h_\rho^0 - 0.114 h_\omega^0 \right)e\vec{I}.
\ee

In a similar spirit to Ref.~\cite{Partanen:2010yt}, Ref.~\cite{Khriplovich:1999qr} calculates the deuteron anapole moment in a zero-range approximation. The result from the sum of individual proton and neutron anapole moments and the ``non-nucleonic'' part is 
$$\vec A_d = -1.2 e h_\pi^1 \vec I \ \ . $$

%%%%%%%%%%%%%%%%%%%%%%%%%%%%%%%%%%%%%%%%%%%%%%%%%%%%%%%%%%%%%

\section{Three-nucleon systems}\label{sec:threeN}

In this section we present PV calculations that have been performed on three-body systems.  
Details on the methods used can be found in Sec.~\ref{sec:methods}.
For a recent comprehensive review of the three-nucleon system in the PC sector see, e.g., Ref~\cite{KalantarNayestanaki:2011wz} as well as references therein.
From the \eftnopi point of view, we note that the two-nucleon system alone does not provide enough experimentally accessible observables to determine the five  PV LECs that appear at leading order in \eftnopi. It is necessary to  consider three-nucleon processes as well, which can also provide further nontrivial tests of that theory. In the PC sector it was shown that in addition to a two-nucleon potential the inclusion of three-nucleon (3N) interactions is necessary to describe not only three- and four-nucleon systems (see, e.g., \cite{Epelbaum:2008ga,KalantarNayestanaki:2011wz}), but also the binding energies and energy spectra in light ($A\le8$) nuclei \cite{Pieper:2001mp}.  In the traditional potential approach a number of 3N interactions have been considered in combination with the various 2N potentials, with the Tucson-Melbourne \cite{Coon:1974vc,Coon:1978gr,Coon:2001pv}, Urbana \cite{Carlson:1983kq,Pudliner:1995wk}, and Illinois \cite{Pieper:2001ap} potentials the most widely used ones. This approach has had considerable success, but the consistency of the two- and three-nucleon forces remains unclear. On the other hand, three- and more nucleon interactions arise naturally in the EFT framework. In addition, the power counting of EFT provides estimates for the expected size of the 3N interactions.

\subsection{Neutron-deuteron spin rotation}\label{sec:ndspinrot}

\subsubsection{\eftnopi results}

Motivated by the renewed interest in hadronic parity violation, a number of calculations of neutron-deuteron spin rotation have recently appeared \cite{Schiavilla:2008ic,Song:2010sz,Griesshammer:2011md,Vanasse:2011nd}.
 Refs.~\cite{Griesshammer:2011md,Vanasse:2011nd} analyzed neutron-deuteron spin rotation treating both PC and PV interactions consistently in the \eftnopi approach. The rotation angle can be written as
\begin{align}
\frac{1}{\rho} \frac{\dd \phi}{\dd l} &= \frac{4M}{27 k} \Re\left[ \calM[\twoS \to \twoPone;k] -2\sqrt{2}\calM[\twoS \to \fourPone;k] - 4\calM[\fourS \to \twoPthree;k] -2\sqrt{5}\calM[\fourS \to \fourPthree;k]  \right] \\
&= c[(\threeS - \oneP)](\Lambda)\gA - c[(\threeS - \threePone)](\Lambda)\gE + c[\calT](\Lambda)(3\gB - 2\gC) \ \ ,
\end{align}
where we use the conventions of Ref.~\cite{Griesshammer:2011md} and $\calT=3\gB - 2\gC$. Note that Ref.~\cite{Vanasse:2011nd} uses a different (sign) convention for the PC dibaryon couplings $y$ and for some of the projections of the scattering amplitudes $\calM[X\to Y;k]$ onto particular partial waves. 
\begin{figure}
\begin{center}
\includegraphics[width=0.9\textwidth]{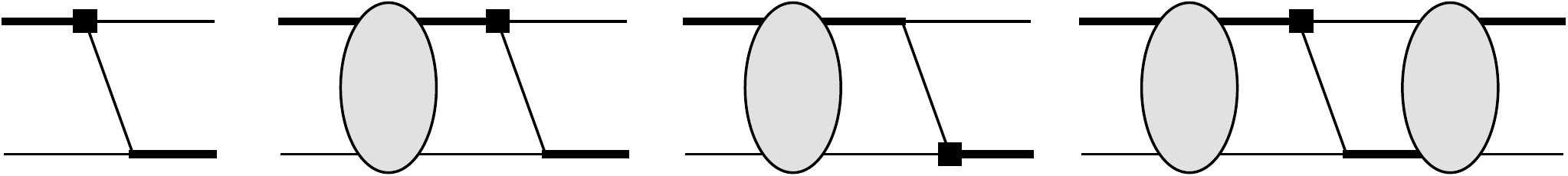}\caption{Diagrams contributing to PV $nd$ scattering in \eftnopi at LO. The square stands for a PV vertex, the thin lines  denote nucleons and the thick lines are dibaryons.  The grey ellipse is the PC LO $nd$ scattering amplitude. \label{fig:3PVndLO}}
\end{center}
\end{figure}
In Ref.~\cite{Vanasse:2011nd} the spin rotation angle was determined  at LO, corresponding to the diagrams in Fig.~\ref{fig:3PVndLO}, in which the grey ellipse stands for the PC LO $nd$ scattering amplitude. Adjusted to the conventions used in Eqs.~\eqref{yDeltaConv} of Sec.~\ref{nopimethod}, the results for the $c[(X - Y)](\Lambda)$ for $\Lambda = 200\,\MeV$ are given by \cite{VanassePhD,Vanasse:2011nd} 
\be\label{eq:ndrotLO}
\begin{aligned}
c[(\threeS - \oneP)](200\,\MeV) & = 10.4\; \text{rad}\, \text{MeV}^{-\half}, & c[(\threeS - \threePone)](200\,\MeV) & = 20.1\; \text{rad}\, \text{MeV}^{-\half},\\ 
c[\calT](200\,\MeV) & = 1.89\; \text{rad}\, \text{MeV}^{-\half}\ \ .
\end{aligned}
\ee
There is very little variation in these values as the cutoff is changed to 1500 MeV, an indication that the results are properly renormalized. Using an estimate of the (unknown) LECs $g^{(X-Y)}$ that is based on the DDH ``best values,'' Ref.~\cite{Vanasse:2011nd} finds a rotation angle of 1.8 $\times 10^{-6}$rad/m. This result is about twice as large as that of the hybrid and model calculations of Refs.~\cite{Schiavilla:2008ic,Song:2010sz} (see below). However, this comparison is based on the translation of the LECs to the DDH parameters. As explained in Sec.~\ref{sec:translate}, these relations are strongly regularization-scale dependent and do not afford more than order of magnitude estimates.

The numbers of Eq.~\eqref{eq:ndrotLO} are in agreement with the LO calculation of Ref.~\cite{Griesshammer:2011md}. In Ref.~\cite{Griesshammer:2011md}, the spin rotation angle is also calculated at one higher order (NLO). No new PV operators have to be taken into account, while NLO corrections to the dibaryon propagators, the 3N interaction, and the dibaryon wave function renormalization appear. For computational reasons, a ``partially re-summed'' approach is employed in Ref.~\cite{Griesshammer:2011md}, which introduces some contributions that are formally of higher order and thus of the order of the theoretical uncertainties.  Theoretical  errors are determined for each of the $c[(X - Y)](\Lambda)$ using the most conservative of three methods: (i) Based on naive power counting arguments, corrections are expected to be about 10 percent; (ii) variation of the cutoff over the range $200\,\MeV \le \Lambda \le 5000\,\MeV$;  (iii) use of different low-energy data to determine the PC LECs. The results are\footnote{The results here correct an error in the original values of Ref.~\cite{Griesshammer:2011md}.}\be\label{eq:ndrotNLO}
\frac{1}{\rho} \frac{\dd \phi}{\dd l} = \left([16\pm1.6]\gA + [34\pm 3.4] \gE + [4.6\pm1.0](3\gB-2\gC)\right) \text{rad}\ \MeV^{-\half}.
\ee
The numbers multiplying the PV couplings do not correspond to a fixed value of the cutoff $\Lambda$; instead they are obtained as the average of the highest and lowest values that are mapped out by varying the cutoff. 
For the coefficients of $\gA$ and $\gE$ the cutoff variation is smaller than the 10\% errors given here, while it is used to estimate the error in the case of $3\gB-2\gC$.
Note that while the size of the coefficients is comparable to that in the $np$ case (see Eq.~\eqref{eq:npspinrotdecomposed}),  the $nd$ spin rotation angle depends upon a different linear combination of the couplings and thus provides an independent constraint. The large shift from the LO results of Eq.~\eqref{eq:ndrotLO} to the NLO ones of Eq.~\eqref{eq:ndrotNLO} is caused by using the so-called Z-parameterization  \cite{Griesshammer:2004pe,Phillips:1999hh}, which takes into account the unnaturally large residue of the deuteron pole. The advantage of this method is that convergence at higher orders is improved, see, e.g.,  Refs.~\cite{Griesshammer:2004pe,Phillips:1999hh}. Using an order-of-magnitude estimate for the PV LECs based on naturalness arguments, a relation to PV $\vec{p}p$ scattering, or the DDH ``best values'' results in rotation angles of the order
\be
\left| \frac{\dd \phi}{\dd l} \right| \approx \left[10^{-7}\ldots10^{-6} \right] \frac{\text{rad}}{\text{m}},
\ee
in agreement with the DDH and hybrid calculations. A more detailed comparison with the hybrid and model results of Refs.~\cite{Schiavilla:2008ic,Song:2010sz} would again require the relation between the DDH parameters and the LECs, and as explained above the strong renormalization-scale dependence of this relation casts doubts on the usefulness of such a comparison.

\subsubsection{Hybrid and model results}

In Ref.~\cite{Schiavilla:2008ic}, both model (see Sec.~\ref{subsec:models}) and hybrid (see Sec.~\ref{subsec:hybrid}) techniques are used to determine the spin rotation angle. 
PC wave functions are calculated in a hyperspherical harmonics approach \cite{Kievsky:2008es} using the Argonne $v_{18}$ (AV18) potential with and without inclusion of the Urbana IX (UIX) three-nucleon potential. 
Parity-violating effects are included perturbatively, using either the DDH potential or a potential based on \eftnopi. 
In both cases, two different sets of regulator values are employed in the PV sector. 
For the DDH case, the result is dominated by pion exchange ($\sim 90\%$ of total). 
It is also largely independent of whether the 3N force is included as well as the different regulator values. 
Using two different sets of values for the DDH couplings, the spin rotation angle is estimated to be $\simeq 0.9\times 10^{-6}$ rad/m for a target density of $0.4\times10^{-23}$ atoms/$\text{cm}^{3}$. 
This rotation angle is about the same size as that found in the neutron-proton case (see Sec.~\ref{sec:npspinrot}).\footnote{At one stage it was believed that the $nd$ rotation angle was enhanced compared to that from $np$, but this turned out to be an erroneous result that was soon corrected. }
For the pionless EFT potential, these authors find that the contributions to the spin rotation angle corresponding to individual operators show a much larger dependence on the values of the regulators and are also more sensitive to the inclusion of the 3N force. 
Due to the lack of reliable determinations for the LECs, the authors of Ref.~\cite{Schiavilla:2008ic} refrain from making any estimates based on the \eftnopi potential.

The calculation of Ref.~\cite{Song:2010sz} is similar in spirit. 
Strong interactions are again taken into account using the AV18 with and without the UIX potentials, while wave functions are calculated by solving Faddeev equations in configuration space. 
In addition to the DDH and \eftnopi potentials, a potential derived from chiral EFT is also considered. 
Again two sets of regulators are employed for each PV potential. 
The results are in general agreement with those of Ref.~\cite{Schiavilla:2008ic}. 
There is only weak dependence on whether the three-nucleon force is included for the DDH case, with a stronger effect in the EFT calculations. 
The results for some of the coefficients multiplying the unknown PV LECs also vary significantly as the regulator is changed. 
This is particularly noteworthy for the \eftnopi interaction. 
The authors of Ref.~\cite{Song:2010sz} argue that this dependence can be absorbed by the running of the LECs. 
It is not clear, however, whether this running is an artifact of the hybrid approach and whether the renormalization group running of the PV LECs actually matches the regulator dependence found in Ref.~\cite{Song:2010sz}. 
Using the ``best values'' for the DDH parameters and the same target density as above, Ref.~\cite{Song:2010sz} finds a rotation angle of $\simeq 0.7\times 10^{-6}$ rad/m. 
However, using two parameter sets obtained from fits to experimental data \cite{Bowman:int07} the rotation angle is $\simeq -0.6\times 10^{-6}$ and $\simeq -0.8\times 10^{-6}$, respectively, and thus of opposite sign. 
Using these values, pion exchange is no longer the dominant contribution, and $\rho$ and $\omega$ exchange contributions are comparable. 
However, the values of Ref.~\cite{Bowman:int07} are encompassed in the DDH reasonable ranges and thus no definite prediction of either the size or sign of the spin rotation angle can be made.

\subsubsection{Experimental status}

No experimental results for the spin rotation of a polarized neutron beam in a deuterium target exist at the moment, but its measurement might be feasible in the future \cite{Snow}. This measurement would provide important experimental input into the determination of the PV LECs. Alternatively, if the LECs can be determined from a different set of observables, measurement of the neutron-deuteron spin rotation would provide an important check on the consistency of the above theories.

\subsection{Nucleon-deuteron scattering}

\subsubsection{\eftnopi results} 
 
The LO \eftnopi calculation of Ref.~\cite{Vanasse:2011nd} finds the longitudinal asymmetry to be 
\be
A_L^{\vec{n}d} = \left[124 \gA + 15.8 \gB - 10.5 \gC + 342 \gE \right] \MeV^\frac{3}{2}
\ee
at an energy of 15 keV and a cutoff $\Lambda = 200\, \MeV$, where we have adjusted the notation to the conventions used in this review. There is very little variation in the coefficients of the LECs with cutoff values from 200 MeV to 1500 MeV. Using the estimates based on the DDH parameters, the asymmetry is $2.2\times 10^{-8}$ at an energy of 15 keV. As a leading-order calculation, errors of roughly 30\% are to be expected. This result is again about twice as large as that of a model calculation in Ref.~\cite{Song:2010sz} (see below). However, the same caveats as in the case of spin rotation apply here, namely that there is a strong scale dependence in the connection between the DDH parameters and LECs as well as in the hybrid results.  Reference \cite{Vanasse:2011nd} also considers the asymmetry for an unpolarized neutron beam on a polarized deuteron target and finds 
\be
A_L^{n\vec{d}} = \left[-77.1 \gA - 14.3 \gB + 9.51 \gC + 516 \gE \right] \MeV^\frac{3}{2}
\ee
at 15 keV neutron energy and $\Lambda = 200\, \MeV$. The result is again practically independent of the cutoff value. Using the DDH-matched values of the LECs, the asymmetry is $A_L^{n\vec{d}}(15\,\text{keV}) = 4.0\times 10^{-8}$.

\subsubsection{Hybrid and model results}

In addition to the calculation of the neutron-deuteron spin rotation angle, Ref.~\cite{Song:2010sz} includes a discussion of the $nd$ longitudinal asymmetry. The AV18 potential with and without inclusion of the UIX three-nucleon potential is combined with the DDH potential and PV potentials derived from \eftnopi and chiral EFT. Again, for each of the PV potentials, two different sets of regularization parameters are used.  For the PV \eftnopi and the chiral EFT potentials some coefficients change significantly (one order of magnitude or more) depending on the choice of the regulator, while the DDH results are much less regulator dependent. Considering a neutron laboratory energy of 15 keV and the DDH ``best values," the asymmetry based on the DDH potential and including the UIX 3N interaction is $A_L \approx 0.8\times10^{-8}$, while the values from Ref.~\cite{Bowman:int07} give $A_L\approx-0.2\times10^{-8}$. Note again the change in sign for the observable.

There exist a number of earlier model calculations for the asymmetries in $\vec{n}d$ and $\vec{p}d$ scattering, see, e.g., Refs.~\cite{Kloet:1977wd,Henley:1977qn,Kloet:1983ta} and references therein.  The estimated asymmetries are on the order of a few times $10^{-8}$, with some results up to a few times $10^{-7}$.  

A first calculation of the neutron-deuteron scattering amplitude including PV interactions in terms of the Danilov parameters was performed in Ref.~\cite{Moskalev:1968a}. The strong interactions are treated in the zero-range approximation, and amplitudes are calculated through integral equations derived from a nonrelativistic Feynman diagram approach. The approximations used in the solutions to the integral equations are estimated to be valid at the order of 20-30\%. Only the general form of the scattering amplitude including PV interactions is derived, which in principle allows a determination of the longitudinal asymmetries by calculating the appropriate ratios of cross sections. No 3N forces are included in this calculation.

\subsubsection{Experimental status}
 
No measurement of the asymmetries in the scattering of polarized neutrons off deuterium has been performed to date. Given the estimated size of the asymmetries of a few times $10^{-8}$ it seems to be difficult to do so even in the near future. An upper limit for the longitudinal asymmetry in $\vec{p}d$ scattering has been found \cite{Nagle:1978vn}, but the theoretical interpretation of the data requires additional care (see, e.g., the  discussion in Ref.~\cite{Holstein:2009zzb}).

\subsection{Radiative capture and breakup reactions}

\subsubsection{Model and hybrid results}

There are three PV observables that can be studied in radiative capture of neutrons on a deuteron target, $nd \rightarrow t \gamma$: the circular photon polarization for unpolarized beam and target $P_\gamma$ ($nd\to t \vec{\gamma}$), 
as well as the angular asymmetries in photon direction for either polarized neutrons, $A_\gamma^n$ ($\vec{n}d\to t \gamma$), or polarized deuterons, $A_\gamma^d$ ($n\vec{d}\to t \gamma$).
At low energies these observables are proportional to different ratios of the $E1$ to $M1$ amplitudes.  (Each amplitude receives contributions from two channels because of capture into either the $J=\half$ or $J=\frac{3}{2}$ states.) Ref.~\cite{BlinStoyle1961395} pointed out that these observables might be enhanced compared to neutron capture on protons due to a suppression of the $M1$ amplitudes. This enhancement is, however, subject to the assumption that the $E1$ amplitudes are not similarly suppressed.

The $nd\to t\gamma$ observables are calculated in Ref.~\cite{Song:2012yx}, using the DDH and hybrid formalisms. For the (purely PC) $M1$ transitions the authors use the result of Ref.~\cite{Song:2008zf}, obtained by combining the meson exchange current derived in heavy baryon chiral perturbation theory up to \NthreeLO with a variety of PC potentials. In order to account for the mismatch in three-body binding energies (or equivalently, lack of three-body forces), the $M1$ transitions are fit as a function of the ratio of calculated and experimental triton binding energies, $B_{model}/B_{exp}$. By setting this ratio equal to one, the total cross section is in good agreement with the data. For calculation of the $E1$ amplitudes, both the positive and negative parity parts of the $nd$ and triton bound state wave functions are determined by solving Fadeev equations. The positive parity part only requires the PC potential. In order to study the model dependence of their results, the authors employ the AV18, Reid, Nijm II, and INOY (``Inside Nonlocal Outside Yukawa'') \cite{Doleschall:2003ha,Doleschall:2004bb} two-body potentials, and also consider the combination of AV18 with the UIX three-body force. The negative parity parts of the wave functions are  calculated using DDH, \eftnopi, and chiral EFT potentials, each of these in combination with two different regulators. The $E1$ amplitudes are then determined as the matrix elements of the one-body charge operator between opposite parity wave functions. 

Using the AV18+UIX and DDH potentials with the ``best values'' for the DDH meson-nucleon couplings, the three observables are of the order of several $10^{-7}$. 
However, the results show a strong dependence on the choice of (i) the values of the PV parameters, (ii) the PC potential, and (iii) the regulator values in the PV potential. 
The authors relate this to $J=\half$ contributions that are sensitive to short-range and three-body forces. In the case of \eftnopi PV interactions, the results for some of the coefficients multiplying the PV constants in the $E1$ amplitudes differ by more than a factor of 10 depending on which form of the regulator function is employed (for the same PC potential, AV18+UIX). 
This also holds, although less pronounced, at the level of the observables. Similar observations are made for the chiral EFT results. A more detailed analysis of particular PV operators performed in Ref.~\cite{Song:2012yx} shows that the results are also strongly dependent on the cutoff value.  While the authors argue that these differences can be absorbed in the renormalization of the PV couplings, the strong dependence indicates the need for a consistent implementation of the EFT formalism in all parts of the calculation.

In order to address a number of aspects   in a series of earlier model calculations, Ref.~\cite{Desplanques:1986cq} performed a calculation using the DDH potential in combination with various PC two-nucleon potentials, including the super-soft-core potential of Ref.~\cite{DeTourreil:1973uj} and the Malfliet-Tjon potential \cite{Malfliet:1970vd}.  In particular, the authors showed that neglecting  intermediate-state interactions (as was done in the earlier calculations \cite{Hadjimichael:1971ux,Hadjimichael19741}) underestimates the contributions to the $E1$ amplitudes. Using the ``best values'' for the DDH parameters the magnitudes of the three observables fall in the range $[0.8,1.6]\times 10^{-6}$.  This  calculation did not take into account 3N interactions and the 2N potentials employed underpredict the triton binding energy. 

In the Danilov approach, a calculation of the PV observables in $nd\to t\gamma$ was performed in Ref.~\cite{Moskalev:1968b}. Both PC and PV interactions are treated as zero-range interactions with the PV effects taken into account using the five S-P amplitudes. 
This calculation disagrees with the zero-range result of Ref.~\cite{Desplanques:1986cq}, which attributes the difference to errors in the $nd$ wave function normalization in Ref.~\cite{Moskalev:1968b} as well as in the deuteron polarization density matrix.

\subsubsection{Experimental status}

An experiment was performed to measure the angular asymmetry $A_\gamma^{n}$ in $\vec{n}d\to \HThree \gamma$  \cite{Avenier:1984is,Alberi:1988fd}, but the errors in the measurement are too large to claim a conclusive result.
$\vec{n}d\to \HThree \gamma$ experiments have been considered as future possibilities at neutron sources such as the SNS and NIST \cite{Snow}.

Other possible observables that could be studied in the future are asymmetries from the breakup reactions $\vec{\gamma}\, \HeThree \to pd$ or $\vec{\gamma} \,\HThree  \to nd$ using circularly polarized photons.  As in the two-body breakup reaction $\vec{\gamma}d\to np$, the requirements on the high-intensity source of photons are  stringent. Apparently, no theoretical determination of the asymmetry in \HeThree or \HThree breakup has been published. In addition to PV NN interactions, in the case of \HeThree such a calculation would also need to take into account Coulomb effects between the protons.

%%%%%%%%%%%%%%%%%%%%%%%%%%%%%%%%%%%%%%%%%%%%%%%%%%%%%%%%%%%%%

\section{Four- and five-nucleon systems}\label{sec:fourN}

There are a number of PV observables in four- and five-nucleon systems that are experimentally accessible. Early calculations of these observable relied on various models to describe the light nuclei. However, advancements in few-body calculations (see, e.g., Ref.~\cite{Leidemann:2012hr} for a recent review as well as references therein) now allow descriptions of these systems starting from 2N and 3N interactions without introducing any uncontrolled approximations. The reactions described below can be used to determine some of the 2N  PV couplings or, if the couplings are determined from two- and three-nucleon systems, they provide important consistency checks of our understanding of hadronic parity violation.

\subsection{Asymmetry from $\vec{n}  \HeThree \to \HThree p $}

In the charge-exchange reaction $\vec{n} + \HeThree \to \HThree + p$ the PV angular asymmetry $A_p$ (see Eq.~\eqref{Agamma} in Sec.~\ref{sec:obs}) is a measure of the correlation between the spin of the incoming neutron beam and the outgoing proton momentum ($\vec{\sigma}_n \cdot \vec{p}_p$). It can be written as
\be
A_p = \left(
\left.\frac{\dd\sigma}{\dd\Omega}\right\vert_\uparrow-\left.\frac{\dd\sigma}{\dd\Omega}\right\vert_\downarrow\right) \Big/ \left(
\left.\frac{\dd\sigma}{\dd\Omega}\right\vert_\uparrow+\left.\frac{\dd\sigma}{\dd\Omega}\right\vert_\downarrow\right)
\ee
where $\left.\frac{\dd\sigma}{\dd\Omega}\right\vert_{\uparrow/\downarrow}$ is the differential cross section for proton momentum parallel/antiparallel to the neutron spin. A calculation of $A_p$ using a number of different combinations of PC and PV potentials was presented in Ref.~\cite{Viviani:2010qt} (where the asymmetry $A_p$ is denoted by $a_p$).
For the PC sector, both two- and three-nucleon interactions were taken into account. The potentials used were either the phenomenological AV18 two-nucleon \cite{Wiringa:1994wb} and UIX three-nucleon \cite{Pudliner:1997ck} interactions, or were derived in chiral EFT up to \NthreeLO for  2N \cite{Entem:2003ft} and \NtwoLO for  3N interactions \cite{Navratil:2007zn,Gazit:2008ma}, referred to as \NthreeLO/\NtwoLO below. The inconsistency between the 2N and 3N interaction  level of precision provides a potential source of theoretical uncertainty. Parity-conserving transitions are calculated using the Kohn variational principle in the hyperspherical harmonics approach \cite{Kievsky:2008es}. The calculation accounts for both the elastic $n\HeThree$ and the $p\HThree$ channels. In the PV sector, both the DDH potential and a potential derived from \eftnopi are employed. These potentials are then evaluated in first-order perturbation theory between wave functions obtained from the PC potentials. 

The angular asymmetry $A_p$ is expressed as a linear combination of PV couplings, 
\begin{equation}
A_p = \sum_n c_n^\alpha I_n^\alpha,
\end{equation}
where $\alpha \in \{\text{DDH}, \text{EFT} \}$, the $c_n^\alpha$ are the various PV couplings appropriate for the choice of $\alpha$, and the coefficients $I_n^\alpha$ are determined numerically.

For the DDH potential, the authors of Ref.~\cite{Viviani:2010qt} consider four different PC potentials: two with only 2N interactions (AV18, \NthreeLO) and two including 3N interactions (AV18/UIX, \NthreeLO/\NtwoLO). In all cases, the PV pion-exchange contribution to $A_p$ is significantly larger than those from the PV vector-meson exchanges. For any choice of PC potential, the coefficients corresponding to isoscalar and isovector operators are roughly of the same size, while the coefficient of the isotensor structure is suppressed. However, for the PC EFT potential \NthreeLO/\NtwoLO the PV pion-exchange contribution is considerably smaller than for the other PC potentials. 
This discrepancy is possibly related to the fact that the wavefunctions for this potential are not fully converged in the  hyperspherical harmonics expansion. 
Varying the values of the DDH parameters within the allowed ranges provided in Ref.~\cite{Desplanques:1979hn}, the coefficient $A_p$ is predicted to fall in the range
\begin{equation}\label{eq:n3He-a}
-2.97 \times 10^{-7}  < A_p < 1.38\times 10^{-7}.
\end{equation}
Using only the ``best values'' for the DDH couplings this translates into asymmetries in the range
\begin{equation}\label{eq:n3He-Az}
-9.44 \times 10^{-8}  < A_p < -2.48\times 10^{-8}
\end{equation}
depending on which PC potential is employed. This strong model dependence of the observable is related to large cancellations between pion and vector-meson contributions. Since these cancellations depend on the exact values of the PV couplings, values that differ from the DDH ``best value'' set could result in an asymmetry that is larger than suggested by Eq.~\eqref{eq:n3He-Az}, as reflected in the ranges given in Eq.~\eqref{eq:n3He-a}.

For the PV EFT potential, the results for the $I_n^\text{EFT}$ differ significantly depending on whether the phenomenological AV18/UIX or the EFT \NthreeLO/\NtwoLO potentials are used. 
Most pronounced, the coefficient for the LEC $\calG_1$ (see Eq.~\eqref{STversion}) differs by an order of magnitude.  
However, in both cases the isotensor contribution is again suppressed. In these calculations, a Yukawa regulator function 
\begin{displaymath}
f(r)=\frac{1}{4\pi r}\text{e}^{-\mu_P r}
\end{displaymath}
is used in the PV potential. Since the PV potential is derived from pionless EFT the authors of Ref.~\cite{Viviani:2010qt} choose $\mu_P=m_\pi$. Note, however, that the PC potential does include pions as dynamical degrees of freedom, and the different choices for regulators in the two potentials exemplifies the resolution mismatch between the PC and PV sectors  in the hybrid approach of Ref.~\cite{Viviani:2010qt}.  As the authors point out, the results for $I_n^\text{EFT}$ depend strongly on the choice of $\mu_P$. Since the LECs in PV \eftnopi are not known, no prediction for the size of $A_p$ is made.

A different method to calculate the asymmetry in  $\vec{n} + \HeThree \to \HThree + p$ was used in Ref.~\cite{Gudkov:2010bq}. In order to avoid reference to a specific potential, a microscopic nuclear reaction theory approach \cite{Bunakov:1982is}  is used that takes advantage of well-measured excitation energies in \HeThree and \HeFour. The PV asymmetry is expressed in terms of matrix elements $\langle s^\prime l^\prime \vert R^J\vert s l\rangle$, where $sl$ ($s^\prime l^\prime$) are the spin and orbital angular momentum in the initial (final) channels, $J$ is the total spin, and the matrix $R$ is related to the $S$ and $T$ matrices through 
\be
R=2\pi i T = 1-S.
\ee
The PV effects are calculated in the distorted-wave Born approximation (DWBA) between wavefunctions of the PC Hamiltonian. Making the approximation that resonance contributions dominate these wavefunctions, and restricting the calculation to two resonances with opposite parity, the PV matrix elements for the neutron-proton reaction are given by 
\be\label{4N:Rmatrix}
\langle s^\prime l^\prime \vert R^J\vert s l\rangle = -\frac{i w\left[ \Gamma_l^n(s)\Gamma^p_{l^\prime}(s^\prime) \right]^\half}{(E-E_l+i \Gamma_l/2)(E-E_{l^\prime}+i\Gamma_{l^\prime}/2)}e^{i(\delta^n_l+\delta^p_l)},
\ee
where $E$ is the neutron energy, $E_k$ ($\Gamma_k$) the energy (total width) of the $k$-th resonance, and $\Gamma_k^i$ ($\delta_k^i$) the partial width (scattering phase) in channel $i$ of the $k$-th resonance. The parameter $w$ encodes the information on the PV interaction,
\be
w = - \int \phi_l W \phi_{l^\prime} d\tau,
\ee 
with $\phi_k$ the wave function of the $k$-th resonance and $W$ the PV potential. Instead of relying on a particular form of the potential $W$, Ref.~\cite{Gudkov:2010bq} uses an extrapolation of a statistical formula for weak matrix elements to the region of few-nucleon systems, $w\approx 0.5\, \text{eV}$. While the partial widths entering Eq.~\eqref{4N:Rmatrix} have been measured, the values and phases of the partial widths for particular spin channels that are required for PV observables are not known. Making certain approximations, Ref.~\cite{Gudkov:2010bq} estimates the PV asymmetry for thermal neutron energies as 
\be
-8\times 10^{-7} < A_p < -1\times 10^{-7},
\ee
which is in agreement with the calculation of Ref.~\cite{Viviani:2010qt}, see Eq.~\eqref{eq:n3He-Az}.

Reference~\cite{Gudkov:2010bq} also considers the PV longitudinal asymmetry that is related to the difference of total cross sections for neutrons with positive/negative helicities ($\vec\sigma_n\cdot \vec{p_n}$), 
\be
P=\frac{\sigma_+ - \sigma_-}{\sigma_+ + \sigma_-}\ \ .
\ee
The asymmetry $P$ of Ref.~\cite{Gudkov:2010bq}  corresponds to an observable of type $A_L$ in Eq.~\eqref{def:AL} of  Sec.~\ref{sec:obs}.
Again considering thermal neutrons, the longitudinal asymmetry lies in the the range
\be
-4\times 10^{-10} < P < -2\times 10^{-10}\ \ ,
\ee
much smaller than the asymmetry in the charge exchange reaction. However, the observable scales with the neutron momentum and is enhanced by several orders of magnitude for neutron energies in the MeV range.

An experiment to measure the asymmetry $A_p$ in $\vec{n} + \HeThree \to \HThree + p$ has been approved for the Fundamental Neutron Physics Beam Line at the SNS, see, e.g., Ref.~\cite{n3HeProp}. The result will provide a constraint on one linear combination of PV  couplings. Given the strong model dependence in the results of Eqs.~\eqref{eq:n3He-a} and \eqref{eq:n3He-Az},  a consistent calculation of $A_p$ treating both PC and PV potentials in the same EFT framework as suggested in Ref.~\cite{Viviani:2010qt} is desirable. It would also be interesting to perform a consistent calculation in \eftnopi. As discussed in Sec.~\ref{sec:fewbody}, few-nucleon systems at low energies can be accurately described without dynamical pion degrees of freedom, and a comparison with the chiral EFT result might allow a more detailed analysis of the importance of one-pion exchange contributions.

\subsection{$\vec{p} \, \HeFour$ scattering}

A calculation of  the longitudinal asymmetry $A_L$ ($\vec{\sigma}_p \cdot \vec{p}_p$, see Eq.~\eqref{def:AL}) in the scattering of polarized protons on a \HeFour target was performed in Ref.~\cite{Roser:1985rs} in terms of the DDH parameters. The PV scattering amplitude was calculated in the  DWBA by evaluating the DDH potential between $p\alpha$ scattering wave functions. The authors use a Gaussian bound state wave function for the alpha particle, and $\p\alpha$ interactions are described in an optical model, relying on scattering data wherever possible. In addition, short-range correlations are taken into account  through use of Jastrow factors. The authors find an asymmetry of
\begin{equation}
A_L(46\,\MeV) = -0.34 h_\pi^1+0.14 h_\rho^0+0.059 h_\omega^0 + 0.047 h_\rho^1 +0.059 h_\omega^1 + 0.009 {h_\rho^1}^\prime \ .
\end{equation}
A similar calculation was performed in Ref.~\cite{Flambaum:1985xu}. Besides use of a different Jastrow factor, the main difference is that Ref.~\cite{Flambaum:1985xu} only considered energies below 15 MeV as the authors argue that above these energies internal excitations of the $\alpha$ particle, which they do not take into account, become important. An earlier calculation of the asymmetry using the phenomenological potential of Desplanques and Missimer \cite{Desplanques:1976mt} can  be found in Ref.~\cite{Henley:1977qn}, while in Ref.~\cite{Avishai:1982qf} a PV nucleon-nucleus interaction is considered to estimate $A_L$.
A direct calculation of $A_L$ in terms of two- and three-nucleon interactions will provide an important improvement. Few-nucleon methods using EFT interactions continue to be developed, putting such a calculation within reach. 

A measurement of the longitudinal asymmetry at an energy of 46 \MeV \cite{Lang:1985jv} yielded
\begin{equation}
A_L(46\,\MeV) = (-3.3 \pm 0.9)\times 10^{-7} \ ,
\end{equation}
constraining the linear combination of PV couplings given above.

\subsection{$n \HeFour$ spin rotation}

The other observable considered in the five-nucleon sector is the neutron spin rotation in a \HeFour target. A calculation using the DDH model for the PV interaction combined with wave functions obtained from an optical potential was performed in Ref.~\cite{Dmitriev:1983mg}. Expressed in terms of the DDH couplings, the rotation angle per unit length is given by 
\begin{equation}
\frac{\dd \phi}{\dd l} = -\left( 0.97 h_\pi^1+ 0.22 h_\omega^0 - 0.22 h_\omega^1 +0.32 h_\rho^0 -0.11 h_\rho^1 -0.02 h_\rho^{\prime 1} \right) \,\frac{\text{rad}}{\text{m}}\ \ ,
\end{equation}
with helium density $\rho = 0.19\times 10^{23}\,\text{cm}^{-3}$.
Using the ``best values'' for the DDH coupling yields
\be
\frac{\dd \phi}{\dd l} =  -0.1\times 10^{-6}\,\frac{\text{rad}}{\text{m}}.
\ee
However, for these particular values a strong cancellation between pion and $\rho$-meson contributions occurs. Considering the preferred ranges results in 
\be
\label{45:spinrotrange}
-1.6  \times 10^{-6}\,\frac{\text{rad}}{\text{m}} < \frac{\dd \phi}{\dd l} <  1.2\times 10^{-6} \,\frac{\text{rad}}{\text{m}}.
\ee
Note that this is roughly of the same order of magnitude as the spin rotation angles in hydrogen and deuterium. It is also in agreement with the earlier estimate of Ref.~\cite{Avishai:1982qf},
\be
\frac{\dd \phi}{\dd l} =  0.73\times 10^{-6}\,\frac{\text{rad}}{\text{m}},
\ee 
which is based on a simple model of a PV neutron-$\HeFour$ force.

The latest measurement at NIST gives the result \cite{Snow:2011zza}
\begin{equation}
\frac{\dd \phi}{\dd l} = \left[ +1.7 \pm 9.1\,(\text{stat.}) \pm 1.4\,(\text{sys.}) \right]\times 10^{-7} \,\frac{\text{rad}}{\text{m}}.
\end{equation}
While the statistical uncertainty seems large, it is smaller than the range of values of Eq.~\eqref{45:spinrotrange}. There are current plans for a second phase of the measurement, which would reduce statistical uncertainties to $2\times10^{-7}$ and also reduce systematic uncertainties \cite{Snow:2011zza}.

%%%%%%%%%%%%%%%%%%%%%%%%%%%%%%%%%%%%%%%%%%%%%%%%%%%%%%%%%%%%%

\section{Summary and Outlook}\label{concl}

In this review we have collected and discussed calculations of parity-violating (PV) processes involving one, two, three, and (briefly) four or five nucleons.  
Because the weak interactions that give rise to PV effects are of such short distance compared to typical hadronic scales, we take the point of view that PV  even in large nuclei is built upon the PV interactions among few nucleons.  
Many-body effects will be necessary to predict the manifestation of PV in a heavy nucleus, but an understanding of few-body PV interactions will also be required.  
To encourage progress towards the latter goal, we detail the state of the present theoretical landscape in the treatment of few-body PV  calculations.  Our focus is on those theories that have the potential to systematically and consistently encompass {\sl all} the PV observables that involve one through five nucleons.  

Before discussing particular results, we first gave an overview of the type of PV asymmetries being considered.  We then discussed aspects of the theoretical tools that have been used to understand these asymmetries, providing detail on the relevant degrees of freedoms and assumptions made.  
This was followed by a presentation of existing calculations for systems containing one through five bodies.  Our clear bias is towards consistent effective field theory (EFT) descriptions, since we believe these have the greatest prospect of unifying our understanding of nuclear PV observables.  

Further progress will require both theoretical and experimental developments. At the moment, EFT calculations of all one- and two-nucleon PV observables are available for existing or feasible one- and two-nucleon experiments.   The three-nucleon system including PV effects has also been ``solved" in EFT, in the sense that major hurdles such as including Coulomb effects and dealing with numerical solutions to the relevant coupled equations have been successfully addressed.   But there are still calculations that remain to be completed.  In four- and five-nucleon systems the calculations are still in the "hybrid" stage, an intermediate step towards a fully consistent EFT treatment.  
However, recent theoretical developments show that the EFT approach can also be successfully applied in four- and five-nucleon systems (and even beyond).

On the experimental side, the development of high-intensity neutron and photon sources provide opportunities to increase the accuracy of existing measurement and to access new observables. 
Since these experiments are very challenging, reliable and accurate theoretical support and guidance play a major role in deciding which observables to address.
We claim that EFTs are best-suited for this task: there are predictable relationships between different observables through the low-energy constants, and the power counting provides a reliable estimate of theoretical errors.

Another important recent development is the application of lattice QCD to the determination of the PV couplings. Lattice QCD has the potential to predict the relevent couplings at the level of Standard Model degrees of freedom, establishing a direct link to the well-tested weak quark-quark interactions.
In combination with EFT calculations in few-body systems, their possible extensions to heavier systems, as well as recent and future experimental developments, we have the opportunity to make considerable progress in our understanding of hadronic parity violation.

\begin{table}[tbp]
\begin{center}
\begin{tabular}{|l|l|l|c|}
\hline
{} & Nonzero result & Upper limit & EFT calculation\\
\hline
$\vec{p}p$ scattering & \checkmark \cite{Eversheim:1991tg,Kistryn:1987tq} & &\checkmark \cite{Phillips:2008hn}\\
\hline
$\vec{n}p$ spin rotation &  & &  \checkmark  \cite{Griesshammer:2011md}\\
\hline
$\vec{n}p\to d\gamma$ & & \checkmark \cite{Cavaignac:1977uk,Gericke:2011zz} &  \hspace{1.1em} \checkmark  \cite{Kaplan:1998xi,Savage:2000iv,Schindler:2009wd}\\
\hline
$np\to d\vec{\gamma}$ & & \checkmark \cite{Knyazkov:1984zz} &  \checkmark \cite{Schindler:2009wd}\\
\hline
$\vec{\gamma} d\to np$ & & \checkmark  \cite{Alberi:1988fd,Earle:1988fc}&  \checkmark \cite{Schindler:2009wd}\\
\hline
$\vec{n}d$ spin rotation & &  & \checkmark  \cite{Griesshammer:2011md}\\
\hline
$\vec{p}d$ scattering &  & \checkmark \cite{Nagle:1978vn} & \\
\hline
$\vec{n}d\to t\gamma$ & &\checkmark  \cite{Avenier:1984is,Alberi:1988fd}  & \\
\hline
$\vec{\gamma}\,{}^3\text{He}\to pd$ & &  & \\
\hline
$\vec{n}\,{}^3\text{He}\to p\, {}^3\text{H}$ & &  &  \\
\hline
$\vec{n}\,{}^3\text{He}\to {}^4\text{He}\gamma$ & &   &  \\
\hline
$\vec{n}\,{}^4\text{He}$ spin rotation &  & \checkmark \cite{Snow:2011zza} & \\
\hline
$\vec{p}\,{}^4\text{He}$ scattering & \checkmark \cite{Lang:1985jv}& & \\
\hline
\end{tabular}
\end{center}
\caption{Few-nucleon experiments on hadronic parity violation and corresponding EFT calculations. }
\label{tab:FewBodyExp}
\end{table}

%%%%%%%%%%%%%%%%%%%%%%%%%%%%%%%%%%%%%%%%%%%%%%%%%%%%%%%%%%%%%

\section{Acknowledgements}

We are grateful for the help, advice, and encouragement of our many PV
friends and colleagues.  Special thanks go to J.~Vanasse for his work on
translating PV LEC parameters from one formalism to another and for providing
details on his three-body calculations.  We thank J.~D.~Bowman and W.~M.~Snow for information on recent and expected experimental results.
We are grateful to H.~W.~Grie{\ss}hammer, V.~Gudkov, and D.~R.~Phillips  for many stimulating discussions and for their feedback on various sections of this review. 
Discussions with W.~Detmold and J.~Wasem on the latest results and prospects from the lattice community are gratefully acknowledged.

MRS acknowledges the hospitality of the Lattice and Effective Field Theory group at Duke University. RPS is supported by DOE grant DE-FG02-05ER41368.

\bibliographystyle{elsarticle-num}
\bibliography{parity,EFT,Nuclear,rps}

\end{document}